\documentclass[12pt, psfig]{report}

\newtheorem{thm}{Theorem}
\newtheorem{crl}{Corollary}
\def\qed{$\Box$}
\newtheorem{defni}{Definition}
\newcommand{\ou}{\rm ~OR~}
\newcommand{\og}{\rm ~AND~}
\newcommand{\ouh}{\rm ~OR}

\usepackage{latexsym}

\newcommand{\be}{\begin{equation}}
\newcommand{\ee}{\end{equation}}

\newcommand{\bex}{\begin{eqnarray}}
\newcommand{\eex}{\end{eqnarray}}
\newcommand{\bmin}{\begin{center}\begin{minipage}{460pt}}
\newcommand{\emin}{\end{minipage}\end{center}}

\usepackage{fontenc,amsfonts,amsmath}
\usepackage[latin1]{inputenc}
\usepackage{graphics}

\everyjob{\typeout{MuTeX.sty: LaTeX Modification by EPM.}}
\immediate\write10{MuTeX.sty: LaTeX Modification by EPM.}

\makeatletter   

\input{ifthen.sty}

\newtheorem {Canham threshold} [theorem] {Canham Threshold}

\def\theoremstyle#1#2{\def\@@theoremheadstyle{#1}
                      \def\@@theorembodystyle{#2}}
\def\@@theoremheadstyle{\sc}
\def\@@theorembodystyle{\rm}
\def\@begintheorem#1#2{\@@theorembodystyle 
                       \trivlist 
		       \item[\hskip 
                             \labelsep{\@@theoremheadstyle #1\ #2}]}
\def\@opargbegintheorem#1#2#3{\@@theorembodystyle 
                              \trivlist 
			       \item[\hskip 
				  \labelsep{\@@theoremheadstyle #1\ #2\ (#3)}]}

 \def\@@pc{\bf}

 \newcommand {\pcodestyle}[1] {\def\@@pc{#1}}  

 {
 \def\PROGRAM		{{\@@pc program\ }}
 
 \def\PROCEDURE		{{\@@pc procedure\ }}
 \def\FUNCTION		{{\@@pc function\ }}
 \def\LOCAL		{{\@@pc local\ }}
 \def\GLOBAL		{{\@@pc global\ }}
 \def\RETURNS		{{\@@pc returns\ }}
 \def\RETURN		{{\@@pc return\ }}
 \def\BEGIN		{{\@@pc begin\ }}
 \def\END		{{\@@pc end\ }}
 \def\IF			{{\@@pc if\ }}
 \def\THEN		{{\@@pc then\ }}
 \def\ELSE		{{\@@pc else\ }}
 \def\REPEAT		{{\@@pc repeat\ }}
 \def\UNTIL		{{\@@pc until\ }}
 \def\WHILE		{{\@@pc while\ }}
 \def\DO			{{\@@pc do\ }}
 \def\FOR		{{\@@pc for\ }}
 \def\TO			{{\@@pc to\ }}
 \def\DOWN		{{\@@pc down\ }}
 \def\NEXT		{{\@@pc next\ }}
 \begin{center}
  \begin{minipage}{.9\textwidth}
   \small
    \begin{tabbing}
 }%
 {  \end{tabbing}%
   \end{minipage}%
  \end{center}%
 }

			   {\end{quotation}}

			 {  \end{minipage}%
			   \end{center}%
			  \end{description}%
			 }

				{ \end{minipage}
				 \end{center}%
				}

\def\thebibliography#1{\section*{References}\list
 {[\arabic{enumi}]}{\settowidth\labelwidth{[#1]}\leftmargin\labelwidth
 \advance\leftmargin\labelsep
 \usecounter{enumi}}
 \def\newblock{\hskip .11em plus .33em minus -.07em}
 \sloppy
 \sfcode`\.=1000\relax}

\newsavebox{\ProofSym}
\savebox{\ProofSym}{%
  \begin{picture}(10,10)
    \put(0,0){\framebox(9,9){}}
    \put(0,3){\framebox(6,6){}}
  \end{picture}}
\newcommand{\eop}{\hfill\usebox{\ProofSym}}

\begin{document}

\title{Combinatorial Approaches in Quantum Information Theory }
\author{Sudhir Kumar Singh \\
9925209\\
Mathematics and Computing \\
Department of Mathematics\\
Indian Institute of Technology, Kharagpur, 721302, India\\
\and Masters Thesis under the supervision of\\
\and Prof. Sudebkumar Prasant Pal and  Prof. Somesh Kumar}

\date{}
\maketitle


%

\begin{center}
CERTIFICATE  \\
\end{center}

This is to certify that the dissertation entitled
{\it "Combinatorial Approaches in Quantum Information theory"}, 
submitted by Mr. Sudhir Kumar Singh to the Department of Mathematics,
 Indian Institute of Technology, Kharagpur, 
in partial fulfillment of the requirements of the degree of 
{\it Master of Science} in {\it Mathematics and Computing}, 
is an authentic record of the work carried out by 
him under our supervision and guidance.\\

In my opinion, this work fulfills the requirements 
for which it has been submitted and has not been 
submitted to any other Institution for any degree.\\

Dr. Sudebkumar Prasant Pal \\

Dr. Somesh Kumar

\newpage

\begin{center}
{\it Acknowledgements}
\end{center}

This dissertation, for all its presentation, owes a lot to many people,
first and foremost Prof. Somesh Kumar and Prof. Sudebkumar Prasant Pal
being my guides, to whom I owe a debt of gratitude for their perseverance
in guiding me throughout the course of this work and entertaining my crazy ideas.
Their perceptual inspiration, encouragement and understanding have been a main 
stay of this work. Needless to say, they have also been 
helpful to me on various non-academic problems.
Particularly,  Prof. S. P. Pal has been a true care taker of mine at IIT KGP, 
to whom I can talk even my personal problems and 
who can take a great pain to help me combat the problems.     \\

My work is solely dedicated to the great teacher Prof. K. R. Parthasarathy, 
ISI Delhi, who taught me flying in the Hilbert Spaces and playing with quantum 
information. It was during summer 2002 in his guidance at IMSc Chennai when my 
interest towards the topic and in general towards research blossomed to the fullest. \\

Thanks are also due to Prof. Partha Ghose and Dr. M. K. Samal, SNBose National Centre
for Basis Sciences, Culcutta and Prof. R. Simon, IMSc Chennai. Interactions with
them really enhanced my knowledge and understanding of quantum information. 
I am indebted to Prof. Anil Kumar and his group, SIF, IISc Bangalore who 
introduced me to the practical aspects of quantum information especially 
using NMR as a quantum computer, during summer 2003. \\

My deepest sense of gratitude is to Dr. R. Srikanth, RRI Bangalore for patiently being
one of my most interacting collaborator in research on quantum information.
I would always be grateful to him for being a good friend as well as an able guide 
during my stay at IISc Bangalore during Summer 2003 and thereafter.    
A significant part of this work has been done in his collaboration.\\

I would also like to thank Prof. P. Panigrahi for generously giving me 
lots of time on stimulating discussions on combinatorics in as much 
I could discuss with my guides.  
I am also thankful to the faculty members and students from various departments 
who attended my lectures on quantum computing at Centre for Theoretical Studies, 
IIT Kharagpur in spring 2003 and provided an 
interactive audience. 
I also thank Prof. S. P. Khastigir, Prof. V. K. Jain, Prof. A. Das,
K. Mitra, S. Brahma, S. Macharala and M. Kumbhat for discussions.
I would also thank Prof. Andrew Yao, Prof. Michele Mosca, Dr. Peter Shor, 
Dr. J. Mueller-Quade and Prof. V. P. Roychowdhury for providing me with 
useful references. \\

Finally, I wish to express my deepest sense of gratitude to my Babujee, 
Chacha jee, Maa, Mausi, Deedee, Gopal and Kanhaiya Bhaiya, Ankit, Sonu,
Prasun, Sunidhi and many friends for their contributions 
which are completely beyond the scope of this thesis.\\

\bigskip

\bigskip

Sudhir Kumar Singh

\newpage

\begin{center}
Publications \\
\end{center}

\begin{enumerate}

\item Sudhir Kumar Singh, Somesh Kumar and Sudebkumar Prasant Pal, 
{\it Characterizing the combinatorics of distributed EPR 
pairs for multi-partite entanglement}, 
eprint quant-ph/0306049 (communicated for publication).\\

\item Sudhir Kumar Singh and R. Srikanth, 
{\it Unconditionally Secure Multipartite Quantum Key Distribution}, 
eprint quant-ph/0306118 (communicated for publication). \\

\item Sudhir Kumar Singh and R Srikanth, 
{\it Generalized Quantum Secret Sharing}, 
eprint quant-ph/0307200 (communicated for publication). \\

\item Sudhir Kumar Singh, Sudebkumar Prasant Pal, Somesh Kumar, and R Srikanth,
{\it A combinatorial approach to study the LOCC transformations of multipartite states},
manuscript under preparation.

\end{enumerate} 	

\begin{abstract}
Quantum entanglement is one of the most remarkable aspects 
of quantum physics. If two particles are in a entangled state, then,
even if the particles are physically separated by a great distance,
they behave in some repects as a single entity. 
Entanglement is a key resource for quantum information processing and
spatially separated entangled pairs of particles have been
used for numerous purposes such as teleportation, 
superdense coding and cryptography based on Bell's 
theorem. \\

Just as two distant particles could be entangled,
it is also possible to entangle three or more separated particles.
A well-known application of 
multipartite entanglement is in testing nonlocality from different 
directions.
Recently, it has also been used for many multi-party computation and 
communication tasks and multi-party cryptography.
One of the major issues in dealing with multi-partite
entangled states is of purification. 
Distilling pure maximally entangled state in this case may not be 
as simple as that of bipartite case. 
But if it is possible to create multi-partite entangled 
states from the bipartite ones then we can first distill pure 
maximal bipartite states and can then prepare the multi-partite ones. \\

To this end, we consider the problem of creating maximally
entangled multi-partite states out of Bell pairs distributed in a
communication network from a physical as well as from a 
combinatorial perspective. We investigate the minimal combinatorics 
of Bell pairs distribution 
required for this purpose and discuss how this combinatorics gives rise to 
resource minimization for practical implementations. 
We present two protocols for this purpose. 
The first protocol enables to prepare
a GHZ state using two Bell pairs shared amongst the three users 
with help of two cbits of communication and local operations.
The protocol involes all the three users dynamically and thus can
find applications in cryptographic tasks. Second protocol 
entails the use of $O(n)$ cbits of communication and local
operations to prepare an $n$ partite maximally entangled state in a
distributed network of bell pairs along a {\it spanning tree} of
{ \it EPR graph} of the $n$ users. We show that this spanning 
tree structure is the minimal combinatorial requirement. 
We also characterize the minimal combinatorics 
of agents in the creation of pure maximal multi-partite 
entanglement amongst the set $N$ of $n$ agents in a network 
using apriori multi-partite entanglement states amongst subsets of 
$N$. \\

Another major and interesting issue is of quantifying 
multi-partite entangled states. Multi-partite entangled states,
unlike the bipartite ones, lack convenient mathematical properties like Schmidt decompostion
and therefore it becomes difficult to characterize them. 
Some approches, essentially using the generalization of Schmidt decomposition,  
have been taken in this direction
; however a general formulation in this case is still an outstanding 
unresolved problem.
State transformations under local operations and classical communication (LOCC)
are very important while quantifying entanglement because LOCC can at the best 
increase classical correlations and therefore a good measure of entanglement is not
supposed to increase under LOCC.
All the current approaches to study the state transformation under LOCC are 
based on entropic criterion. We present an 
entirely different approach  based on  nice combinatorial properties of 
graphs and set systems. We introduce a technique called {\it bicolored merging} 
and obtain several results about  such transformations.
We demostrate a partial ordering of multi-partite states and 
various classes of incomparable multi-partite states. 
We utilize these results to establish the 
impossibility of doing {\it selective teleportation} in a case
where the apriori entanglement is in the form of a GHZ state.
We also discuss the minimum number of copies of a state required to prepare another 
state by LOCC and present bounds on this number in terms of {\it quantum distance} between
the two states.
The ideas developed in this work continues the combinatorial setting 
mentioned above  
and can been extended to incorporate other new kinds of multi-partite states.
Moreover, the idea of {\it bicolored merging} may also be appropriate to 
some other areas of information sciences.  \\ 

Key distribution is a fundamental problem in secure communication and 
quantum key distribution (QKD) protocols for key distribution
between two parties on the account of quantum uncertainty and no-cloning
principles was realized two decades ago, however the more rigorous and
comprehensive proofs of this task, 
 taking into consideration source, device and channel noise
as well as an arbitrarily powerful eavesdropper,
have been only recently studied by various authors.
We consider QKD between two parties
extended to that between $n$ trustful parties, that is, how the $n$ parties may share
an identical secret key among themselves. We propose a protocol for this purpose
and prove its unconditional security.
The protocol is simple in the sense that the proof of its security is
established on the basis of the already proven security of the bipartite case. 
Our protocol works in two broad steps. In
the first step, the $n$-partite problem is reduced to a two-party problem.
In the second step, the Lo-Chau protocol or Modified
Lo-Chau protocol is invoked to prove the unconditional security
of sharing nearly perfect EPR pairs between two parties. 
The first step essentially utilizes the spanning tree combinatorics mentioned above. \\

An other interesting aspect of multi-party cryptography is to split information 
through secret sharing and splitting the quantum information has also been 
studied recently. In conventional quantum secret sharing (QSS) schemes, 
it is often implicitly assumed that 
all share-holders carry quantum information. This requirement can be relaxed,
whereby some share holders may carry only classical information and no
quantum information. 
Such a hybrid (classical-quantum) QSS that combines classical and
quantum secret sharing
brings a significant improvement to the
implementation of QSS, in as much as quantum information is much more fragile than
classical information.  
The practical implementation of a quantum secret sharing scheme is facilitated if 
it can be compressed to an equivalent scheme with fewer quantum information carrying players,
 the reduction compensated by players who carry only classical information. Conversely, 
a given quantum secret sharing scheme may be inflated by adding only classical information 
carrying players. 
To this end, 
we explore some generalizations of such quantum secret sharing (QSS).
We study an extended QSS scheme, wherein some shares may be retained by
the share dealer, that enables the construction of access structures with disjoint
authorized sets, similar to classical secret sharing.
We also propose a hybrid (classical-quantum) generalization of the
threshold scheme.
Our schemes are based on two interesting ideas. First one is the idea of qubit encryption
and encryption key as the relevant classical information. 
However, in principle any classical data
whose suppression leads to maximal ignorance of the secret
is also good. The second, while
more restricted, is interesting
because it is not directly based on quantum erasure correction, but on
information dilution via homogenization, in contrast to current proposals of QSS. \\

QKD involves sharing a random key amongst trustworthy parties 
where as QSS splits quantum information amongst untrusted parties. 
We discuss situations where some kind of mutual
trust may be present between sets of parties while parties being individually mistrustful.
This way we take a step towards combining the essential features of 
QKD and QSS. We discuss two problems where this idea is applicable.
The first problem is of secure key distribution between two trustful groups
where the invidual group members may be mistrustful. The two groups 
retrieve the secure key string, only if all members should cooperate with one
another in each group. In the second case, 
we consider several such groups. Members of the same group
trust each other whereas members from different groups do not and the problem is to establish
a common shared random key amongst the $n$ untrustful parties. 
We present protocols for these cases and discuss the proof of their unconditional security.
Finally, we conclude brifly with some open research directions based on our research.

\end{abstract} 

\tableofcontents{}

\chapter{Introduction}

\section{ Stepping in to the Quantum World}

{ \it The most incomprehensible thing about the world is that it is 
comprehensible! --- Albert Einstein} \\

{ \it I am not happy with all the analyses that go with 
just classical theory, because 
Nature isn't classical, dammit, and if you want to
make a simulation of Nature,
you'd better make it quantum mechanical, and by golly 
it's a wonderful problem! 
--- Richard Feynman } \\

This wonderful observation of Feynman in the early 
1980's \cite {feynman82, feynman96} that no classical 
computer could simulate quantum mechanical systems without 
incurring exponential slowdown but it
might be possible provided the simulator was itself 
quantum mechanical, stemmed the widespread
interest in the field of quantum computation and quantum information. 
There are three
very important features of quantum mechanics which provides us 
the way to exploit such
powerful processors:
\begin{enumerate}
\item A quantum particle can exist simultaneously in many 
incompatible states.

\item We can operate on a quantum particle while it is in a 
superimposed state and
affect all the states at once.

\item One quantum system can influence another far 
away quantum system instantaneously

\end{enumerate}

This is the kind of parallelism inherent in quantum world and 
combined with the present
day information processing gives birth to exciting developments: 
Quantum computation,
quantum error correction, quantum entanglement  and 
teleportation and quantum
cryptography. Quantum computers are able to solve some problems 
intractable to
conventional computation (problems like prime factorization 
and discrete logarithm).
Quantum error correcting techniques enable us to do quantum 
computation and
communication in present of noise. 
Quantum cryptosystems provide guaranteed secure
communication using no-cloning theorem and 
uncertainty principle and quantum
teleportation provides the way to do quantum 
communication in absence of a quantum
channel using prior quantum entanglement and classical communication.

\section{ Quantum Mechanical Model of Computation }

In any model of information processing there are at least 
three requirements:
\begin{enumerate}

\item The representation of the information

\item The operations to be applied on the information

\item The way for extracting result after the operation

\end{enumerate}

In the quantum mechanical model of information processing 
the information is 
mathematically represented by a ray
in a Hilbert Space, operations are the unitary operators 
in the space and result is the
measurement of an observable described by a hermitian 
operator in the space.

\subsection{ Representing Quantum Information: Qubits and 
Quantum Registers}

The first postulate of quantum mechanics sets up the arena in which quantum
mechanics takes place. The arena is our familiar friend from linear algebra,
Hilbert space. \\

{\it Postulate 1:}  
Associated to any isolated physical system is a Hilbert space
known as the {\it state space} of the system. The system is completely described by its 
{ \it state vector}, which is a unit (ray) in the systems's state space. \\

The simplest non-trivial Hilbert space is of dimension two and a state vector in the
state space of dimension two is called a { \it qubit} (stands for a quantum bit). 
Suppose $|0\rangle$ and $|1\rangle$ form an orthonormal basis for that state space.
Then an arbitrary state vector in the state space can be written 

$|\psi \rangle = a|0\rangle + b|1\rangle $,

where a and b are complex numbers. The condition that $|\psi \rangle$
be a unit (ray), $< \psi | \psi > = 1$, is therefore equivalent to
$|a|^2 + |b|^2 = 1$.   \\

Nature is not so simple and a qubit is not sufficient to deal with its complexity.
We must be interested in composite system made of two (or more) distinct 
physical systems and we must also have a mathematical way of playing around
with them. In analogy to the classical terminology, a composite quantum
system i.e. a set of qubits is called a { \it quantum register}.
The following postulate describes how the state space of a 
composite system (quantum register)is built up from the state spaces 
of the component systems (the qubits). \\

{\it Postulate 2:} 
The state space of a composite physical system is the tensor product 
of the state spaces of the component physical systems. Moreover, if we 
have systems numbered $1$ through $n$, and a number $i$ is prepared in the 
state $|\psi_{i} \rangle $, then the joint state of the total system is 
$|\psi_{1} \rangle \bigotimes |\psi_{2} \rangle \bigotimes \ldots \bigotimes 
|\psi_{n} \rangle$, where $ \bigotimes$ denotes tensor product. 

\subsection{ Unitary Evolution: Quantum Gates}

Time evolution of a quantum state is unitary; it is generated by 
a self-adjoint (Hermitian) operator, called the {\it Hamiltonian} of the system.
In the { \it Schrodinger picture} of dynamics, the vector describing 
the system moves in time as governed by the { \it Schrodinger equantion}

$\frac{d}{dt} |\psi (t) \rangle = - i \bf{H} |\psi (t) \rangle $

where $\bf{H}$ is the Hamiltonian. We may express this equation, to 
first order in the infinitesimal quantity $dt$, as 

$|\psi (t+dt) \rangle = (1- i \bf{H} dt)|\psi (t) \rangle $.
  
Clearly, the operator $\bf{U}(dt) \equiv 1- i \bf{H} dt$ is unitary.
Thus the time evolution over a finite interval is unitary given by

 $|\psi (t) \rangle = \bf{U}(t)|\psi (0) \rangle $. \\

{\it Postulate 3:}
The evolution of a { \it closed } quantum system is described by a { \it unitary
transformation}. That is, the state $|\psi \rangle $ of the system at time $t_{1}$ 
is related to the state $|\psi^{'} \rangle $ of the system at time $t_{2}$ by a unitary 
operator $U$ which depends only on the times $t_{1}$ and $t_{2}$,

  $|\psi^{'} \rangle  = U|\psi \rangle $.

Now that a quantum system evolves according to a unitary operator which is 
always invertible, {\it quantum gates } must be reversible. Infact, quantum 
gates are nothing but these unitary operations. Following are some commonly
used one qubit and two qubit gates in terms of their unitary operations represented
by matrices in the computational basis.

{ \it Pauli's Gates (Operators) :}

$\bf{X} \equiv $ $$\bordermatrix{\text{}&|0\rangle &|1\rangle \cr
                |0\rangle & 0 &  1\cr
                |1\rangle & 1 &  0}$$

$\bf{Y} \equiv $ $$\bordermatrix{\text{}&|0\rangle &|1\rangle \cr
                |0\rangle & 0 &  -i\cr
                |1\rangle & i &  0}$$

$\bf{Z} \equiv $ $$\bordermatrix{\text{}&|0\rangle &|1\rangle \cr
                |0\rangle & 1 &  0\cr
                |1\rangle & 0 &  -1}$$

{ \it Hadamard Gate :}

$\bf{H} \equiv $ $$\bordermatrix{\text{}&|0\rangle &|1\rangle \cr
                |0\rangle & 1/ \sqrt{2} &  1/ \sqrt{2}\cr
                |1\rangle & 1/ \sqrt{2} &  -1/ \sqrt{2}}$$

{ Controlled- NOT (CNOT) Gate: A two qubit gate} 

$$\bordermatrix{\text{}&|00\rangle &|01\rangle &|10\rangle &|11\rangle \cr
                |00\rangle & 1 & 0 & 0 & 0 \cr
                |01\rangle & 0 & 1 & 0 & 0 \cr
                |10\rangle & 0 & 0 & 0 & 1 \cr
                |11\rangle & 0 & 0 & 1 & 0 }$$ \\

The Postulate $3$ requires that the system being described be closed.
That is, it is not interactive in any way with other systems.
In reality, of course, all systems (except the Universe as a whole) 
interact at least somewhat with the other systems. 
Nevertheless, there are interesting systems which can be described 
by unitary evolution to some good approximation.
Furthermore, at least in principle every open system can be described 
as part of a larger closed system (the Universe) which is undergoing 
unitary evolution.

\subsection{ Quantum Measurement: Observables }

An observable is a property of a physical system that in principle
can be measured. In quantum mechanics, an observable is
a hermitian operator. We also know that a hermitian operator
in a Hilbert space $ \bf{H}$ has a spectral decomposition- it's eigenstates
form a complete orthonormal basis in $ \bf{H}$. We can express a hermitian 
operetor $\bf{A}$ as 

$ \bf{A} = \sum_{n} a_n \bf{P}_n$ . 

Here each $a_n$ is an eigen value of $\bf{A}$, and $\bf{P}_n$ is the 
corresponding orthogonal projection onto the space of eigenvectors with
eigenvalues $a_n$. ( If $a_n$ is non-degenerate, 
then $\bf{P}_n = |n\rangle \langle n| $; it is the projection onto
the corresponding eigenvector.) The $\bf{P}_n$ satisfy

$\bf{P}_n \bf{P}_m = \delta_{n,m} \bf{P}_n $

$\bf{P}_n^{\dagger} = \bf{P}_n $.    \\

{\it Postulate 4:}
In quantum mechanics, the numerical outcome of a measurement
of the observable $\bf{A}$ is an eigenvalue of $\bf{A}$; right
after the measurement, the quantum state is an eigenstate of $\bf{A}$
with the measured eigenvalue. If the quantum state just prior to the 
measurement is $|\psi \rangle$, then the outcome $a_n$ is obtained with
the { \it probability} 

$ \bf{Prob} (a_n) = {\parallel \bf{P}_n |\psi \rangle \parallel}^2 = 
\langle\psi|\bf{P}_n|\psi\rangle$ ;   

If the outcome attained is $a_n$, then the (normalized) quantum state becomes

$\frac{\bf{P}_n |\psi \rangle}{\sqrt{(\langle\psi|\bf{P}_n|\psi\rangle)}}$.

(Note that if the measurement is immediately repeated, then according to this
rule the same outcome is attained again, with probability one.)

\section{Quantum Entanglement}

Quantum mechanics builds systems out of subsystems in a remarkable,
holistic way. The states of the subsystems do not determine the 
state of the system. Schrodinger, commenting on the EPR paper\cite{epr1935}
in 1935, the year it appeared, coined the term {\it entanglement} for this
aspects of quantum mechanics. \\

Consider a system consisting of two subsystems. Quantum mechanics associates
to each subsystem a Hilbert space. Let $\bf{H}_A$ and $\bf{H}_B$ denote these
two Hilbert spaces; let $|i_A\rangle$ (where i=1,2,...) represent a 
complete orthonormal basis for $\bf{H}_A$, and $|i_B\rangle$ (where i=1,2,...) represent a 
complete orthonormal basis for $\bf{H}_B$. Quantum mechanics asociates to the system-i.e.
the two subsystem taken together-the Hilbert space $\bf{H}_A \bigotimes \bf{H}_B$, namely
the Hilbert space spanned by the states $|i_A\rangle \bigotimes |i_B\rangle$. 
In the following we will drop the tensor product symbol $\bigotimes$ and write 
$|i_A\rangle \bigotimes |i_B\rangle$ as $|i_A\rangle |i_B\rangle$. \\

Any linear combinations of the basis states $|i_A\rangle |i_B\rangle$ is a state
of the system, any state $|\psi\rangle_{AB}$ of the system can be written as 

$|\psi\rangle_{AB} = \sum_{i,j} c_{i,j} |i_A\rangle |i_B\rangle$,

where the $c_{i,j}$ are complex coefficients, we take $|\psi\rangle_{AB}$ to be 
normalized, hence 
$\sum_{i,j} |c_{i,j}|^2 = 1$. \\

A special case of the above state is a {\it direct product} state in which
$|\psi\rangle_{AB}$ factors into (a tensor product of) a normalized state
$|\psi^{(A)}\rangle_A = \sum_i c_i^{(A)} |i\rangle_A$ in  $\bf{H}_A$ and a normalized state
$|\psi^{(B)}\rangle_B = \sum_j c_j^{(B)} |j\rangle_B$ in  $\bf{H}_B$.

$|\psi\rangle_{AB} = |\psi^{(A)}\rangle_A |\psi^{(B)}\rangle_B = 
(\sum_i c_i^{(A)}|i\rangle_A) (\sum_j c_j^{(B)}|j\rangle_B)$

Not every state in $\bf{H}_A \bigotimes \bf{H}_B$ is a product state. Take,
 for example, the state $(|1\rangle_A |1\rangle_B + |2\rangle_A|2\rangle_B)/\sqrt{2}$;
if we try to write it as a direct product of states of $\bf{H}_A$ and $\bf{H}_B$, we 
will find that we can {\it not}.

If $|\psi\rangle_{AB}$ is not a product state, we say that it is {\it entangled}. 

Thus when two quantum subsystems are entangled, we may have a complete knowledge of the 
composite system as a whole but not of the individual subsystems. Technically speaking,
the system as a whole may be in a {\it pure} state while individual subsystems 
still being in { \it mixed} states. \\

Entanglement is a key resource for quantum information processing and
spatially separated entangled pairs of particles have been
used for numerous purposes like teleportation \cite{bennett93}, 
superdense coding \cite{Bwiesner92} and cryptography based on Bell's 
Theorem \cite{ekert91}, to name a few. We shall see some
novel and interesting characterization and applications of 
entanglement in this thesis.

\chapter{Characterizing the Combinatorics of Distributed EPR Pairs for 
Multi-partite Entanglement }

\section{Introduction}
\label{intro}

\noindent Quantum entanglement is one of the most remarkable aspects 
of quantum physics. Two particles in an entangled state
behave in some respects as a single entity 
even if the particles are physically separated by a great distance.
The entangled particles
exhibit what physicists call non-local effects. Such non-local effects
were alluded to in the famous 1935 paper by Einstein, Podolsky, and Rosen
\cite{epr1935} and were later referred to as spooky actions at a distance
by Einstein. In 1964, Bell \cite{bell64} formalized the notion of 
two-particle non-locality in terms of correlations amongst probabilities in a
scenario where measurements are performed on each 
particle. He showed that the results of the measurements that occur 
quantum physically can be correlated in a way that cannot occur 
classically unless the type of measurement selected to be performed on
one particle affects the result of the measurement performed on the 
other particle. The two particles thus correlated maximally are called
EPR pairs or Bell pairs. 
Non-local effects, however, without being supplemented by additional quantum or
classical communication, do not convey any signal and therefore
the question of faster than light 
communication does not arise.\\  

Entanglement is a key resource for quantum information processing and
spatially separated entangled pairs of particles have been
used for numerous purposes like teleportation \cite{bennett93}, 
superdense coding \cite{Bwiesner92} and cryptography based on Bell's 
Theorem \cite{ekert91}, to name a few. An EPR channel (a bipartite 
maximally entangled distributed pair of entangled particles) can 
be used in conjunction with a classical communication channel to
send an unknown quantum state of a particle to a distant particle.
The original unknown quantum state is destroyed in the process 
and reproduced at the other end. The process does not copy the original
state; it transports a state and thus does not violate the quantum 
no-cloning theorem \cite{wooters82}. 
This process is called teleportation and was proposed by 
Bennett et al. in their seminal work \cite{bennett93}. 
In teleporation, a quantum communication channel is 
simulated using a classical channel and an EPR channel. A classical 
channel can also be simulated using a quantum channel and an EPR channel
using {\it superdense coding} as 
proposed by Bennett and Wiesner \cite{Bwiesner92}.
Two cbits (classical bits) are compressed in to a qubit and 
sent through an EPR channel to the distant party, who then
recovers the cbits by local operations.  \\

The use of EPR pairs for cryptography was first proposed by Ekert in 1991 
\cite{ekert91}. He proposed a protocol based  
on generalized Bell's theorem for quantum key distribution between 
two parties. The two parties share an EPR pair in advance. They do a
computational basis measurement on their respective qubits and the 
mesurement result is then used as the one bit shared key. 
While the measurement result is maximally uncertain, the 
correlation between their results is deterministic. 
Based on similar principles, a multiparty quantum key 
distribution protocol using EPR pairs in a distributed network 
and its proof of unconditional security has been proposed by 
Singh and Srikanth \cite{sinsrik031} (Chapter 4). 
Apart from these applications, entanglement 
has been used in several other
applications such as cheating bit commitment \cite{lochau97}, 
broadcasting of entanglement \cite{buzek97} and testing Bell's 
inequalities \cite{bell64,clauser69,gisin92}.\\
 
Just as two distant particles could be entangled forming an EPR pair,
it is also possible to entangle three or more separated particles. One 
example (called GHZ state) is due to Greenberger, Horne and Zeilinger 
\cite{ghz89}; here
three particles are entangled. A well-known manifestation of 
multipartite entanglement is in testing nonlocality from different 
directions \cite{ghz89, mermin90, home95, peres97}.
Recently, it has also been used for many multi-party computation and 
communication tasks \cite{buhrman01,buhrman99,brassard03,grover97,pal03}
and multi-party cryptography 
\cite{hil00, scarani01, bose98}.
Buhrman, Cleve
and van Dam \cite{buhrman01}, make use of  three-party entanglement 
and demostrate the existence of a function whose computation requires 
strictly lesser classical communication complexity compared to the 
scenario where no entanglement is used. Brassard et al. in 
\cite{brassard03}, show that prior multipartite 
entanglement can be used by
$n$ agents to solve a multi-party distributed problem, whereas no 
classical deterministic protocol succeeds in solving the same 
problem with probability away from half by a fraction that is larger
than an inverse exponential in the number of agents. Buhrman et al.
\cite{buhrman99} solves an $n$-party problem which shows a separation 
of $n$ versus  $\Theta (n\log n)$ bits between quantum and 
classical communication complexity. For one round, three-party problem
this article also proved a difference of $(n+1)$ versus $((3/2)n+1)$ bits
between communication with and without intial entanglement. 
In \cite{pal03}, Pal et al. present a {\it fair} and {\it unbiased} 
leader election protocol using maximal multi-partite entanglement. \\

Quantum teleportation strikingly underlines the peculiar features of
the quantum world. All the properties of the quantum state being 
transfered are retained during this process. So it is natural to ask 
whether one qubit of an entangled state can be teleported while 
retaining its entanglement with the other qubit. The answer is, not
surprisingly, in the affirmative. This is called {\it entanglement swapping}.
Yurke and Stoler \cite{yurke92} and Zukowski et al \cite{zukowski93}
have shown that through entanglement 
swapping one can entangle particles that do not even share any 
common past. This idea has been generalized to the tripartite case by
Zukowski et al. in \cite{zukowski95} and later to the multipartite
case by Zeilinger et al. \cite{zeilinger97} and 
Bose et al. \cite{bose98}. 
Zeilinger et al. presented a general scheme and realizable procedures
for generating GHZ states out of two pairs of entangled particles 
from independent emissions. They also proposed a scheme for observing 
four-particle GHZ state and their scheme can directly be generalized
to more particles. To create a maximally entangled state of $(n+m-1)$ 
particles from two groups, one of $n$ maximally entangled particles 
and the other of $m$ 
maximally entangled particles, it is enough to perform a controlled 
operation between a particle from the first group and 
a particle from the 
second group and then a measurement of the target particle. We observe that 
if particles are distributed in a network then a single 
cbit of communication is required
for broadcasting the mesurement result to construct the 
desired $(n+m-1)$ maximally entangled state using local operations.
In \cite{bose98},
Bose et al. have generalized the entangled swapping scheme of 
Zukowski et al. in a different way. In their scheme, the basic 
ingredient is a projection onto a maximally entangled state of $N$
particles. Each of the $N$ users needs
to share a Bell pair with a central exchange. The central exchange 
then projects the $N$ qubits with it on to an $N$ particle maximally 
enatngled basis, thus leaving the $N$ users in an $N$ partite maximally 
entangled state. However, we note that in order 
to get a desired state, the measurement
result obtained by the central exchange must be broadcast so that
the end users can appropriatly apply requisite local operations. This 
involves $N$ cbits of communication.\\   

In this chapter, we consider the problem of creating pure maximally
entangled multi-partite states out of Bell pairs distributed in a
communication network from a physical as well as a 
combinatorial perspective. We 
investigate and characterize
the minimal combinatorics of the distribution
of Bell pairs and show how this combinatorics gives rise to 
resource minimization in long-distance quantum communication.
We present protocols for creating maximal multi-partite entanglement. 
The first protocol (see Theorem \ref{ghztheorem}) enables us to prepare
a GHZ state using two Bell pairs shared amongst the three agents
with the help of two cbits of communication and local operations with 
the additional feature that this 
protocol involves all the three agents {\it dynamically}. Such a protocol with
local dynamic involvement in creating entanglement  
may find applications in cryptographic tasks.
The second protocol (see Theorem \ref{qhtheorem3}) 
entails the use of $O(n)$ cbits of communication and local
operations to prepare a pure $n$-partite maximally entangled state in a
distributed network of Bell pairs; the requirement here is
that the pairs of nodes sharing EPR pairs must form a connected graph. 
We show that a spanning 
tree structure (see Theorem \ref{qhtheorem2}) 
is the minimal combinatorial requirement for creating
multi-partite entanglement. 
We also characterize the minimal combinatorics 
of agents in the creation of pure maximal multi-partite 
entanglement amongst the set $N$ of $n$ agents in a network 
using apriori multi-partite entanglement states amongst subsets of 
$N$. This is done by generalizing the {\it  EPR graph} representation 
to an {\it entangled hypergraph} and the requirement here is that the 
entangled hypergraph representing the entanglement structure must be
connected.\\

The chapter is organized as follows. In 
Section \ref{ghzsection}, we present
our protocol I to prepare a GHZ state from two Bell pairs involving all
the three agents dynamically and compare our protocol 
in the light of existing schemes. Section \ref{eprsection} is devoted to
characterizing the spanning tree combinatorics 
of Bell pairs for preparing a pure $n$-partite
maximally entangled state. We develop our protocol II for this purpose. 
In Section \ref{hypersection} we generalize the results of 
Section \ref{eprsection} to the setting where subsets 
of agents in the network
share apriori pure multi-partite maximally entangled states.
Finally in Section \ref{conclusion} we compare our scheme of 
Section \ref{eprsection} with
the multipartite entanglement swapping scheme of Bose et al. \cite{bose98},
observe the similarity
between Helly-type theorems and the combinatorics developed
in Sections \ref{eprsection} and \ref{hypersection} and conclude with 
a few remarks on open research directions.

\section{Preparing a GHZ State from Two EPR Pairs Shared
amongst Three Agents}
\label{ghzsection}

In this section we consider the preparation of a 
GHZ state from two EPR pairs shared amongst three agents in a 
distributed network. We establish the following theorem. 


\begin{figure}
\resizebox*{0.9\textwidth}{0.35\textheight}{\includegraphics{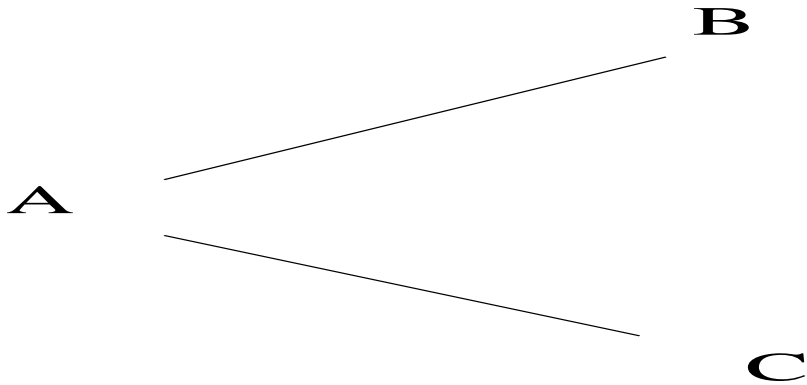}} 
\caption{$A$ shares an EPR pair with each of $B$ and $C$.}
\label{figure1}
\end{figure}

\begin{thm}
If any two pairs of the three agents $A$ (Alice), $B$ (Bob) and $C$
(Charlie) share EPR pairs (say the state $(|00\rangle +
|11\rangle)/\surd{2}$) then we can prepare a GHZ state
$(|000\rangle + |111\rangle)/\surd{2}$ amongst them with 
two bits of classical communication, while involving all the 
three agents dynamically.
\label{ghztheorem}
\end{thm}

\noindent {Proof:}
The proof follows from the Protocol I.

\noindent {\it The Protocol I:} Without loss of generality let
us assume that the sharing arrangement is as in Figure \ref{figure1}.
$A$ shares an EPR pair with $B$ and another EPR pair with $C$
but $B$ and $C$ do not share an EPR pair. This means that we have the
states $(|0_{a1}0_b\rangle + |1_{a1}1_b\rangle)/\surd{2}$ and
$(|0_{a2}0_c\rangle + |1_{a2}1_c\rangle)/\surd{2}$ where
subscripts $a1$ and $a2$ denote the first and second qubits with $A$ and
subscripts $b$ and $c$ denote qubits with $B$ and $C$, respectively.


\begin{figure}
\resizebox*{0.9\textwidth}{0.35\textheight}{\includegraphics{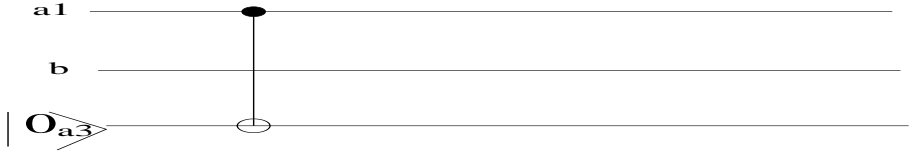}} 
\caption{Entangling qubits $a3$ with the EPR pair between $A$ and $B$.}
\label{figure2}
\end{figure}

Our aim is to prepare $(|0_{a1}0_b0_c\rangle +
|1_{a1}1_b1_c\rangle)/\surd{2}$ or $(|0_{a2}0_b0_c\rangle +
|1_{a2}1_b1_c\rangle)/\surd{2}$. We need three steps to do so.

\noindent {\it Step 1:} $A$ prepares a third qubit in the state
$|0\rangle$. We denote this state as $|0_{a3}\rangle$ where the
subscript $a3$ indicates that this is the third qubit of $A$.

\noindent {\it Step 2:} $A$ prepares the state
$(|0_{a1}0_b0_{a3}\rangle +
|1_{a1}1_b1_{a3}\rangle)/\surd{2}$ using 
the circuit in Figure \ref{figure2}.

\noindent {\it Step 3:} $A$ sends her third qubit to $C$ with the help
of the EPR channel $(|0_{a2}0_c\rangle + |1_{a2}1_c\rangle)/\surd{2}$.
A straight forward way to do this is through teleportation. This 
method however, does not involve both of $B$ 
and $C$ dynamically. By a party being dynamic we mean that the party is 
involved in applying the local operations for the completion of 
the transfer of the state of third qubit to create 
the desired GHZ state.


\begin{figure}
\resizebox*{0.9\textwidth}{0.9\textheight}{\includegraphics{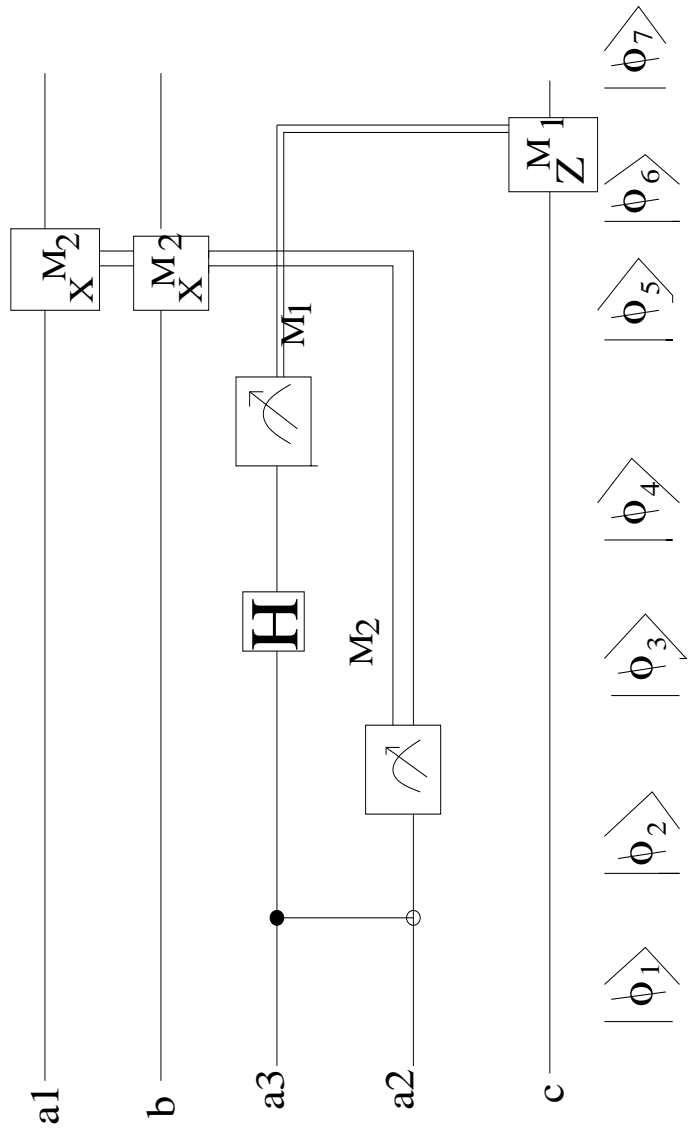}} 
\caption{Circuit for creating a GHZ state from two EPR pairs 
with dynamic involment of both $B$ and $C$.}
\label{figure3}
\end{figure}

We use our new and novel teleportation circuit 
as shown in Figure \ref{figure3} where both $B$ and $C$ are dynamic.
The circuit works as follows.
$A$ has all her three qubits with her and can do any
operation she wants to be performed on them. Initially the five
qubits are jointly in the state $|\phi_1\rangle$. $A$ first applies
a controlled NOT gate on her second qubit controlling it from her
third qubit changing $|\phi_1\rangle$ to $|\phi_2\rangle$. Then
she measures her second qubit yielding measurement result $M_2$
and bringing the joint state to $|\phi_3\rangle$. She then applies
a Hadamard gate on her third qubit and the joint state becomes
$|\phi_4\rangle$. A measurement on the third qubit is then done by
her yielding the result $M_1$ and bringing the joint state to
$|\phi_5\rangle$. She then applies a NOT (Pauli's X operator) on
her first qubit, if $M_2$ is 1.  Now she sends the measurement
results $M_2$ to $B$ and $M_1$ to $C$. $B$ applies an $X$ gate on his
qubit if he gets 1 and $C$ applies a $Z$ gate (Pauli's $Z$ operator) if
he gets 1. The order in which $B$ and $C$ apply their operations does
not matter. The final state is $|\phi_7\rangle$. 
The circuit indeed produces the GHZ state between $A$, $B$ and
$C$ as can be seen from the detailed mathematical explanation 
of the circuit given below. It can be noted that this protocol 
requires two cbits of communication.

The above circuit can be explained as
follows:

$$|\phi_1\rangle  = (|0_{a1}0_b0_{a3}\rangle +
|1_{a1}1_b1_{a3}\rangle) (|0_{a2}0_c\rangle +
|1_{a2}1_c\rangle)/2,$$

$$|\phi_2\rangle  = [|0_{a1}0_b0_{a3}\rangle
(|0_{a2}0_c\rangle +
|1_{a2}1_c\rangle) + |1_{a1}1_b1_{a3}\rangle
(|1_{a2}0_c\rangle +
|0_{a2}1_c\rangle)]/2.$$

\noindent {\it Case 1:} $M_2 = 0$
\begin{eqnarray*}
|\phi_3\rangle & = & (|0_{a1}0_b0_{a3}0_{a2}0_c\rangle +
|1_{a1}1_{b}1_{a3}0_{a2}1_c\rangle)/\surd{2}\\
& = & (|0_{a1}0_b0_{a3}0_c\rangle +
|1_{a1}1_b1_{a3}1_c\rangle)
|0_{a2}\rangle /\surd{2},
\end{eqnarray*}
\begin{eqnarray*}
|\phi_4\rangle & = & (|0_{a1}0_b0_{a3}0_c\rangle +
|0_{a1}0_b1_{a3}0_c\rangle + |1_{a1}1_b0_{a3}1_c\rangle -
|1_{a1}1_b1_{a3}1_c\rangle) |0_{a2}\rangle /2\\
& = & [(|0_{a1}0_b0_c\rangle + |1_{a1}1_b1_c\rangle)
|0_{a3}\rangle + (|0_{a1}0_b0_c\rangle -
|1_{a1}1_b1_c\rangle)
|1_{a3}\rangle ] |0_{a2}\rangle /2.
\end{eqnarray*}
\noindent When $M_1 = 0$,

$$|\phi_5\rangle  = (|0_{a1}0_b0_c\rangle +
|1_{a1}1_b1_c\rangle)|0_{a3}\rangle |0_{a2}\rangle
/\surd{2},$$

$$|\phi_6\rangle  = (|0_{a1}0_b0_c\rangle +
|1_{a1}1_b1_c\rangle)
|0_{a3}\rangle |0_{a2}\rangle /\surd{2},$$

$$|\phi_7\rangle  = (|0_{a1}0_b0_c\rangle +
|1_{a1}1_b1_c\rangle)|0_{a3}\rangle |0_{a2}\rangle /\surd{2}.$$

\noindent When $M_1 = 1$,

$$|\phi_5\rangle  = (|0_{a1}0_b0_c\rangle - |1_{a1}1_b1_c\rangle)
|1_{a3}\rangle |0_{a2}\rangle /\surd{2},$$

$$|\phi_6\rangle  = (|0_{a1}0_b0_c\rangle - |1_{a1}1_b1_c\rangle)
|1_{a3}\rangle |0_{a2}\rangle /\surd{2},$$

$$|\phi_7\rangle  = (|0_{a1}0_b0_c\rangle + |1_{a1}1_b1_c\rangle)
|1_{a3}\rangle |0_{a2}\rangle /\surd{2}.$$

\noindent {\it Case 2:} $M_2 =1$
\begin{eqnarray*}
|\phi_3\rangle & = & (|0_{a1}0_b0_{a3}1_{a2}1_c\rangle +
|1_{a1}1_b1_{a3}1_{a2}0_c\rangle)/\surd{2}\\
& = & (|0_{a1}0_b0_{a3}1_c\rangle +
|1_{a1}1_b1_{a3}0_c\rangle)|1_{a2}\rangle /\surd{2},
\end{eqnarray*}
\begin{eqnarray*}
|\phi_4\rangle & = & (|0_{a1}0_b0_{a3}1_c\rangle +
|0_{a1}0_b1_{a3}1_c\rangle + |1_{a1}1_b0_{a3}0_c\rangle -
|1_{a1}1_b1_{a3}0_c\rangle) |1_{a2}\rangle /2\\
& = & [(|0_{a1}0_b1_c\rangle + |1_{a1}1_b0_c\rangle)
|0_{a3}\rangle + (|0_{a1}0_b1_c\rangle - |1_{a1}1_b0_c\rangle)
|1_{a3}\rangle ] |1_{a2}\rangle /2.
\end{eqnarray*}
\noindent When $M_1 = 0$,

$$|\phi_5\rangle  = (|0_{a1}0_b1_c\rangle +
|1_{a1}1_b0_c\rangle)|0_{a3}\rangle |1_{a2}\rangle /\surd{2},$$

$$|\phi_6\rangle  = (|1_{a1}1_b1_c\rangle + |0_{a1}0_b0_c\rangle)
|0_{a3}\rangle |1_{a2}\rangle /\surd{2},$$

$$|\phi_7\rangle  = (|1_{a1}1_b1_c\rangle + |0_{a1}0_b0_c\rangle)
|0_{a3}\rangle |1_{a2}\rangle /\surd{2}.$$

\noindent When $M_1 =1$,

$$|\phi_5\rangle  = (|0_{a1}0_b1_c\rangle -
|1_{a1}1_b0_c\rangle)|1_{a3}\rangle |1_{a2}\rangle /\surd{2},$$

$$|\phi_6\rangle  = (|1_{a1}1_b1_c\rangle - |0_{a1}0_b0_c\rangle)
|1_{a3}\rangle |1_{a2}\rangle /\surd{2},$$
\begin{eqnarray*}
|\phi_7\rangle & = & (- |1_{a1}1_b1_c\rangle
-|0_{a1}0_b0_c\rangle)|1_{a3}\rangle |1_{a2}\rangle /\surd{2}\\
& = & (|1_{a1}1_b1_c\rangle + |0_{a1}0_b0_c\rangle)|1_{a3}\rangle
|1_{a2}\rangle /\surd{2}.
\end{eqnarray*}

The roles of $B$ and $C$ are
symmetrical. Nevertheless, there is a condition on what operations
they should perform when they get a single cbit from $A$. $B$ performs
an $X$ and $C$ performs a $Z$ operation, as required. We set a
cyclic ordering $A \rightarrow B \rightarrow C \rightarrow A$. Let
$A$ be the one sharing EPR pairs with the other two; $A$ is the first
one in the ordering. The second one is $B$, and he must perform an $X$
operation when he gets a single cbit from $A$. The third one is $C$,
and he must perform a $Z$ operation on his qubit when he gets a
single cbit from $A$.  If $B$ is the one sharing EPR pairs with the
other two then $C$ applies an $X$ on his qubit after getting a cbit from
$B$ and, $A$ applies $Z$ on her qubit after getting a cbit from $B$ and
so on.

As we mentioned in the introduction, methods for creating a GHZ
state from Bell pairs have also been discussed by Zukowski et al. 
\cite {zukowski95} and  
Zeilinger et al. \cite {zeilinger97}. First one uses three Bell pairs 
for this purpose, therefore our protocol seems better than theirs in 
the sense that it uses only two Bell pairs. The later, however, uses
only two Bell pairs and only one cbit of communication and seems to
be better than our method at first sight. However, the most interesting 
fact and the motivation for developing our protocol is the dynamic
involvement of both $B$ and $C$ which was lacking in the above
methods. It might me highly desired in many multi-party interactive 
quantum protocols and multi-party cryptograhy (viz. secret sharing)
that both $B$ and $C$ take part actively, say for fairness. 
By fairness we mean that every party has an equal chance for 
participating and effecting the protocol in probabilistic sense. 
It should be interesting to implement this in practical situation.

\section{Preparing a Pure $n$-partite Maximally Entangled State from
EPR Pairs Shared amongst $n$ Agents}
\label{eprsection}

\begin{defni}
EPR graph: 
Suppose there are $n$ agents. We denote them as
$A_1, A_2, ..., A_n$. Construct an undirected graph $G = (V, E)$
as follows:

$$V = \{A_i: i=1, 2, 3,..., n\},$$
$$E = \{\{A_i , A_j \} : A_i ~\mbox{and} ~A_j, ~\mbox{share an EPR pair,}
1\leq i, j\leq n; i \neq j\}.$$

\noindent We call the graph $G = (V, E)$, thus formed, the {\it EPR
graph} of the $n$ agents.

\end{defni}

Our definition should not be confused with the {\it entangled graph}
proposed by Plesch and Buzek \cite {plesch02, plesch03}. In entangled 
graph edges represent any kind of entanglement and not neccessarily
maximal entanglement and therefore there is no one to one 
correspondence between graphs and states. EPR graph is unique up to
different EPR pairs. Moreover, we are not concerned with classical
correlations which are also represented by different kind of 
edges in entangled graphs. In an 
EPR graph, two vertices are connected
by an edge if and only if they share an EPR pair. \\

\begin{defni}
Spanning EPR Tree:
We call an {\it EPR Graph} $G = (V, E)$ as
{\it spanning EPR tree} when the undirected graph $G = (V, E)$ is a
spanning tree \cite{clr1990}.
\end{defni}

We are now ready to develop protocol-II to create the $n$-partite
maximally entangled state 
$(|000...0\rangle + |111...1\rangle)/\surd{2}$ along a spanning 
EPR tree. The protocol uses only $O(n)$ cbits of communication and 
local operations. We do not use any qubit communication after
the distribution of EPR pairs to form a spanning EPR tree.



\begin{figure}
\resizebox*{0.9\textwidth}{0.35\textheight}{\includegraphics{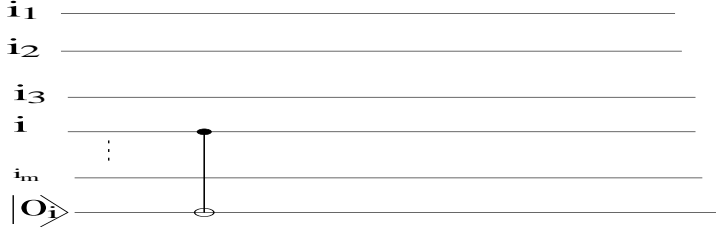}} 
\caption{Entangling a new qubit in agent $i$.}
\label{figure5}
\end{figure}

\noindent {\it Protocol II:} Let $G = (V, E)$ be the spanning EPR
tree. Since $G$ is a spanning EPR tree, it must have a vertex say $T
= A_t$ which has degree one (the number of edges incident on a
vertex is called its degree). Note that vertices of $G$ are 
denoted $A_i$, where $1\leq i\leq n$. Let $S = A_s$ be the unique vertex
connected to $T$ by an edge in $G$ and $L$ be the set of all vertices of
$G$ having degree one.  
The vertex $T$ and its only neighbor $S$ are vertices
we start with. We eventually prepare the $n$-partite maximally entangled
state using the three steps summarized below. 
In the first step, one cbit (say 1) is broadcasted by $S$ to
signal other $(n-1)$ parties that the 
protocol for the preparation of the 
$n$-partite entangled state is about to commence.
The second step creates the GHZ state between $S=A_s$, $T=A_t$ 
and another neighbour $R=A_r$ of $S$. The third step 
is the main inductive step where multi-partite entanglement states are
created in a systematic manner over the spanning EPR tree. At the end
of step 3,
when the $n$-partite entangled state is ready, one cbit of 
broadcating from
all the $(k-1)$ elements of $L\setminus \{T\}$ (terminal or degree one 
vertices of $G$) is expected. The $(k-1)th$ such cbit indicates that  
the protocol is over and that maximally entangled state is ready. 
The details are stated below.

\noindent {\it Step 1:} $S$ broadcast one classical bit to signal
the other $(n-1)$ agents that the 
preparation of an $n$-partite entangled state is going to be started
and they must not use their EPR pairs for a qubit teleportation
amongst themselves. In other words, they must save their EPR pairs in
order to use them for the preparation of the $n$-partite entangled
state.

\noindent {\it Step 2:} Clearly $S$ must be connected to a vertex $R$
(say $A_r$) other than $T$ by an edge in $G$, otherwise $G$ will not be
a spanning EPR tree.  A GHZ state among $S$, $T$ and $A_r$ is created.
The GHZ state can be prepared either by using 
usual teleportation circuit, 
the symmetric circuit of protocol I or by the Zeilinger et al. scheme.
Thus using the EPR pairs $(|0_{s,t}0_{t,s}\rangle +
|1_{s,t}1_{t,s}\rangle)/\surd{2}$ and $(|0_{s,r}0_{r,s}\rangle +
|1_{s,r}1_{r,s}\rangle)/\surd{2}$, we prepare the GHZ state
$(|0_{s,t}0_{t,s}0_{r,s}\rangle +
|1_{s,t}1_{t,s}1_{r,s}\rangle)/\surd{2}$. Here the double
subscript $i,j$ denotes that in preparing the given state, EPR pairs
among the agents $A_i$ and $A_j$ have been used.
Here, $A_s=S, A_r=R$ and $A_t=T$.

\noindent {\it Step 3:} Suppose we are currently at vertex $A_i$
and we have already prepared the $m$-partite maximally entangled
state, say

$$(|0_{i1, j1}0_{i2, j2} 0_{i3, j3}...0_{im, jm}\rangle +
|1_{i1, j1}1_{i2, j2}1_{i3, j3}...1_{im,jm}\rangle)/\surd{2},$$

\noindent where $i1 = s, j1 = t, i2 = t, j2 = s, i3 = r, j3 = s$ and
$i = ir$ for some $1 \leq r \leq m$.

\noindent The vertex $A_i$ starts as follows. As soon as he gets
two cbits from one of his neighbors, he completes the operations
required for the success of teleportation and starts processing as
follows. If $A_i \in L$ then $A_i$ broadcast a single cbit.
Otherwise,
(when $A_i \notin L$) let $A_{k1}, A_{k2},..., A_{kp}$
be the vertices connected to $A_i$ by an edge in $G$ such that {\it  k1,
k2,..., kp} are not in the already entangled set with vertex indices 
$\{i1, i2, i3,...,im\}$. $A_i$ takes an extra 
qubit and prepares this qubit in
the state $|0\rangle$ denoted by $|0_i\rangle$. He then prepares
the state

$$(|0_{i1, j1}0_{i2, j2}0_{i3, j3}... 0_{im, jm}0_i\rangle +
|1_{i1, j1}1_{i2, j2}1_{i3, j3}...1_{im, jm}1_i\rangle)/\surd{2}$$

\noindent using the circuit in  Figure \ref{figure5}.
Finally, he teleports his extra qubit to $A_{k1}$ using the EPR
pair $(|0_{i, k1}0_{k1, i}\rangle + |1_{i, k1}1_{k1,
i}\rangle)/\surd{2}$, thus enabling the preparation 
of the $(m+1)$-partite maximally entangled state:

$$(|0_{i1, j1}0_{i2, j2} 0_{i3, j3}...0_{im, jm}0_{k1, i}\rangle +
|1_{i1, j1}1_{i2, j2}1_{i3, j3}...1_{im, jm}1_{k1,
i}\rangle)/\surd{2}.$$

\noindent $A_i$ repeats this until no other vertex, which is
connected to it by an edge in $G$, is left.

\noindent Step 3 is repeated until one cbit each from the
elements of $L$ (except for $T$) is broadcasted, 
indicating that all vertices in $L$ as well
as in $V\setminus L$ have got entangled. 

Note that more than one vertex might be
processing Step 3 at the same time. This however does not matter
since local operations do not change the reduced density matrix of
other qubits. Moreover, while processing the Step 3 together, such vertices will no longer  
be directly connected by an edge in $G$. 

Now we determine the communication complexity of protocol II, 
the number of cbits used in creating 
the $n$-partite maximally entangled state. 
Step 1 involves one cbit broadcast by $S$ to signal the 
initiation of the protocol. To create the GHZ state in Step 2,
atmost $2$ cbits is required. In Step 3, teleportation is 
used to create an $(m+1)$-partite maximally entangled state
from that of $m$-partite. Such $(n-3)$ teleportation steps are
used in this Step entailing $2(n-3)$ cbits of communication.
Finally, $(k-1)$ cbits are broadcast by terminal 
vertices (except $T$). Thus the total cbits used in 
protocol II is 
$1+2+2(n-3)+k-1 = 2n+k-4 \leq 2n+n-1-4 =3n-5=O(n)$.  

Protocol-II leads to the following interesting theorem.

\begin{thm}
If the combinatorial arrangement of distributed EPR pairs 
amongst $n$ agents forms a spanning EPR tree, then the $n$-partite
maximally entangled state $(|000...0\rangle + |111...1\rangle)
/\surd{2}$
can be prepared amongst them with $O(n)$ bits of classical
communication.
\label{qhtheorem1}
\end{thm}

Theorem \ref{qhtheorem1} thus gives a sufficient condition for preparing  
a maximally entangled $n$-partite state in a distributed network
of EPR pairs. In order to prove this sufficiency, 
we have also developed two more protocols 
which require $O(n)$ cbits of communication. 
The first two steps of these protocols are essentially the 
same as that of Protocol II.
The first protocol involves all the 
agents already entangled in each iteration in Step 3, where, a circuit
very similar to the symmetric teleportation circuit 
(Figure \ref{figure3}) of Protocol I is used. The classical communication 
cost is $(2n-4)$ bits. The second protocol 
uses a generalization of the method of  
Zeilinger et al. in each iteration of Step 3 and requires   
$(2n-3)$ cbits of communication. In this 
paper we have presented only Protocol II 
instead of these two protocols because 
of simplicity and the direct use of teleportation.\\

The question of interest now is that of determining 
the minimal structure 
or combinatorics of the distribution of EPR pairs neccessary for
creating the $n$-partite maximally entangled state. 
In other words, we wish to characterize 
neccessary properties to be satisfied by the EPR graph
for this purpose. We argue below that the EPR graph, indeed,
must contain a spanning EPR tree, and must therefore be connected. 
We assume for the sake of contradiction that the EPR graph $G$ is
not connected. 
Then, it must have at least two components,
say $C_1$ and $C_2$. No member of $C_1$ is connected to any member of 
$C_2$ by an edge in $G$. This means that no member of $C_1$ is
sharing an EPR pair with any member of $C_2$. 
Suppose a protocol $P$ can create a pure $n$-partite maximally 
entangled state starting from the disconnected EPR graph $G$
of $n$ agents. 
If we are able
to create an $n$-partite maximally entangled state using
protocol $P$ with this 
structure using only classical communication and local operations,
it is easy to see that we will also be able to create 
an EPR pair between two parties that were not earlier sharing 
any EPR pair, using just local operations and classical communication.
This can be done as follows. 
Let $A$ be the first party that 
posseses all the qubits of his group (say $C_1$)
and $B$ be the second party that possesses all the qubits
of his group (say $C_2$). Now the protocol $P$ is run on this
structure to create the $n$-partite maximally entangled state. 
Then, $A$ ($B$) disentangles all of his 
qubits except one by reversing the circuit in Figure \ref{figure5};
this leaves $A$ and $B$ sharing an EPR pair.
This means that two parties which were never sharing an EPR pair
are able to share it just by local operations 
and classical communication (LOCC).
This is forbidden by  
fundamental laws in quantum information theory (LOCC cannot increase the
expected entanglement \cite{vedral02}), hence $G$ must be connected. 
Note that no qubit communication is permitted 
after the formation of EPR graph $G$.  
We present this neccessary condition in the following theorem.

\begin{thm}
A necessary condition that the $n$-partite maximally entangled state 
$(|000...0\rangle + |111...1\rangle)/\surd{2}$ be prepared
in a distributed 
network permitting only EPR pairs for pairwise entanglement between agents 
is that the EPR graph of the $n$ agents must be connected.
\label{qhtheorem2}
\end{thm}

It can be noted that after the preparation of the state 
$(|000...0\rangle + |111...1\rangle)/\surd{2}$, any other
pure $n$-partite maximally entangled state can also be
prepared by just using local operations.
We also know that any connected undirected graph contains a 
spanning tree \cite{clr1990}. Thus a connected EPR graph 
will contain a spanning EPR tree.
With this observations,
we combine the above two theorems in the following theorem.

\begin{thm}
Amongst $n$ agents in a communication network permitting only 
pairwise entanglement in the form of EPR pairs, a pure $n$-partite
maximally entangled state can be prepared if and only if the 
EPR graph of the $n$ agents is connected. 
\label{qhtheorem3}
\end{thm}

\section{Entangling a Set of Agents from Entangled States of Subsets: Combinatorics
of General Entanglement Structure }
\label{hypersection}

In the previous section we have presented the necessary and sufficient
condition for preparing a pure multi-partite maximally entangled state in a 
distributed network of EPR pairs (see Theorem \ref{qhtheorem3}). 
However, agents 
may not be connected by EPR pairs in a general network. We assume that 
subsets of agents may be sharing pure maximally entangled states. So, some 
triples of agents may be GHZ entangled, some pairs of agents may share EPR 
pairs and some subsets of agents may share even higher dimensional entangled
states. \\

Now we develop the combinatorics of multi-partite entanglement within subsets 
of agents required to 
prepare multi-partite entanglement between all the agents. When we were 
dealing only with EPR pairs in the case of EPR graphs or spanning EPR trees,
we used the simple graph representation. Now subsets of the set of all 
agents may be in multi-partite entangled states and therefore we use a natural 
representation for such entanglement structures with hypergraphs as follows.\\

Let $S$ be the set of $n$ agents in a 
communication network. Let $E\subset S$, 
$|E|=k$. Suppose $E$ is such that the $k$ agents in $E$ are 
in a $k$-partite
pure maximally entangled state.
Let $E_1$, $E_2$, ..., $E_m$ be such subsets of $S$, each having a 
pure maximally entangled shared state amongst its agents.
Note that the sizes of these subsets may be different.
Consider the
hypergraph $H=(S,F)$ \cite{hbc1, berge} such 
that $F=\{E_1,E_2,...,E_m\}$. We call
such a
hypergraph $H$, an {\it entangled hypergraph}
of the $n$ agents. In standard hypergraph notation the elements of $F$ are 
called {\it hyperedges} of $H$.
Now we present the necessary and sufficient condition for
preparing a $n$-partite pure maximally entangled state in such 
networks, given entanglements as per the entangled hypergraph.
We need the definition of a {\it hyperpath} in a hypergraph: a sequence of
$j$ hyperedges $E_1$, $E_2$, ..., $E_j$ in a hypergraph
is called a {\it hyperpath} from a vertex $a$ to a vertex $b$ 
if (i) $E_i$ and $E_{i+1}$ have a common vertex (agent)
for all $1\leq i\leq j-1$ (ii) $a$ and $b$ are agents in $S$ 
(iii) $a\in E_1$ and (iv) $b\in E_j$. If there is a hyperpath between every
pair of vertices of $S$ in a hypergraph $H$ then we say that $H$ is 
connected.  

\begin{thm}

Given $n$ agents in a communication network and an entangled hypergraph, a
pure $n$-partite maximally entangled state can be
prepared amongst the $n$ agents if and only if the entangled hypergraph 
is connected.

\label{entanhyperiff}

\end{thm}

The proof of this theorem is based on the following Protocol-III.

\noindent {\it The Protocol III:}
We assume without loss of generality that $n>|E_1|\geq |E_2|...\geq |E_m|$.
We maintain the set $F$ and $R=S\setminus F$ where $F$ contains 
the agents already entangled in the $|F|$-partite pure maximally 
entangled state. Initially, $F=E_1$ and $R=\{E_2, E_3,..., E_m\}$.
We repeat the following steps until $F=S$. 
Choose $E_i\in R$ with minimum $i$ such that $F$ and $E_i$ have at least one
common agent and $E_i$ is not in $F$; 
let the smallest index common element between $E_i$ and $F$ 
be the agent $A_j$. (Since the entangled hypergraph 
is connected, there is always such a hyperedge $E_i$.) We 
can now use the method of Zeilinger et al. to create an 
$(N+M-1)$-partite maximum entangled state from two groups, one 
containing $N=|F|$ agents
and the other containing $M=|E_i|$ agents. 
The measurement is processed by $A_j$. 
So, an $(|F|+|E_i|-1)$-partite
entanglement state is prepared from amongst the members of $F$ and $E_i$.
If $F$ and $E_i$ share only one common agent then we are done. Otherwise, each 
member common to $F$ and $E_i$ other than $A_j$ will have two qubits each
from the $|F|+|E_i|-1$ entangled qubits.
These qubits must be disentangled using a circuit same as the reverse of
circuit in Figure \ref{figure5}. Now the members of $F$ and $E_i$ 
remain entangled in $(|F|+|E_i|-|F \bigcap E_i|)$-partite state, 
each holding 
exactly one qubit. Finally, we set $F=F \bigcup E_i$ and $R=R\setminus E_i$.

The proof of necessity is similar to that of
the proof of necessity in Theorem \ref{qhtheorem2}. 
For the sake of contradiction assume that the entangled hypergraph $H$
is not connected. Then there is no hyperpath between two agents (say 
$a$ and $b$), implying the existence 
of at least two components $C_1$ and $C_2$
in $H$, with no member of $C_1$ sharing 
a hyperedge of entanglement with any 
member of $C_2$. 
Suppose a protocol $P$ can create a pure $n$-partite maximally 
entangled state starting from the disconnected entangled hypergraph 
$H$ of $n$ agents. 
If we are able
to create an $n$-partite maximally entangled state using
protocol $P$ with this 
structure using only classical communication and local operations,
it is easy to see that we will also be able to create 
an EPR pair between two parties that were not earlier sharing 
any EPR pair, using just local operations and classical communication.
This is forbidden by 
fundamental laws in quantum information theory 
(LOCC cannot increase the
expected entanglement \cite{vedral02, hord01}). Hence $H$ must be connected. 
This completes the proof of Theorem \ref{entanhyperiff}.

\section{Concluding Remarks}
\label{conclusion}

We compare our method (Protocol-II) of
generating multipartite maximally entangled states
with that of Bose et al. \cite{bose98}.
The scheme of Bose et al. works as follows.
Each agent needs to share a Bell pair with a 
central exchange 
in the communication network of $n$ agents.
The central exchange then projects the $n$-qubits 
with him, on to the $n$-partite maximally entangled basis.
This leaves the $n$ agents in a $n$-partite maximally 
entangled state. Thus, the two basic requirements 
of their scheme are a central exchange and a projective 
measurement on a multi-partite maximally entangled basis.
The central exchange essentially represents a star topology 
in a communication network and allows certain degree of 
freedom to entangle particles belonging to any set of users
only if the necessity arises. 
However, a real time communication network may not
always be a star network, in which case, we may need to
have several such central exchanges. 
Of course, one will also be interested in setting up
such a network with minimum resources, especially in 
the case of a 
long distance communication network. 
Issues involved in the design of such 
central exchanges such as 
minimizing required resources, are of vital interest
while dealing with real communication 
networks. Such networks may be called 
{\it Quantum Local Area Network} (Q-LAN or Non-LAN, a Non-Local LAN)
or {\it Quantum Wide Area Network} (Q-WAN or Non-WAN).
Our scheme presented in Section \ref{eprsection} adresses these  
issues.
We have shown in Theorem \ref{qhtheorem3}
that the {\it spanning EPR tree} is 
the minimal combinatorial requirement for this purpose.
The star topology is a special case of the spanning EPR tree where
the central exchange is one of the agents.
It is therefore clear that 
the star network requirements of the scheme of
Bose et al. provides a sufficient condition where as the 
requirement in our spanning EPR tree
scheme is the most general and minimal possible structure.\\

Our scheme also helps in minimizing resources.
Our spanning tree topology has been used by 
Singh and Srikanth \cite{sinsrik031} (Chapter 4) for this purpose.
They assign weights to the edges of the EPR graph
based on the resources (such as quantum repeaters, etc.)
needed to build that particular edge. 
Then, a {\it minimum spanning EPR tree} represents 
the optimized requirement.
They also use this topology for multi-party quantum cryptography
to minimize the size of the sector that can be potentially
controlled by an evesdropper. 
Thus our topology seems to be a potential candidate for
building a long distance quantum communication network
(such as in a Non-LAN or Non-WAN).\\

The second basic ingredient of the scheme of Bose et al. is 
the projection on a multipartite maximally entangled basis.
As they point out, the circuit for such a measurement 
is an inverse of the circuit that generates a maximally 
entangled state from a disentangled input in the 
computational basis. 
In a communication network involving a large number
of agents, this entails a lot of work to be done
on part of the central exchange while the agents are
idle.
In our scheme, work is distributed 
amongst the agents.
Moreover, the $n$-qubit joint measurement on the entangled basis 
in the scheme of Bose et al. seems to be well high impossible from 
a practical standpoint given the current technology, 
whereas all the practical requirements of our scheme (Protocol II)  
can be met using current technology (using telecom cables to distribute
entanglement etc).\\  

The projection used by the central exchange in the 
scheme of Bose et al. may lead to any of the 
$2^n$ possible $n$-partite maximally entangled states.
For practical purposes, one might be more interested
in a particular state. 
To get the desired state, the measurement result 
must be broadcast by the central exchange.
The $2^n$ possible states can be represented by 
a $n$ bit number and thus the communication 
complexity involved in their scheme is $n$ 
cbits, essentially the same as that of ours asymptotically.
Therefore, our scheme is comparable to their scheme also
in terms of communication complexity.
It can also be noted at this point that,
in our topology, even the method of
Zeilinger et al. for creating $(m+1)$-partite
maximally entangled state from a $m$-partite
maximally entangled state becomes applicable.
The use even reduces the communication complexity
by some cbits but still requires $2n-3$ cbits which is $O(n)$.
As it can be observed, all these schemes require 
$O(n)$ cbits of communication. 
Whether there is an $\Omega (n)$ lower bound 
on the cbit communication complexity for preparing 
$n$-partite a pure maximally entangled state given a 
spanning EPR tree
remains open for further research.\\

The results in Theorem \ref{qhtheorem3} and 
Theorem \ref{entanhyperiff} are similar to the
classical theorem by Helly \cite{v1964} in convex geometry.  Helly's
theorem states that a collection of closed convex sets in the
plane must have a non-empty intersection if each triplet of the
convex sets from the collection has a non-empty intersection. In
one dimension, Helly's theorem ensures a non-empty intersection of
a collection of intervals if each pair of intervals has a
non-empty intersection. In our case (Theorems \ref{qhtheorem3} and 
\ref{entanhyperiff}), there is
similar combinatorial nature; if $n$ agents are such that each pair
has a shared EPR pair, then (with linear classical communication
cost) a pure $n$-partite state with maximum entanglement can be created
entangling all the $n$ agents. As stated in Theorem \ref{qhtheorem1},
the case is
stronger because just $(n-1)$ EPR pairs suffice. Due to this
similarity in combinatorial nature, we call our results in Theorem
\ref{qhtheorem3} and Theorem \ref{entanhyperiff} 
quantum Helly-type theorems.\\

\chapter{A Combinatorial Approach to Study the LOCC 
Transformations of Multipartite States}

\section{Introduction}

Given the extensive use of quantum entanglement as a 
resource for quantum information processing \cite{ekert2000,nielsen2000},
its quantification has become one of the central topics of 
quantum information theory and, of late, a lot of research has been going on 
in this direction. However, apart 
from simple cases (for example, low-dimensions, few particles, pure states etc.) 
the mathematical structure of entanglement is not yet fully 
understood. In particular, the entanglement properties of 
bipartite states have been widely explored (see \cite{brub02,hord01} for a comprehensive review). 
Fortunately, bipartite states possess a nice mathematical property
in the  form of the Schmidt decomposition \cite{nielsen2000} which encompasses
their all non-local peroperties. However, the entangled states 
involving more than two parties lack such convenient form
and so it is difficult to characterize them. 
Some approches, essentially using the generalization of Schmidt decomposition,  have been taken in this direction \cite{bennet2000,kempe99,partovi04}
; however a general formulation in this case is still an outstanding 
unresolved problem. \\ 

State transformations under local operations and classical communication (LOCC)
are very important while quantifying entanglement because LOCC can at the best 
increase classical correlations and therefore a good measure of entanglement is not
supposed to increase under LOCC. A necessary and sufficient condition for such transformation
to be possible with certainty in the case of bipartite states was given
by Nielsen \cite{neilsen99} and an immediate consequence of his result was 
the existence of {\it incomparable} states (the states which can not be 
obtained by LOCC from one another). Bennett et al. \cite{bennet2000}
formalized the notions of reducibility, equivalance and imcomparability to
multi-partite states and gave a sufficient condition for incomparability
based on {\it partial} entropic criteria.\\

All the current approaches to study the state transformation under LOCC are 
based on entropic criterion. In this work, we present a 
entirely different approach  based on  nice combinatorial properties of 
graphs and set systems. We introduce a technique called {\it bicolored merging} 
and obtain several results about  such transformations.
We demostrate a partial ordering of multi-partite states and 
various classes of incomparable multi-partite states. 
We utilize these results to establish the 
impossibility of doing {\it selective teleportation} in a case
where the apriori entanglement is in the form of a GHZ state.
We also discuss the minimum number of copies of a state required to prepare another 
state by LOCC and present bounds on this number in terms of {\it quantum distance} between
the two states.
The ideas developed in this work continues the combinatorial setting developed 
in Chapter 2 
and can been extended to incorporate other new kinds of multi-partite states.
Moreover, the idea of {\it bicolored merging} may also be appropriate to 
some other areas of information sciences.  \\  
    
\section{The Combinatorial Framework}

In this section we first revise the combinatorics developed in Chapter 2 
and introduce the framework for deriving our results.\\

\begin{defni}
 EPR Graph: For $n$ agents $A_1, A_2, \cdots , A_n$ an undirected graph $G = (V,E)$
is constructed as
follows:

$ V = \{ A_i: i=1, 2, \cdots , n \}$ ,
$ E = \{ \{ A_i, A_j \}: A_i $ and  $A_j ~\mbox{share an EPR pair}, 1 \leq i, j \leq n; i \neq j \}$.

The graph $G = (V, E)$ thus formed is called the EPR graph of the $n$ agents.
\end{defni} 

\begin{defni}
Spanning EPR Tree: An EPR graph $ G = (V, E) $ is called a spanning EPR tree if the undirected graph 
$ G = (V, E) $ is a spanning tree. 
\end{defni}

\begin{defni}
Entangled Hypergraph: Let $S$ be the set of $n$ agents and $F= \{E_1, E_2, \cdots , E_m \}$, where 
$ E_i \subset S; i = 1, 2, \cdots , m$ and $E_i$ is such that its elements (agents) are in 
$|E_i|$-partite pure maximally entangled state. The hypergraph (set system) $ H = (S, F) $  
is called an entangled hypergraph of the $n$ agents.
\end{defni} 

\begin{defni}
Connected Entangled Hypergraph: A sequence of
$j$ hyperedges $E_1$, $E_2$, ..., $E_j$ in a hypergraph $H=(S,F)$
is called a {\it hyperpath} (path) from a vertex $a$ to a vertex $b$ 
if 
\begin{enumerate}
\item  $E_i$ and $E_{i+1}$ have a common vertex
for all $1\leq i\leq j-1$,
\item $a$ and $b$ are agents in $S$, 
\item $a\in E_1$, and 
\item $b\in E_j$. 
\end{enumerate}
If there is a hyperpath between every
pair of vertices of $S$ in the hypergraph $H$, we say that $H$ is 
connected. 
\end{defni} 

\begin{defni}
Entangled Hypertree: A connected entangled hypergraph $H = (S, F) $ is called an entangled hypertree if it contains 
no cycles, that is, there do not exist any pair of vertices from $S$ such that there are two paths between them.
\end{defni}

\begin{defni}
$r$-Uniform Entangled Hypertree: An entangled hypertree is called a $r$-uniform entangled hypertree 
if all of its hyperedges are of size $r$.
\end{defni}

\begin{thm}
\label{skp1}

If any two pairs of the three agents $A$, $B$ and $C$ share EPR pairs 
(say the state $(|00\rangle + |11\rangle)/\sqrt{2}$) then we can prepare a GHZ state ($|000\rangle + |111\rangle)/\sqrt{2}$) 
amongst them with two bits of classical communication, while involving all the three agents dynamically.
\end{thm}

We would henceforth refer the protocol developed in \cite{suds} (Chapter 2)
in order to establish Theorem \ref{skp1} as 
{\it SKP-1}. 

\begin{thm}
\label{skp2}
Amongst $n$ agents in a communication network permitting only pairwise entanglement
in the form of EPR pairs, a pure $n$-partite maximally entangled state can be prepared if and 
only if the EPR graph of the $n$ agents is connected.
\end{thm}

We shall use the name {\it SKP-2} to refer to the protocol suggested in \cite{suds}
(Chapter 2) to prove 
the sufficiency of Theorem \ref{skp2}. 

\begin{thm}
\label{skp3}
Given $n$ agents in a communication network and an entangled hypergraph, a 
pure $n$-partite entangled state can be prepared amongst the $n$ agents if and only if the entangled 
hypergraph is connected.
\end{thm}

We shall use the name {\it SKP-3} to refer to the protocol suggested in \cite{suds} (Chapter 2)
to prove
the sufficiency of the above theorem.

\section{Bicolored Merging}

Monotonicity is easily the most natural characteristic that ought to be
satisfied by all entanglement measures \cite{hord01}. This means that
any appropriate measure of entanglement must not change by local
unitary operations and more generally the expected entanglement 
must not increase under LOCC. We should note here that
in LOCC, LO involves unitary transformations, additions of ancillas (that is, enlarging the 
Hilbert Space), measurements, and throwing away parts of the system, each of these 
actions performed by one party on his or her subsystem.
CC 
between the parties allows local actions by one party to be conditioned 
on the outcomes of the earlier measurements performed by the other parties. \\

\begin{figure}
\resizebox*{0.9\textwidth}{0.9\textheight}{\includegraphics{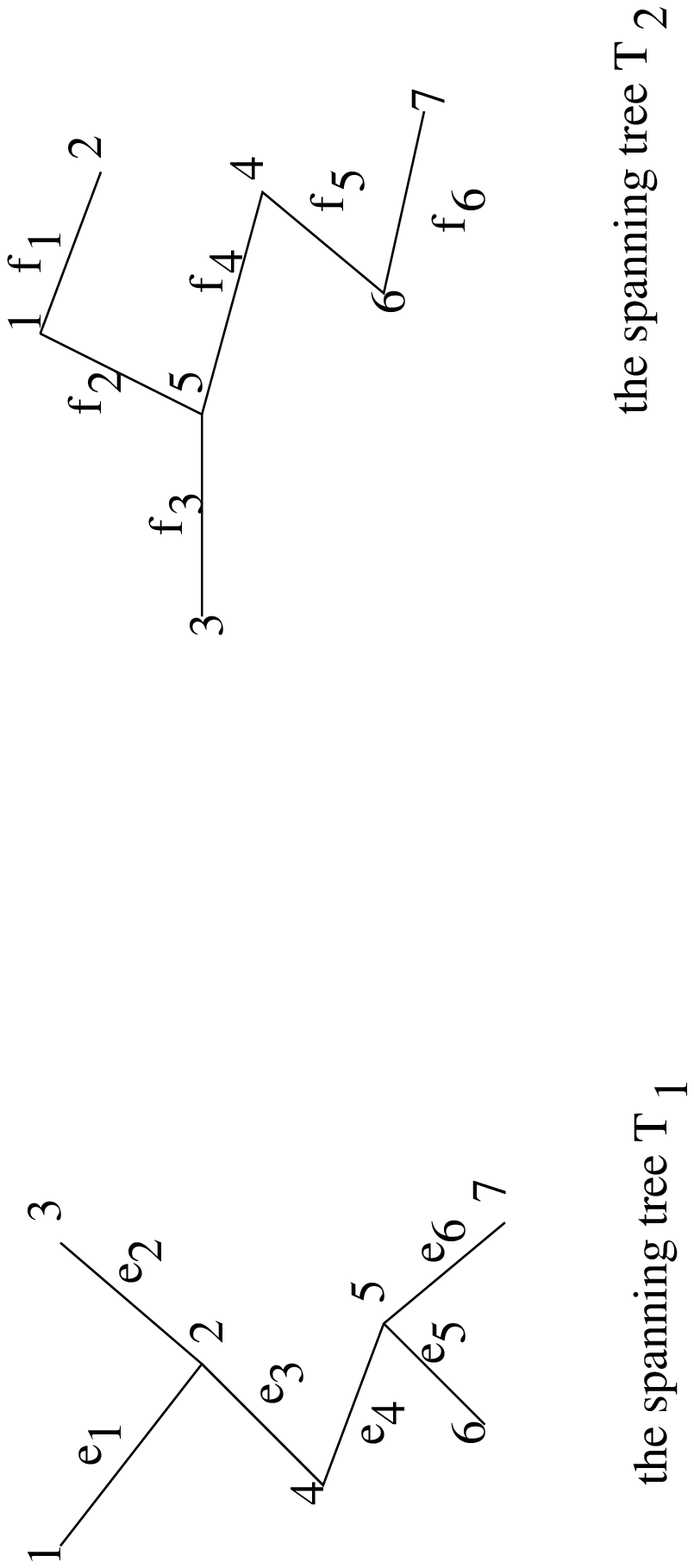}} 
\caption{The spanning EPR trees $T_1$ and $T_2$}
\label{figure30A}
\end{figure}

\begin{figure}
\resizebox*{0.9\textwidth}{0.9\textheight}{\includegraphics{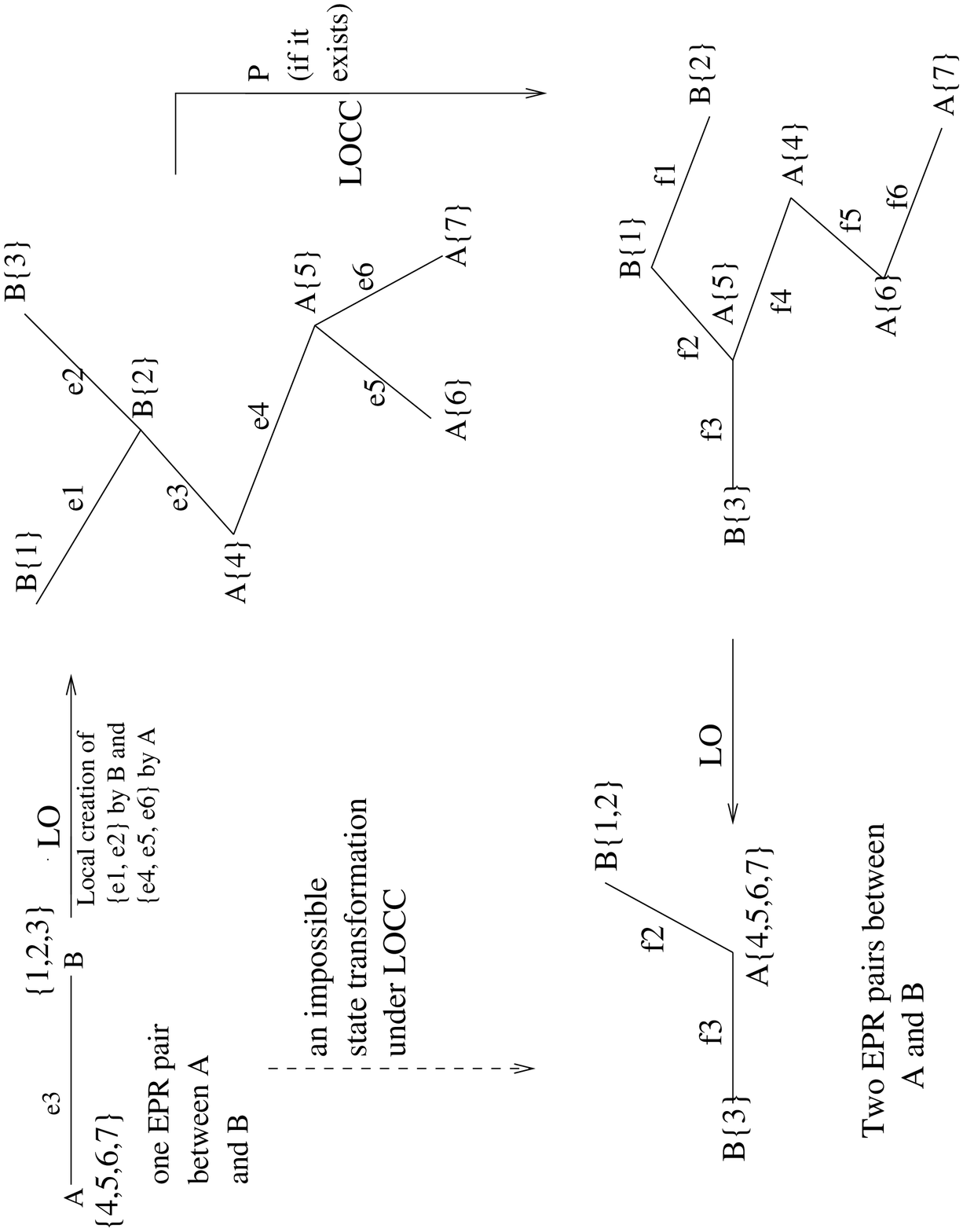}}
\caption{Converting $T_1$ to $T_2$ under LOCC through $P$}
\label{figure30C}
\end{figure}

Apart from monotonicity, there are certain other characteristics required to be satisfied 
by the entanglement measures. It is interesting to note (as we show in this work)
that monotonicity itself restricts a large number of state transformations
and gives rise to several classes of incomparable (multi-partite) states.
Thus, to study the possible state transformations of (multipartite) states 
under LOCC, it would be interesting to look at the kind of state transforms under LOCC 
which monotonicity does not allow. 
We can observe that monotonicity does not allow 
the preparation of $n+1$ or more  EPR pairs between two parties 
starting from only $n$ EPR pairs between them. 
In particular, it is not possible
to prepare two or more EPR pairs between two parties starting only with 
a single EPR pair and only LOCC. 
This is an example of impossible state transformation in bipartite case 
as dictated by the monotonicity postulate. 
Thus, we might anticipate that a large class of multi-partite states 
could also be shown to be incomparable
just by using impossibility results in bipartite case through 
a suitable reduction.
For example, consider transforming (under LOCC) the state represented by a spanning EPR tree, say $T_1$,
to that of the state represented by 
the spanning EPR tree, say $T_2$. (See the Figure \ref{figure30A}).
This transformation can be shown to be impossible by reducing to the bipartite case as follows:
Let us assume that there exists a protocol $P$ which 
can perform the required transformation. It is easy to see that the protocol 
$P$ is also applicable in the case when a party $A$ possesses all the qubits of 
parties $4, 5, 6, $ and $7$ and all the qubits of the parties $1, 2, $ and $3$ are
possessed by another party $B$. This means that party $A$ is playing the role of 
all the parties $4, 5, 6, $ and $7$ and $B$ is playing the role of 
all the parties $1, 2, $ and $3$. Therefore for the protocol $P$,
these two parties represent the complete EPR spanning tree $T_1$.
It is indeed reasonable as any LOCC actions done amongst $\{1,2,3\}$ ($\{4,5,6,7\}$) 
is reduced to just LO done by $B$ ($A$) and any CC done between one party from $\{1,2,3\}$
and the other from $\{4,5,6,7\}$ is managed by CC between $B$ and $A$. 
Thefore, starting only with one edge ($e_3$) they eventually construct $T_1$
just by LO (by local creation of EPR pairs representing the edges $e_1, e_2, e_4, e_5, $ and $e_6$;
$\{e_1, e_2\}$  by $B$ and $\{e_4, e_5, e_6\}$ by $A$). They then apply protocol $P$ to obtain $T_2$
with the edges $f_1, f_2, f_3, f_4, f_5$ and $f_6$. (Refer to the Figure \ref{figure30C}).
All edges except $f_2$ and $f_3$ are local EPR pairs (that is, both qubits are with the same party).
 Now the parties $A$ and $B$ share two EPR pairs in the form of 
the edges $f_2$ and $f_3$, though they started with sharing only one EPR pair.
This is actually an impossible state transformation under LOCC in the bipartite case. 
Hence, we can conclude that such a protocol $P$ can not exist! 
The complete reduction process is shown in Figure \ref{figure30C}  below. \\   

In general, suppose we want to show that the  multi-partite state $|\psi\rangle$ 
can not be converted to the multi-partite state $|\phi\rangle$ by LOCC.
This can be done by showing an assignment of the qubits (of all parties) 
only to two parties such that $|\psi\rangle$ can be obtained from $n$ ($n = 0, 1, 2, \cdots $)
EPR pairs between the two parties by LOCC while $|\phi\rangle$ can be converted 
to more than $n$ EPR pairs between the two parties by LOCC.
This is equivalent to saying that each party is given either of two colors (say $A$ or $B$).
Finally all qubits with parties colored with color $A$ are assigned to
the first party (say $A$) and that with parties colored with second color to
the second party (say $B$). This coloring is done in such a way that the state $|\psi\rangle$ 
can be obtained by LOCC from less number of EPR pairs between $A$ and $B$ than that can be obtained from
$|\phi\rangle$ by LOCC.
Local preparation (or throwing away) of EPR pairs is what we call merging in combinatorial sense.  
Keeping this idea in mind, we now formally introduce the idea of 
bicolored merging for such reductions in the case of 
the multi-partite states represented by EPR graphs and entangled hypergraphs. \\

Suppose that there are two EPR graphs $G_1=(V, E_1)$ and $G_2=(V, E_2)$ on the same 
vertex set $V$ (meaning that the two multi-partite states are shared amongst the 
same set of parties) and we want to show the impossibility of transforming 
$G_1$ to $G_2$ under LOCC, then this is reduced to a bipartite LOCC transformation 
which violates monotonicity, as follows:

\begin{enumerate}
\item Bicoloring: Assign either of the two colors $A$ or $B$ to every vertex, that is, 
each element of $V$. 
\item Merging: For each element $\{v_i, v_j\}$ of $E_1$, merge the two vertices $v_i$ and 
$v_j$ if and only if they have been assigned the same color during the bicoloring stage and assign 
the same color to the merged vertex. Call this graph obtained from $G_1$ as BCM (Bicolored-Merged) EPR graph of 
$G_1$ and denote it by $G_1^{bcm}$. Similarily, obtain the BCM EPR graph $G_2^{bcm}$ of 
$G_2$.
\item The bicoloring and merging is done in such a way that the graph $G_2^{bcm}$ has 
more number of edges than that of $G_1^{bcm}$. 
\item Give all the qubits possessed by the vertices with color $A$ to the first party (say, party $A$)
and all the qubits possessed by the vertices with color $B$ to the second party (say, party $B$).
Combining this with the previous steps, it is ensured that in the bipartite reduction of the multi-partite state
represented by $G_2$, the two parties $A$ and $B$ share more number of EPR pairs (say, state $|\psi_2\rangle$)
than that for 
$G_1$ (say, state $|\psi_1\rangle$).  
\end{enumerate}

\begin{figure}
\resizebox*{0.9\textwidth}{0.9\textheight}{\includegraphics{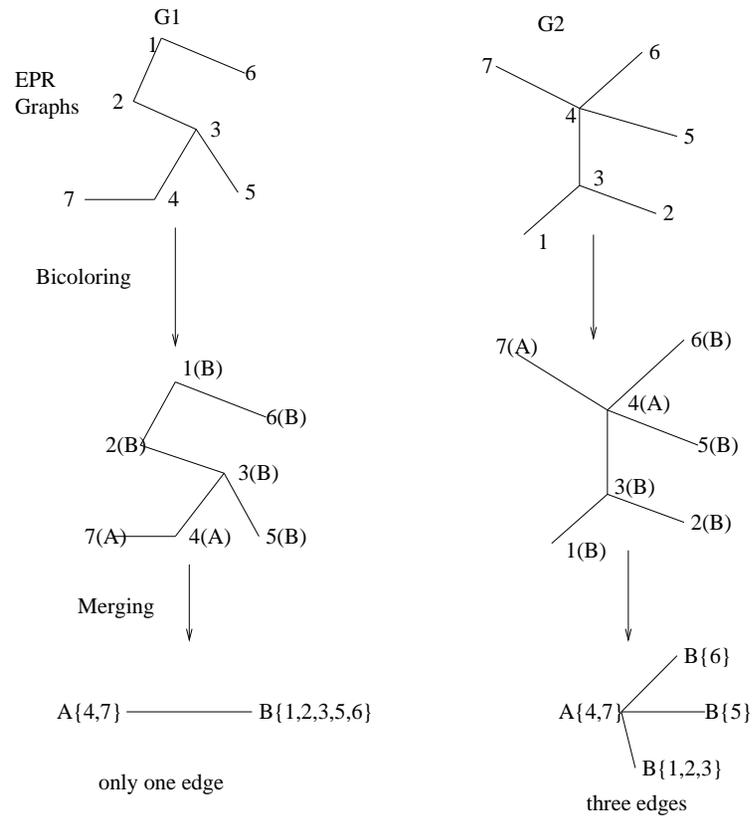}}
\caption{Bicolored Merging of EPR Graphs}
\label{figure31}
\end{figure}

Now if there exits a protocol $P$ which can transform $G_1$ to $G_2$ by LOCC, then 
$P$ can also transform $|\psi_1\rangle$ to $|\psi_2\rangle$ just by LOCC as follows:
$A$ ($B$) will play the role of all vertices in $V$ which were colored as $A$ ($B$).
The edges which were removed due to merging can easily be cretated by local operations 
(local preparation of EPR pairs) by the party $A$ ($B$) if the color of the merged end-vertices 
of the edge was assigned color $A$ ($B$). This means that starting from $|\psi_1\rangle$
and only LO, $G_1$ can be created. This graph is virtually amongst $|V|$ parties even though 
there are only two parties. The protocol $P$ then, can be applied to $G_1$ to obtain $G_2$ by LOCC.
Subsequently $|\psi_2\rangle$ can be obtained by the necessary merging of vertices by LO, that is by 
throwing away the local EPR pair represented by the edges between the vertices being merged.
Since the preparation of $|\psi_2\rangle$ from $|\psi_1\rangle$ by LOCC violates monotonocity postulate,
such a protocol $P$ can not exist! An example of bicolored merging for EPR graphs has been 
illustrated in Figure \ref{figure31}. \\

The bicolored merging in the case of entangled hypergraphs is essentially the same as that for 
EPR graphs. For the sake of completeness, we present it here.
Suppose there are two entangled hypergraphs $H_1=(S, F_1)$ and $H_2=(S, F_2)$ on the same 
vertex set $S$ (that is, the two multi-partite states are shared amongst the 
same set of parties) and we want to show the impossibility of transforming 
$H_1$ to $H_2$ under LOCC. Transformation of $H_1$ to $H_2$ can be reduced to a bipartite LOCC transformation 
which violates monotonicity thus proving the impossibility. The reduction is done as follows:

\begin{enumerate}
\item Bicoloring: Assign either of the two colors $A$ or $B$ to every vertex, 
that is, each element of $S$. 
\item Merging: For each element $E=\{v_{i1}, v_{i2}, \cdots ,  v_{ij}\}$ of $F_1$, merge 
all vertices with color $A$ to one vertex and those with color $B$ to another vertex and give 
them colors $A$ and $B$ respectively. This merging collapses each hyperedge to either a simple 
edge or a vertex and thus the hypergraph reduces to a simple graph with vertices assigned with either 
of the two colors $A$ or $B$.  
Call this graph obtained from $H_1$ as BCM EPR graph of 
$H_1$ and denote it by $H_1^{bcm}$. Similarily obtain the BCM EPR graph $H_2^{bcm}$ of 
$H_2$.
\item The bicoloring and merging is done in such a way that the graph $H_2^{bcm}$ has 
more number of edges than that of $H_1^{bcm}$. 
\item Give all the qubits possessed by the vertices with color $A$ to the party one (say party $A$)
and all the qubits possessed by the vertices with color $B$ to the second party (say party $B$).
\end{enumerate}
Rest of the discussion goes exactly as in the EPR graph case. In the Figure \ref{figure32} below,
we demostrate the bicolored merging of entangled hypergraphs. \\

\begin{figure}
\resizebox*{0.9\textwidth}{0.9\textheight}{\includegraphics{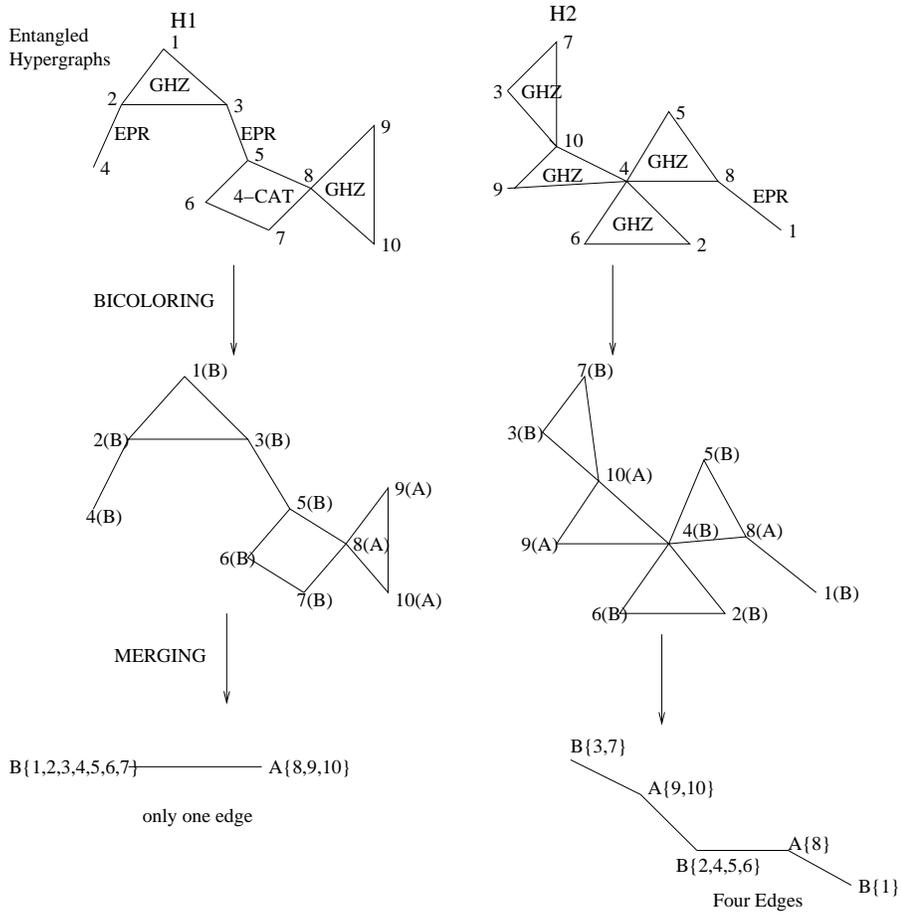}}
\caption{Bicolored Merging of Entangled Hypergraphs}
\label{figure32}
\end{figure}

It is interesting to note at this point that the LOCC incomparability shown
by using the method of bicolored merging is in fact {\it strong incomparability} \cite{vwani02}.
We would also like to stress that any kind of reduction (in particular, 
various possible extensions of bicolored merging) which leads to the violation of {\it any} of
the properties of a potential entanglement measure, is pertinent to show the impossibility
of many multi-partite state transformations under LOCC. Since the bipartite case has been extensively studied,
such reductions can potentially provide many ideas about multi-partite case by just exploiting 
the results from bipartite case. In particular, the definitions of EPR graphs and entangled hypergraphs
could also be suitably extended to capture more types of multi-partite pure states and even mixed states 
and a generalization of the idea of bicolored merging as a suitable reduction for this case could also
be worked out. It would be interesting to investigate such issues. \\

\section{Irreversibility of SKP-1 and Selective Teleportation}

We know that a GHZ state amongst three agents $A$, $B$ and $C$ 
can be prepared from EPR pairs shared between any two pairs of
the three agents using only LOCC \cite{suds}. 
We consider the problem of {\it reversing} 
this operation, that is, whether it is possible
to construct two EPR pairs between 
any two pairs of the three agents from a GHZ state amongst the three 
agents, using only LOCC. By using the method of bicolored merging, we show below that this is not possible.\\

\begin{figure}
\resizebox*{0.9\textwidth}{0.9\textheight}{\includegraphics{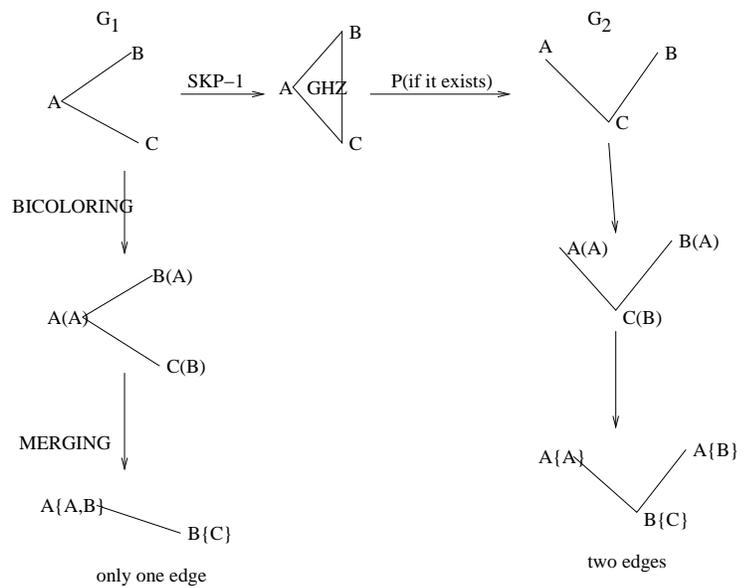}}
\caption{Irreversibility of SKP-1}
\label{figure33}
\end{figure}

Suppose there exists a protocol $P$ for {\it reversing} a GHZ state  
into two EPR pairs using only LOCC. 
More precisely, protocol $P$ starts with a GHZ 
state amongst the agents 
$A$, $B$ and $C$, and prepares EPR pairs between any two pairs of  
$A$, $B$ and $C$ (say, $A$ and $C$, and $B$ and $C$).
Since we can prepare the GHZ state from EPR pairs between any two pairs
of the three agents, we can prepare the GHZ state starting from EPR pairs
between $A$ and $B$, and $A$ and $C$. Once the GHZ state is prepared, we 
can apply protocol $P$ to construct EPR pairs between $A$ and $C$ and between $B$
and $C$ using only LOCC. So, we can use only LOCC to convert a 
configuration where EPR pairs exist between $A$ and $C$ and between $A$ and $B$, to
a configuration where EPR pairs are 
shared between $A$ and $C$ and between $B$ and $C$. 
It can be noted that the configuration where the two EPR pairs are 
shared between
${A,B}$ and ${A,C}$ (say, EPR graph $G_1$) is symmetrical with respect to the GHZ state amongst 
$A, B$ and $C$ to the configuration where the two EPR pairs are shared 
between ${A,C}$ and ${B,C}$ (say, the EPR graph $G_2$).
Apply the bicolored merging by giving the color $A$ to parties $A$ and $B$ and
the color $B$ to the party $C$.
We can observe that 
$A$ and $C$ start with a single EPR pair between themselves and (by only LOCC)
end up sharing two EPR pairs between themselves (Figure \ref{figure33}).
The same result could also be achieved by similar bicolored merging
directly applied on the GHZ state and any of $G_1$ or $G_2$ but
we prefer the above proof for stressing the argument on the symmetry 
of $G_1$ and $G_2$ with respect to the GHZ. Moreover,
this proof gives an intuition about possibility of incomparability
amongst spanning EPR trees as $G_1$ and $G_2$ are two
distinct spanning EPR trees on three vertices.
We prove this general result in the Theorem \ref{twoeprtrees}  \\ 

By repeatedly applying the protocol $P$ (if possible),
we can indeed prepare as many EPR pairs bewteen 
$A$ and $C$ (using only LOCC) as we
wish, starting from a single shared EPR pair.
This is impossible and so our assertion is proved. 
We summarize this result in the following theorem.\\

\begin{thm}
\label{skpghzreverse}
Starting from a GHZ state shared amongst three parties in a 
communication network, two EPR pairs can not be created 
between any two sets of two parties using only LOCC.

\end{thm}

The above theorem motivates us to think of some kind of comparison 
between 
a GHZ state and two pairs of EPR pairs in terms of the 
non-local correlations they possess. In this sense, therefore, 
a GHZ state is stictly less than two EPR pairs.
This is also easy to see that an EPR pair between
any two parties can be obtained starting only 
from a GHZ state shared amongst the three parties and 
LOCC. The third party will just do a measurement  and send
the result to other two. By applying the corresponding 
suitable operations they get the required EPR pair.
Using the necessary conditions presented in \cite{suds} (Chapter 2)
to prepare a multi-partite
entangled state starting from only bi-partite entanglement 
in a distibuted network, we observe that an EPR pair between 
any two of the three parites is not sufficient for preparing 
a GHZ state amongst the three parties using only LOCC.
These arguments can be summarised in the following 
theorem.\\

\begin{thm}
\label{eprghz}

1-EPR pair $<_{LOCC}$ a GHZ state $<_{LOCC}$ 2-EPR pairs 

\end{thm}

An interesting problem in quantum information theory is that of
{\it selective teleportation} \cite{mks}. Given three agents $A$, $B$ 
and $C$, and two qubits of unknown quantum states 
$|\psi_1\rangle$ 
and
$|\psi_2\rangle$ with $A$, the problem is to send 
$|\psi_1\rangle$ to $B$ and  
$|\psi_2\rangle$ to $C$ selectively, using only LOCC and
apriori entanglement between the three agents.
A simple solution to this problem is by applying standard 
teleportation \cite{bennett93}, in the case where $A$ shares 
EPR pairs with both $B$ and $C$.
An interesting question is whether any other 
form of apriori entanglement
can help achieving selective teleportation. In particular,
is it possible to perform selective teleportation where 
the apriori entanglement is in the form of a GHZ state amongst the 
three agents. Following theorem answers this question using the 
result of the Theorem \ref{skpghzreverse}. \\

\begin{thm}
\label{seltel}
With a prior entanglement given in the form of a
GHZ state shared amongst
three agents, two qubits can not be
selectively teleported by ane of the three
parties to the other two parties. 

\end{thm}

Proof: Suppose there exists a protocol $P$ which 
can enable one of the three parties (say $A$)
to teleport two qubits $|\psi_1\rangle$ and 
$|\psi_2\rangle$ selectively to the other two parties
(say $B$ and $C$). 
Now $A$ takes four qubits; she prepares two EPR pairs 
one from the first and second qubits and the other from
the third and fourth qubits. He then teleports the first
and third qubits selectively to $B$ and $C$ using $P$
( consider first qubit as $|\psi_1\rangle$ and the third  
qubie as $|\psi_2\rangle$ ). We can note here that 
in this way $A$ is able to share one EPR pair each 
with $B$ and $C$. But this is impossible for it enables
$A$ to prepare two EPR pairs starting from a GHZ state 
and only using LOCC which already we have proved
(Theorem \ref{skpghzreverse}) not to be possible.
Hence follows the result. \hfill \qed

\section{A Partial Ordering of Entangled Hypergraphs}

In this section, we investigate whether some kind 
of comparison and ordering can be made between various 
multi-partite entangled states based on the 
non-local content contained in them and establish the
following theorems. 

\begin{thm}
\label{skp2rev}

None of the inductive steps of SKP-2 can be reversed.
Hence the non-local content contained in the multi-partite
state represented by the structure in a step is stictly 
less than that of the multi-partite states represented by the 
structure in the subsequent steps. \\

\end{thm}

\begin{figure}
\resizebox*{0.9\textwidth}{0.9\textheight}{\includegraphics{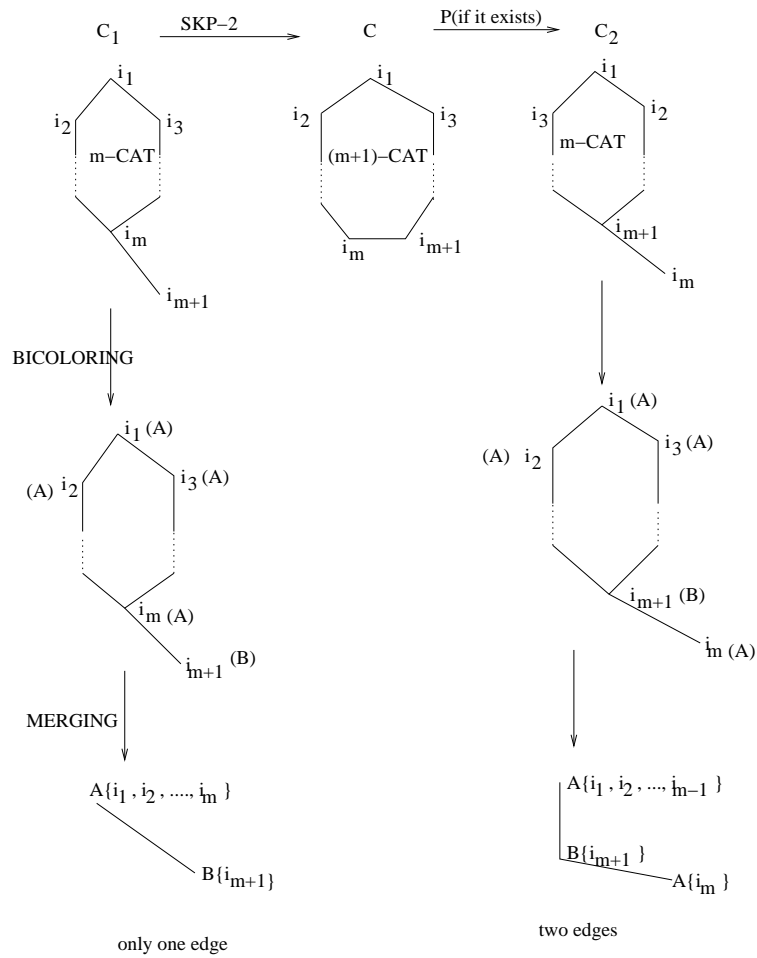}}
\caption{Inductive steps of SKP-2 are irreversible}
\label{figure34}
\end{figure}

Proof: The proof goes much similar as the proof of irreversibility of SKP-1.
Suppose in an inductive step the $m$ parties $\{i_1, i_2, \cdots ,i_m\}$ already entangled 
in a $m$-partite maximally entangled state, with help of EPR pair shared between $i_m$
 and $i_{m+1}$
(call this configuration $C_1$) prepare an $(m+1)$-partite maximally entangled state amongst $\{i_1, i_2, \cdots , i_m, i_{m+1} \}$
( call this configuration $C$).
We are interested in proving the impossibility of reversing this step.
For the sake of contradiction assume that such a protocol $P$ for this reversing exists.
Then it is also easy to see that starting with the $(m+1)$-partite maximally entangled state
amongst $\{i_1, i_2, \cdots , i_m, i_{m+1} \}$ and using LOCC one can get a configuration (call $C_2$)
where the parties $\{ i_1, i_2, \cdots , i_{m+1} \}$ are in $m$-partite maximally entangled state
and parties $i_m$ and $i_{m+1}$ are sharing an EPR pair (use the symmetry argument as in the proof of 
irreversibility of SKP-1).      
Thus it is possible to convert state represented by $C_1$ to that by $C_2$ just using LOCC by first 
converting $C_1$ to $C$ by the inductive step of SKP-2 and subsequently applying $P$ to $C$.
This state transformation under LOCC from $C_1$ to $C_2$ is shown impossible by bicolored merging 
where the color $A$ is asigned to the parties $\{i_1, i_2, \cdots , i_m \}$ and the color $B$ to
the party $i_{m+1}$. The detail is suggested in the Figure \ref{figure34}. 
\hfill \qed \\

\begin{thm}
\label{skp3rev}
None of the inductive steps of SKP-3 can be reversed.
Hence the non-local content contained in the multi-partite 
state represented by the entangled hypergraph in a step is 
strictly less than that of the multi-partite state represented 
by the entangled hypergraphs in the subsequent steps.
\end{thm}

\begin{figure}
\resizebox*{0.9\textwidth}{0.9\textheight}{\includegraphics{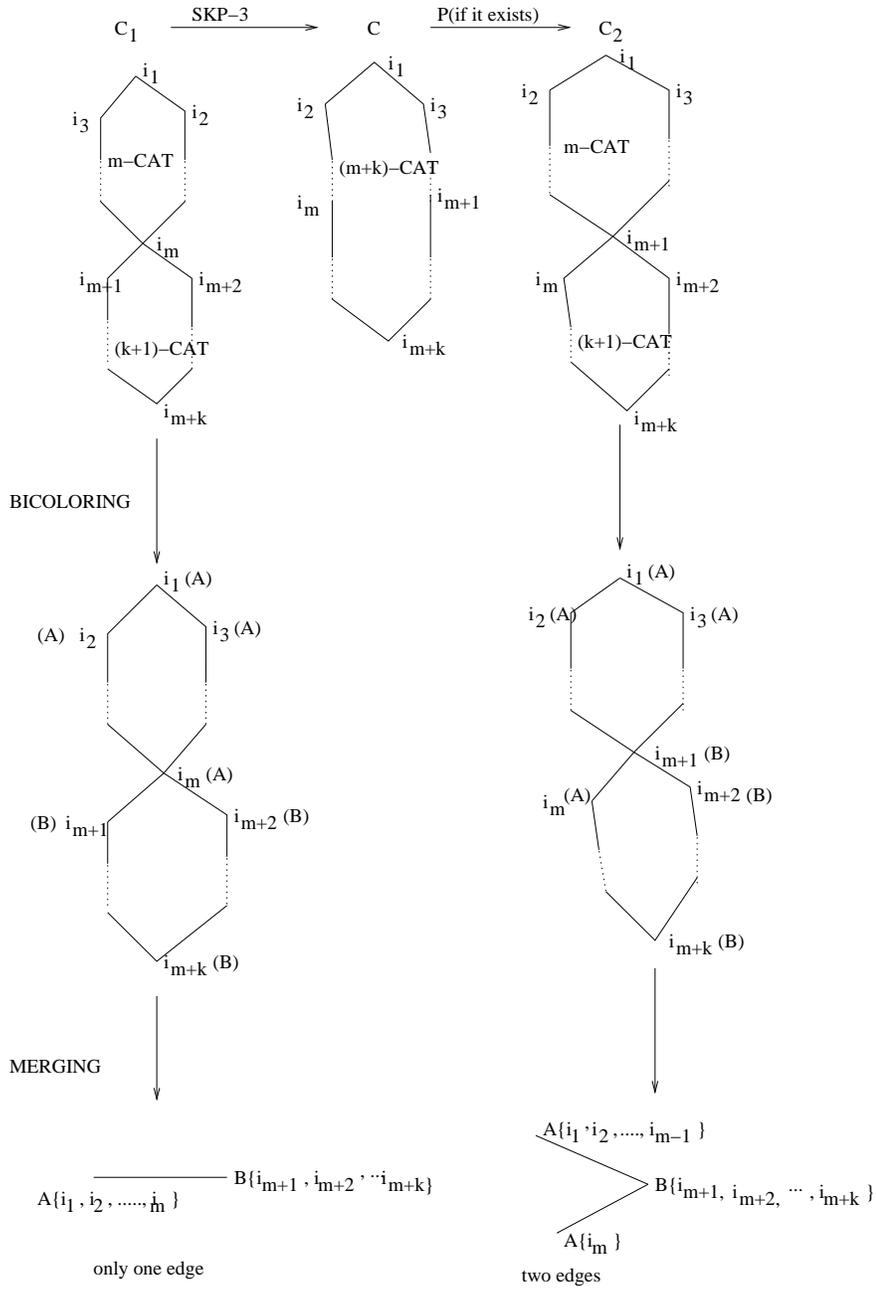}}
\caption{Inductive steps of SKP-3 are irreversible}
\label{figure35}
\end{figure}

Proof: The proof is exactly same as the proof of the last theorem using bicolored merging and 
we just depict it by Figure \ref{figure35}. \hfill \qed \\

It is worth noting at this point that the above theorems gives partial orderings 
of multi-partite states. This order thus relatively quantifies the entanglement 
for these states.

\section{Classifying Multi-partite Entanglement.}  
An immediate result comparing an n-CAT state with EPR pairs follows from 
Theorem \ref{eprghz} and Theorem \ref{skp2rev}. \\

\begin{thm}
\label{catepr}

$ 1-EPR  <_{LOCC} n-CAT <_{LOCC} (n-1)-EPR$.

\end{thm}

On similar lines we can argue that an $n$-CAT 
state amongst $n$-parties can not, by just using LOCC,
be converted to any form of entanglement structure which possess EPR pairs 
between any two or more different sets of two parties.
Let this be possible then the two edges could be in either of the two forms:

(1) $\{i_1,i_2\}$ and $\{j_1, j_2\}$

(2) $\{i_1,i_2\}$ and $\{i_2, j_2\}$

where $i_1, i_2, j_1, j_2$ are all distinct.

In bicolor-merging assign the colors as follows:

In case (1), give color $A$ to $i_2$ and $j_2$ and give the color $B$ to the 
rest of the vertices.

In case (2), give color $A$ to $i_2$ and color $B$ to the rest of the vertices. 

Thus our above assertion follows.  
Moreover, from the necessity condition in \cite{suds} (Chapter 2) for preparing 
an $n$-CAT state, no disconncted EPR graph would be able to yield 
$n$-CAT just by LOCC. These two observations combined together 
leads to the following theorem which signifies the fact that these 
two multi-partite states can not be compared. \\
  
\begin{thm}
\label{catgraph}

A CAT state amongst $n$ agents in a communication network
is LOCC incomparable to any disconnected EPR graph associated 
with the $n$ agents having more than one edge.

\end{thm}

The above result indicates that there are many possible form of
entanglement structures (multi-partite states) which can not be compared 
at all in terms of non-local contents they deserve to have and this
simple result was just an implication of the necessity combinatorics 
required for the preparation of the CAT states. 
One more interesting question while still in the domain of that combinatorics 
is to compare an spanning EPR tree and a CAT state. 
An spanning EPR tree is a sufficient combinatorics to prepare the CAT state and
thus seems to entail more non-local content than in a CAT state but whether in 
a strict sense is still required to be investigated. 
It is easy to see that an EPR pair between any two parites can be obtained 
starting from a CAT state shared amongst the $n$ agents just by LOCC (theorem\ref{catepr}).
Therefore, given $n-1$ copies of the CAT state we can build all the $n-1$ 
edges of any spanning EPR tree just by LOCC. 
But whether this is the lower bound on the number of copies of 
$n$-CAT required to obtain an spanning EPR tree is even more interesting.
The following theorem shows that this indeed is the lower bound.  \\  

\begin{thm}
\label{treecat}  

Starting with only $n-2$ copies of $n$-CAT state shared amongst its 
$n$ agents, any spanning EPR tree of the $n$ agents can not be 
obtained just by LOCC. 

\end{thm}

\begin{figure}
\resizebox*{0.9\textwidth}{0.9\textheight}{\includegraphics{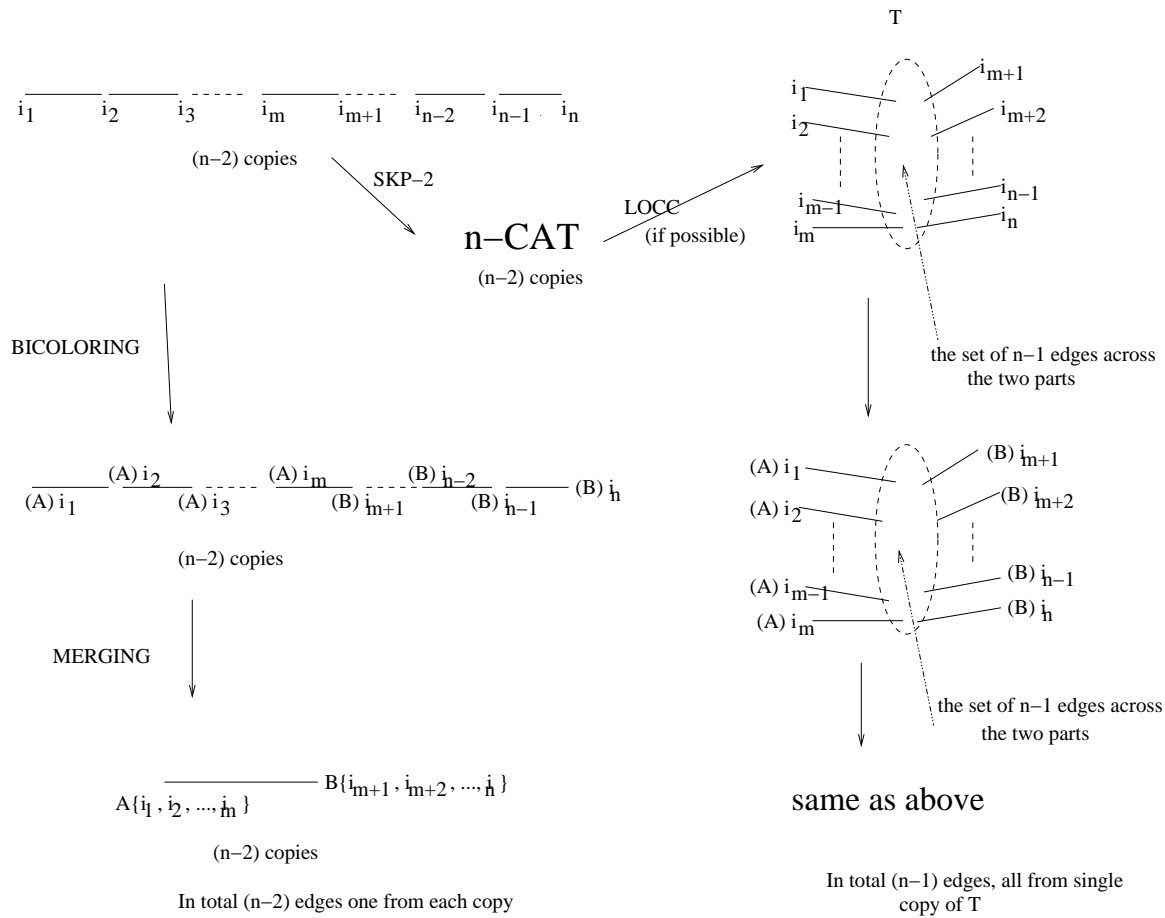}}
\caption{$n-2$ copies of $n$-CAT are not sufficient to prepare an spanning EPR tree }
\label{figure36}
\end{figure}

{\it Proof:} 
Suppose it is possible to create a spanning EPR tree T 
from $(n-2)$ copies of n-CAT states.
As we know, an $n$-CAT state can be prepared from any spanning
EPR tree by LOCC (use SKP-2).
Thus if $(n-2)$ copies of $n$-CAT can be converted to $T$ then $(n-2)$ 
copies of any spanning EPR tree can be converted to $T$ just by LOCC. 
In particular, $(n-2)$ copies of a chain EPR graph 
(which is clearly a spanning EPR tree) can be converted to $T$ just by LOCC.  
Now, we know that any tree is a bipartite connected graph with $n-1$ 
edges across the two parts. Let $i_1,i_2, \cdots ,i_m$ be the members of 
first group and the rest are in the other group. 
Construct a chain EPR 
graph where the first $m$ vertices are $i_1, i_2,...,$ and $ i_m$
in a sequence and the rest 
of vertices are from the other group in the sequence (Figure \ref{figure36}). 
As in our usual proofs,
 we give the color $A$ to the parties $\{i_1, i_2, \cdots ,i_m\}$
and the rest of the parties are given the color $B$.
This way we are able to create $(n-1)$ EPR pairs 
(Note that there are $n-1$ edges in $T$ across the two groups) 
between $A$ and $B$ starting only from
 $(n-2)$ EPR pairs. 
Therefore, we conclude that $(n-2)$ 
copies of $n$-CAT can not be converted to any spanning 
EPR tree just by LOCC. See Figure \ref{figure36} for illustration of required bicolored merging.
The proof could also be acheived by similar kind of bicolored merging directly
applied on $n$-CAT and $T$. \hfill \qed \\

In the preceding results we tried to compare an spanning EPR trees with 
CAT states. What about two different spanning EPR trees? 
Are they comparable ? 
The following theorem
targets to answer these questions.    

\begin{thm}
\label{twoeprtrees}
 Any two distinct spanning  EPR trees are LOCC-incomparable. 
\end{thm}

\begin{figure}
\resizebox*{1.0\textwidth}{1.0\textheight}{\includegraphics{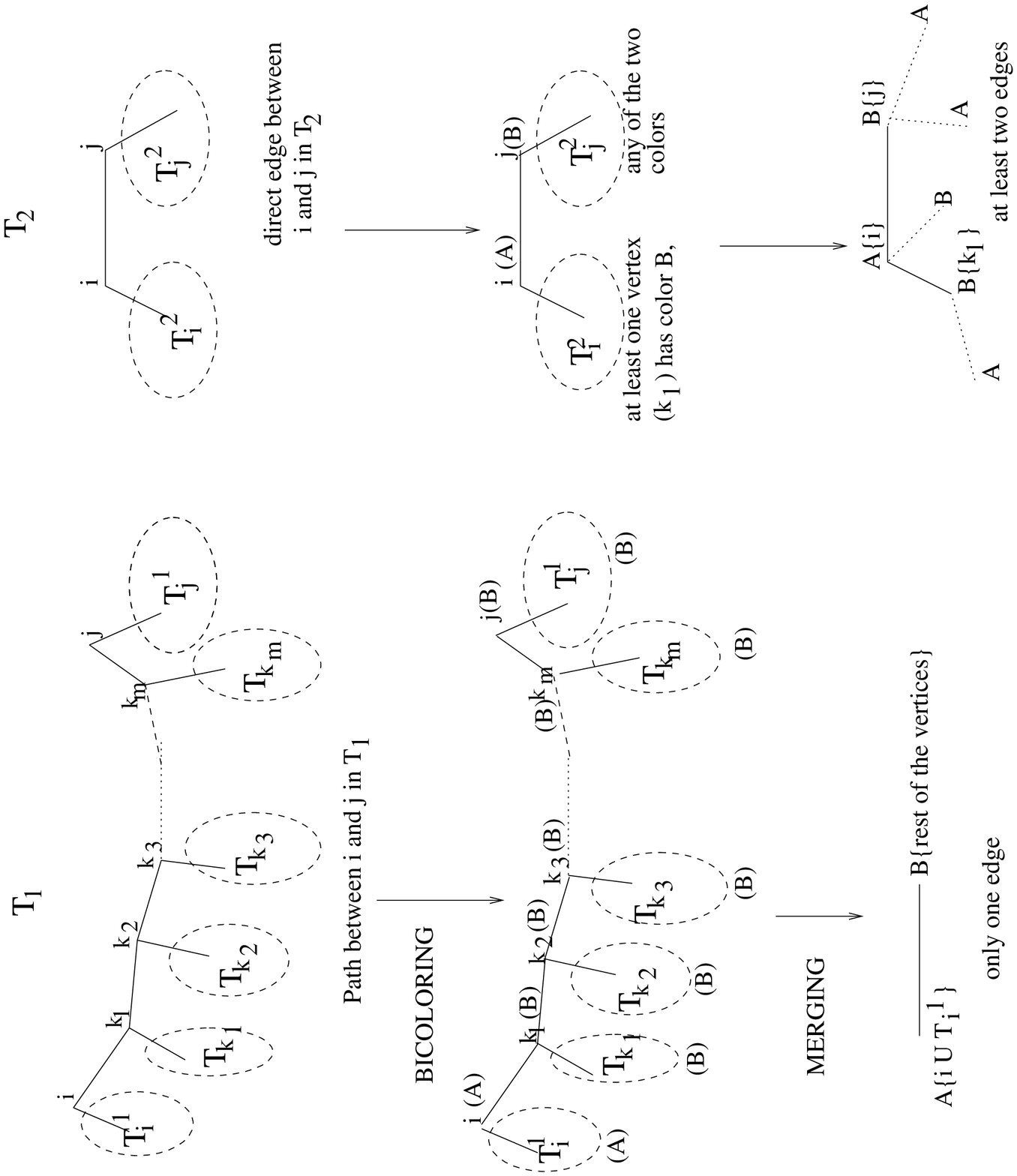}}
\caption{ Spanning EPR trees are LOCC incomparable}
\label{figure37}
\end{figure}

{\it Proof:}
Let $T_1$ and $T_2$ be the two respective spanning EPR trees.
By the way of number of edges ($n-1$) in the spanning tree,
there exists two vertices (say $i$ and $j$) which are 
connected by an edge in $T_2$ but not in $T_1$.
Also by virtue of connectedness of spanning trees,
there will be a path between $i$ and $j$ in $T_1$.
Let this path be $i k_1 k_2 \cdots k_m j$ with
$m > 0$ (See figure \ref{figure37}). 
Since $m > 0$, $k_1$ must exist.

Let $T_i^{1} =$ subtree in $T_1$rooted at $i$ except for the branch which contains 
the edge $\{i ,k_1\}$. \\

$T_j^{1} =$ subtree in $T_1$ rooted at $j$ except for the branch which contains 
the edge $\{j ,k_m\}$. \\

$T_{k_r} =$ subtree in $T_1$ rooted at $k_r$ except for the branches which contain 
either of the edges $\{K_{r-1} ,k_r\}$ and $\{K_r ,k_{r+1}\}$ ($k_0=i, k_{m+1}=j$). \\

Let $T_i^{2} =$ subtree in $T_2$rooted at $i$ except for the branch which contains 
the edge $\{i ,j\}$. \\

$T_j^{2} =$ subtree in $T_2$ rooted at $j$ except for the branch which contains 
the edge $\{i ,j\}$. \\

It is easy to see that the set $T_i^{2} \bigcup T_j^{2}$ is not empty for
$T_1$ and $T_2$ being distint must contain more than two vertices.
Also $T_i^{2}$ and $ T_j^{2}$ are disjoint otherwise there will be
a path between $i$ and $j$ in $T_2$ which does not contain the edge $\{i,j\}$,
and there will be thus two paths between $i$ and $j$ contradicting the fact 
that $T_2$ is a spanning EPR tree (Figure \ref{figure37}).   
With these two charactistics of $T_i^{2}$ and $ T_j^{2}$,
it is clear that $k_1$ will lie either in $T_i^{2}$ or in $ T_j^{2}$.
Without loss of generality let us assume that $k_1 \in T_i^{2}$.
Now we do bicolored merging where the color $A$ is assigned to
$i$ and all vertices in $T_i^{1}$ and rest of the vertices are
assigned the color $B$. Refer to Figure \ref{figure37} for 
illustraion.   
Since $T_1$ and $T_2$ were choosen arbitrarily, 
the same arguments also imply that there can not exist 
a protocol which can convert $T_2$ to $T_1$.
Hence we lead to conclusion that any two distinct spanning EPR trees are 
LOCC imcomparable.\\
  
\begin{crl}
\label{treenum}
There are exponentially many LOCC-incomparable pure multi-partite 
entangled states.

\end{crl}
{\it Proof:}
We know from results in graph theory \cite{ndeo} that on a labelled 
graph on $n$ vertices, there are $n^{n-2}$ different spanning trees
possible. Hence there are $n^{n-2}$ different spanning EPR trees 
in a network of $n$ agents.
From the theorem \ref{twoeprtrees} all these spanning 
EPR trees are LOCC imcomparable. Hence the result.\\

Since entangled hypergraphs represent more general entanglement structures 
than that represented by the EPR graphs (in particular spanning EPR trees are nothing 
but 2-uniform entangled hypertrees), it is likely that there will be even more 
classes of incomparable multi-partite states and this motivates us for
generalizing the theorem \ref{twoeprtrees} for entangled hypertrees, however 
remarkably this intuition does not work directly and there are
entangled hypertrees which are not incomparable.
An immediate contradiction comes from the partial ordering of 
entangled hypergraphs dictated by the theorem \ref{skp3rev}.   
However, there are still a large number of entangled 
hypertrees which do not fall under any such partial ordering and
thus remains incomparable. We investigate such states here below.
We need this important definition.

\begin{defni}
Pendant Vertex: A vertex of a hypergraph $H=(S,F)$ such that it belongs 
to only one hyperedge of $F$ is called a pendant vertex in $H$.
Vertices which belong to more than one hyperedge of $H$ are 
called non-pendant. 
\end{defni}

We are now ready to present our first imcomparability result 
on entangled hypergraphs.\\ 

\begin{thm}
\label{pendhypincomp}
Let $H_1=(S,F_1)$ and $H_2=(S,F_2)$ be two entangled hypertrees.
Let $P_1$ and $P_2$ be the set of pendant vertices of $H_1$ 
and $H_2$ respectively. If the sets $P_1 \setminus P_2$ and $P_2 \setminus P_1$ are both 
non-empty then the multi-partite states represented by $H_1$ and $H_2$ are
necessarily LOCC-incomparable.
\end{thm} 

Proof: First we show by using bicolored merging that 
$H_1$ can not be converted to $H_2$ under LOCC.
Impossibility of the reverse conversion will also be immediate.
From the hypothesis, $P_1 \setminus P_2$ is non-empty, therefore 
there exists $u \in S$ such that $u \in P_1 \setminus P_2$.
This means to say that $u$ is pendant in $H_1$ but non-pendant 
in $H_2$.

\begin{figure}
\resizebox*{0.9\textwidth}{0.9\textheight}{\includegraphics{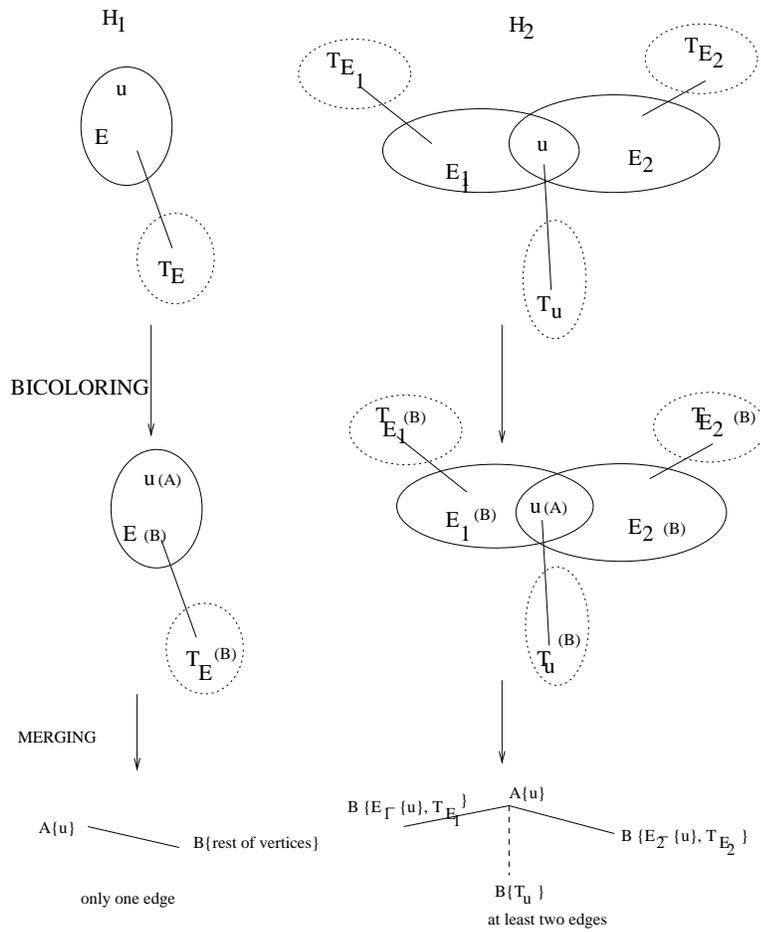}}
\caption{Entangled hypergraphs with $P_1 \setminus P_2$ non-empty}
\label{figure30D}
\end{figure}

The bicolored merging is then done where the color $A$
is assigned to the vertex $u$ and all other vertices are assigned 
the color $B$. This way $H_1$ reduces to a single EPR pair shared 
between the two parties $A$ and $B$ where as 
$H_2$ reduces to two EPR pairs shared between $A$ and $B$.
The complete becolored merging is shown in figure \ref{figure30D}. \hfill \qed \\

We can note that this proof does not utilize the fact that
$H_1$ and $H_2$ are entangled hypertrees, and thus 
the theorem is indeed true even for entangled hypergraphs 
satifying the conditions specified on the set of pendant vertices.\\

The conditions specified on the set of pendant vertices in the theorem
\ref{pendhypincomp} cover a very small fraction of the entangled hypergraphs.
However this condition is not necessary and thus there may be other 
classes and characterization which can decipher such incomparable classes 
of entangled hypergraphs. In passing we first give some examples,
for whenever the above conditions are not satisfied $H_1$ and $H_2$ may or 
may not be incomparable. \\   

\begin{figure}
\resizebox*{0.9\textwidth}{0.9\textheight}{\includegraphics{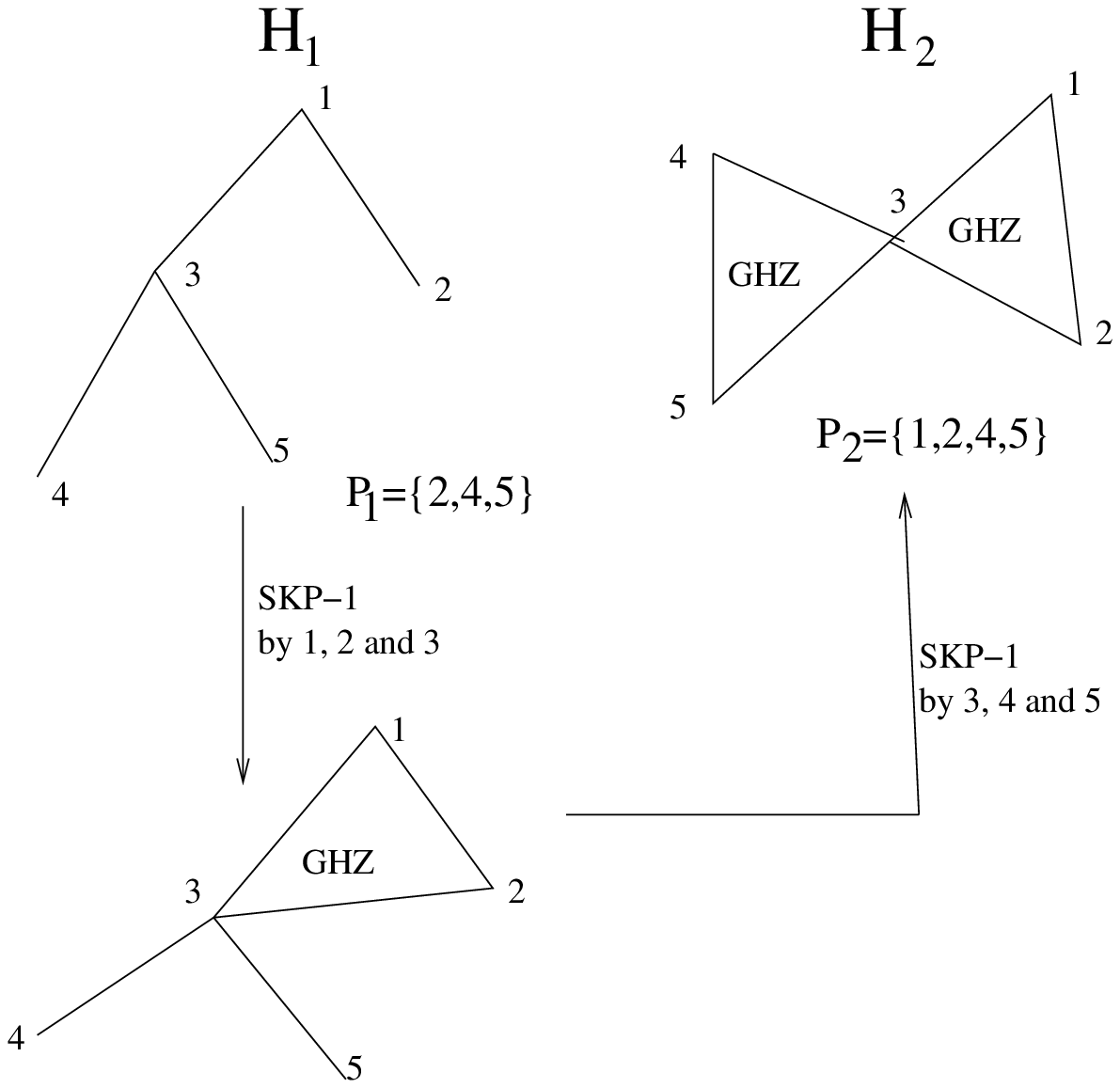}}
\caption{Comparable with $P_1 \neq P_2$ and $P_1 \subset P_2$}
\label{figure38}
\end{figure}

\begin{figure}
\resizebox*{0.9\textwidth}{0.9\textheight}{\includegraphics{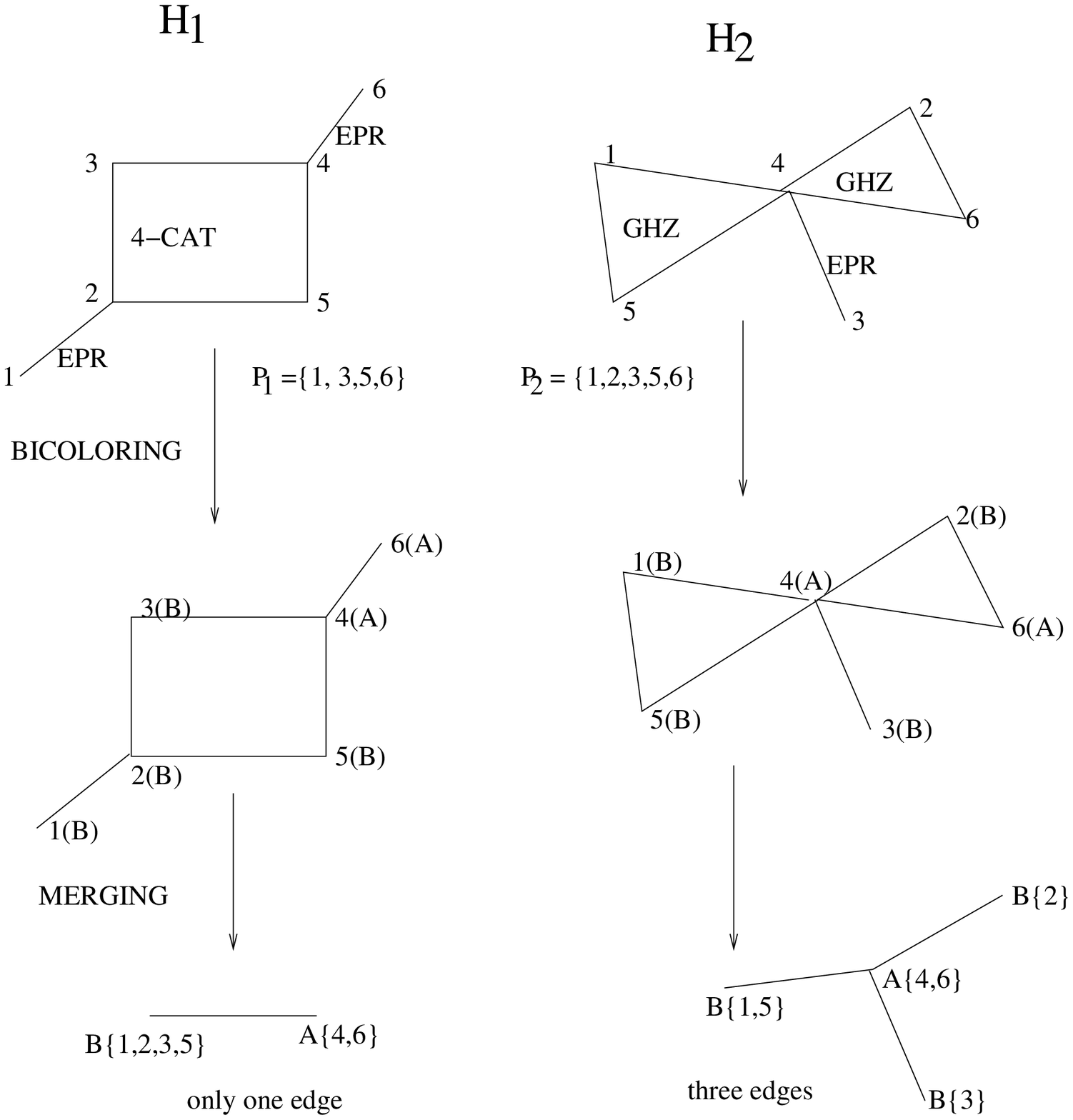}}
\caption{InComparable with $P_1 \neq P_2$ and $P_1 \subset P_2$}
\label{figure39}
\end{figure}

{\it Example-1}:(Figures \ref{figure38} and \ref{figure39})
$P_1 \neq P_2$ but either $P_1 \subset P_2$ or $P_2 \subset P_1$.\\

\begin{figure}[htbp]
\resizebox*{0.9\textwidth}{0.9\textheight}{\includegraphics{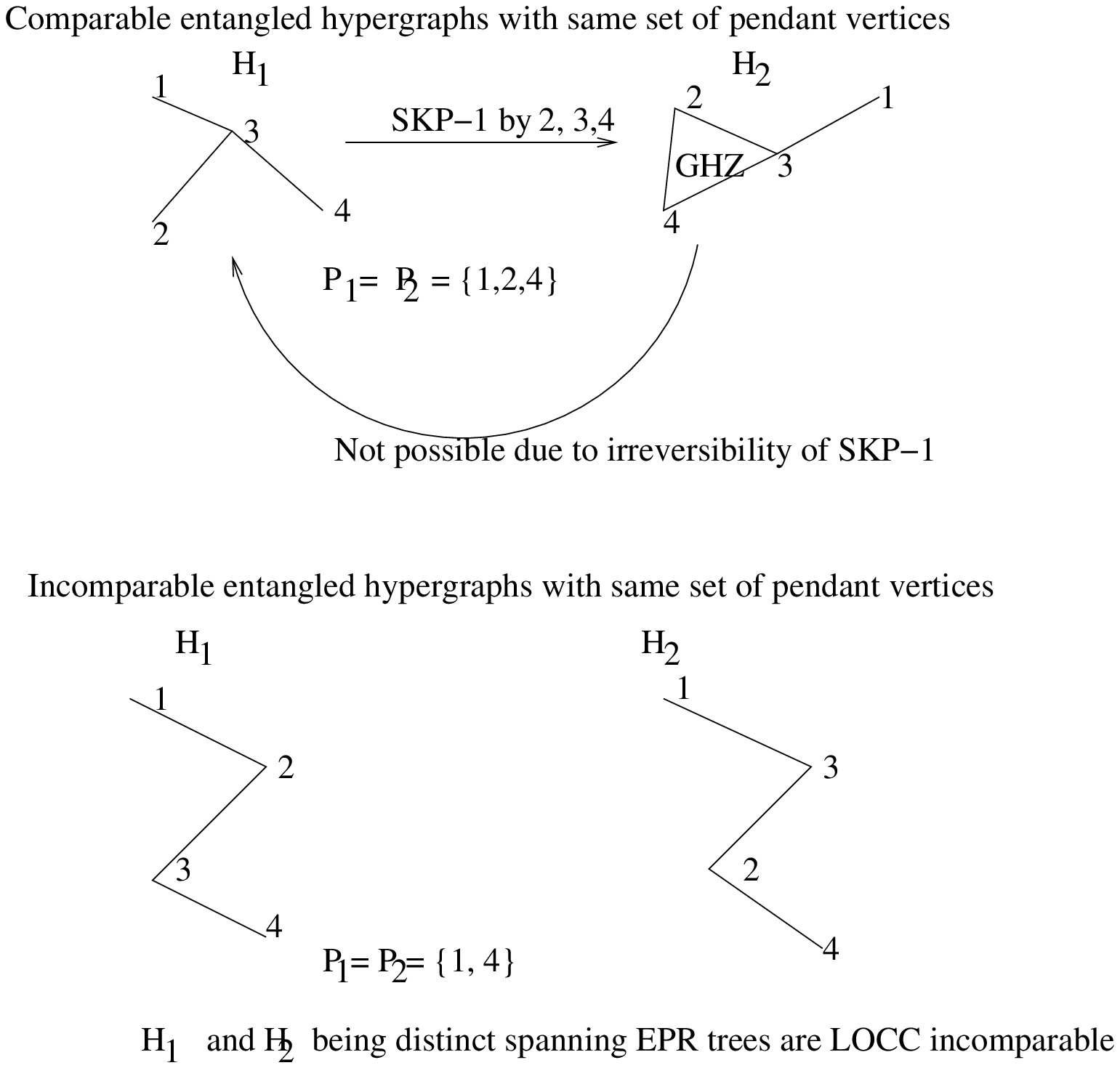}}
\caption{$P_1 = P_2$}
\label{figure310}
\end{figure}

{\it Example-2}: (Figure \ref{figure310}) $P_1 = P_2$\\ 

\begin{figure}
\resizebox*{0.9\textwidth}{0.9\textheight}{\includegraphics{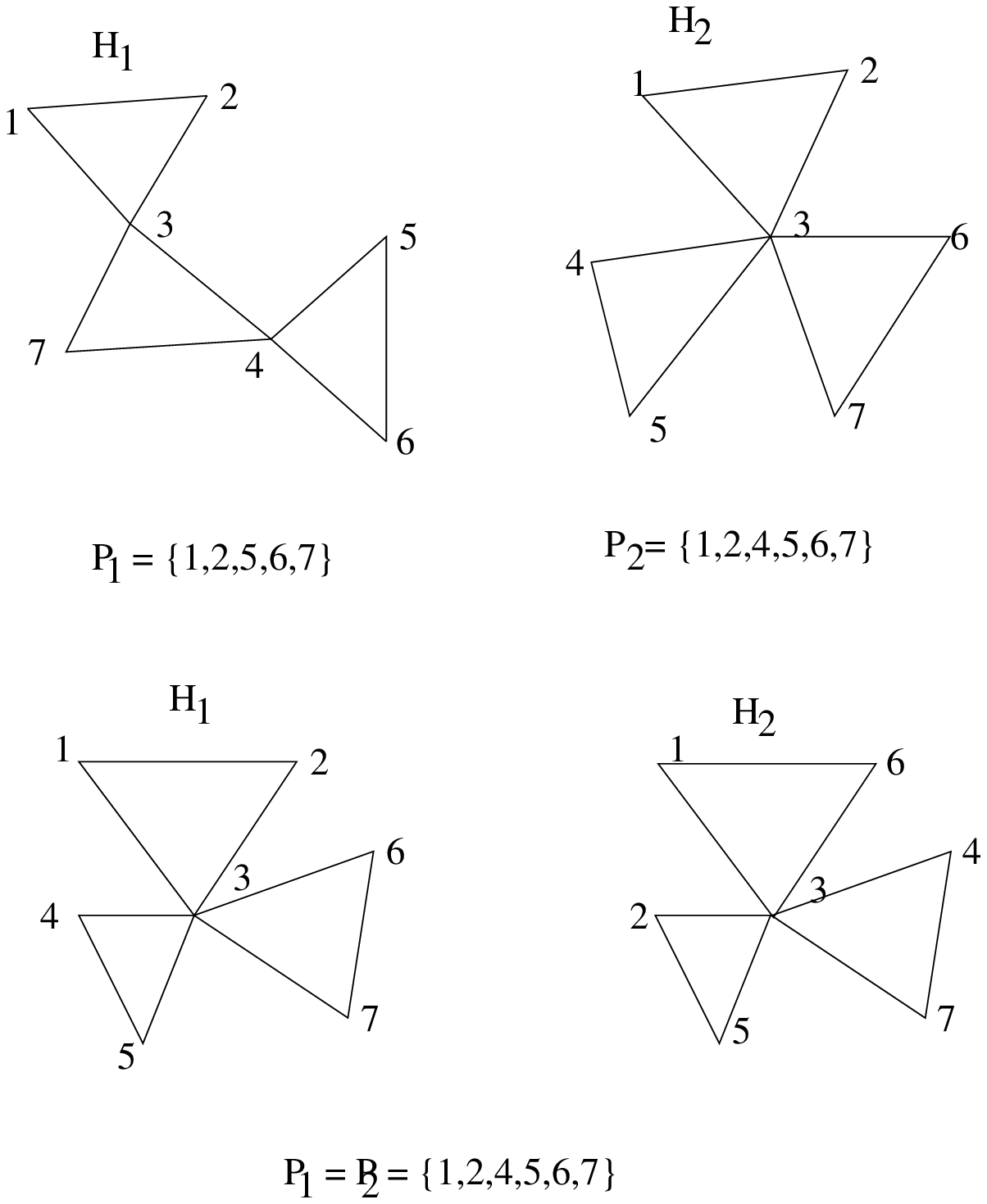}}
\caption{$r$-uniform entangled hypertrees not captured in theorem \ref{pendhypincomp}}
\label{figure311}
\end{figure}

Theorem \ref{twoeprtrees} shows that two EPR spanning trees are LOCC incomparable and the spanning EPR trees are nothing but $2$-uniform entangled hypertrees.
Therefore, a natural generalization of this theorem would be to $r$-uniform entangled 
hypertrees for any $r \geq 3$.
As we show below the generalization indeed holds.
It should be noted that the theorem \ref{pendhypincomp} does not
necessarily capture such entanglement structures (multi-partite states) (Figure \ref{figure311}).\\

However, while proving the fact that two different $r$-uniform 
entangled hypertrees are LOCC incomparable we shall need an important 
result about $r$-uniform hypertrees. We state this result in the 
following theorem for sake of continuity and completeness,
however we defer its proof to the apendix. \\

\begin{thm}
\label{lemmaruht}

Given two different $r$-uniform hypertrees $H_1=(S,F_1)$ and $H_2=(S,F_2)$ with $r \geq 3$,
there exists vertices $ u, v \in S$ such that $u$ and $v$ belong to same hyperedge in $H_2$
but necessarily in different hyperedges in $H_1$. 
\end{thm}

Now we prove our one of the main result on LOCC incomparability of
multi-partite entangled states and establish the following theorem.\\

\begin{thm}
\label{hyptree}

Any two distinct $r$-uniform entangled hypertrees are LOCC-imcomparable.
\end{thm}
  
Proof: Let $H_1=(S,F_1)$ and $H_2=(S,F_2)$ be the two $r$-uniform entangled hypertrees.
If $r=2$ then $H_1$ and $H_2$ happens to be two different 
spanning EPR trees and the proof follows from the theorem \ref{twoeprtrees}.
Therefore, let $r \geq 3$.\\

Now from theorem \ref{lemmaruht}, there exits $u,v \in S$ such that 
$u$ and $v$ belong to same hyperedge in $H_2$
but necessarily in different hyperedges in $H_1$.
Let the same hyperedge in $H_2$ be $E \in F_2$. 
Also, since $H_1$ being hypertree is {\it connected}, 
there exists a path between $u$ and $v$ in $H_1$.
Let this path be $u E_1 E_2 \cdots E_{k+1} v$.
Clearly $ k > 0$ because $u$ and $v$ necessarily do not belong 
to the same hyperedge in $H_1$. \\

\begin{figure}
\resizebox*{0.9\textwidth}{0.9\textheight}{\includegraphics{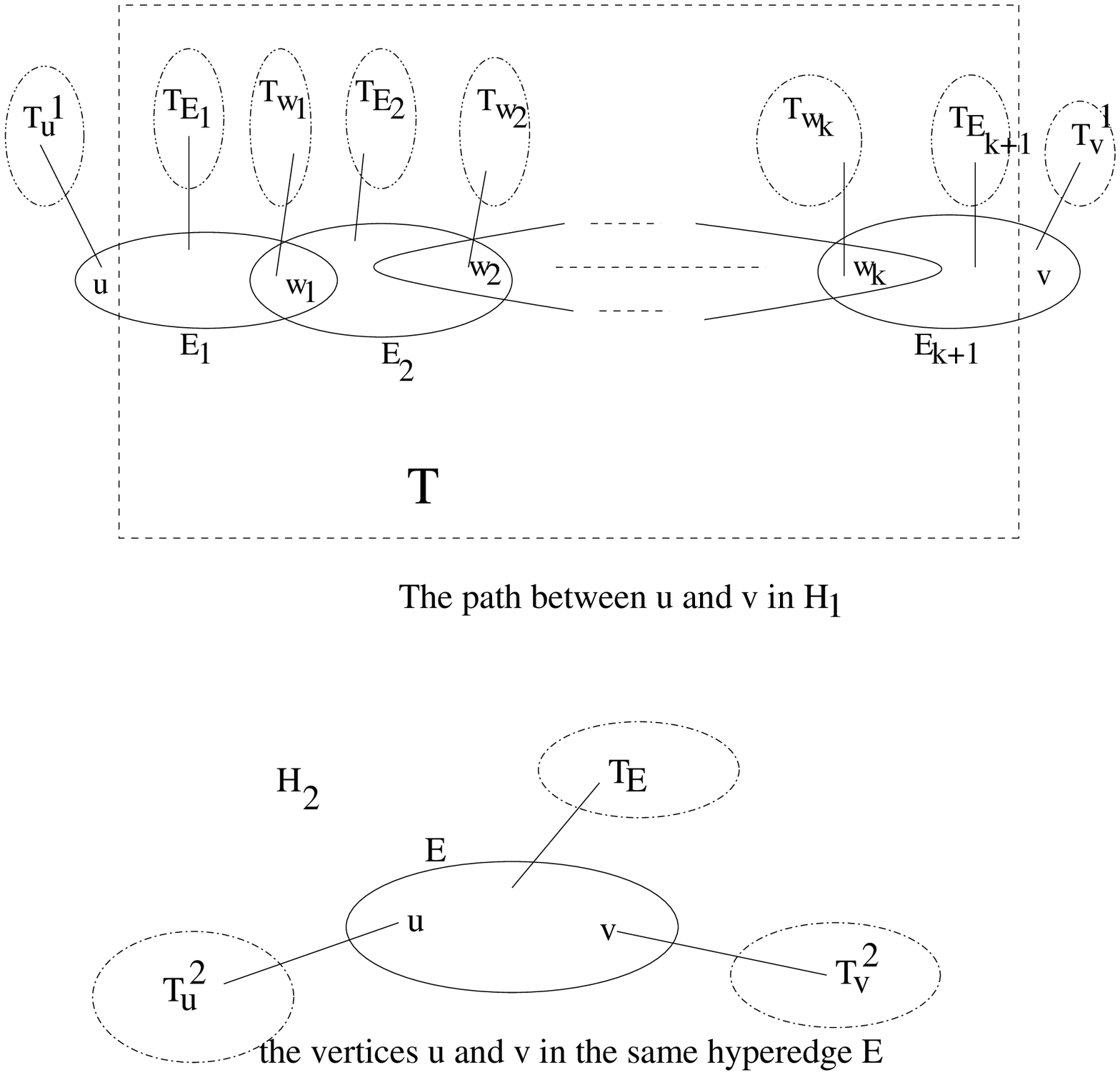}}
\caption{Two distinct $r$-uniform entangled hypertrees}
\label{figure312}
\end{figure}

We keep the following notations (figure \ref{figure312}). \\ 

$T_u^{1} :$ sub-hypertree rooted at $u$ in $H_1$ except that branch which contains $E_1$.\\

$T_v^{1} :$ sub-hypertree rooted at $v$ in $H_1$ except that branch which contains $E_{k+1}$.\\ 

$T_{w_i}:$ sub-hypertree rooted at $w_i$ in $H_1$ except that branches which contain
           $E_i$ and $E_{i+1}$.\\

$T_{E_i}:$ Collection of all sub-hypertrees in $H_1$ rooted at some vertices in $E_i$ 
        other than $w_{i-1}$ and $w_i$ (where $w_0 = u$ and $w_{k+1} = v$) 
        except for the branches which contain $E_i$.\\

$T = (( E_1 \bigcup E_2 \bigcup \cdots \bigcup E_{k+1}) \bigcup 
     ( T_{E_1} \bigcup T_{E_2} \bigcup \cdots \bigcup T_{E_{k+1}}) \bigcup
     ( T_{w_1} \bigcup T_{w_2} \bigcup \cdots \bigcup T_{w_k})) \setminus \{u,v\}$

   $=$ set of all vertices from $S \setminus \{u,v\}$ which are not contained in
   $T_u \bigcup T_v$. \\

$T_u^{2} :$ sub-hypertree rooted at $u$ in $H_2$ except that branch which contains $E$.\\

$T_v^{2} :$ sub-hypertree rooted at $v$ in $H_2$ except that branch which contains $E$.\\

$T_E:$ Collection of all sub-hypertrees in $H_2$ rooted at some vertices in
       $E \setminus \{u,v\} $ except for the branches which contain $E$. \\

We break the proof in to the various cases:\\

{\it CASE $S_1$}: $ \exists w \in T$ such that $w \in (T_u^{2} \bigcup T_v^{2})$ \\

Without loss of generality let us take $w \in T_u^{2}$.
Now since $w \in T$, $ w \in$ exactly one of $E_i$, $T_{w_i}$, or $T_{E_i}$ for some $i$. 
Accordingly there will be three subcases.\\
{\it Case $S_{1_1}$}: $w \in E_i$ for some $i$.

Do bicolored merging where the vertex $u$ along with all the vertices in 
\bigskip

$T_u, E_1, E_2, \cdots , E_{i-1}, T_{w_1}, T_{w_2}, \cdots , T_{w_{i-1}}, 
T_{E_1}, T_{E_2}, \cdots , T_{E_{i-1}}$ 

\bigskip

are given the color $A$ and the rest of the vertices are
given the color $B$.\\

{\it Case $S_{1_2}$ } $w \in T_{w_i}$ for some $i$. 

Do the bicolored merging while assigning the colors as in the above case.\\

{\it Case $S_{1_3}$} $w \in T_{E_i}$ for some $i$.

Bicolored merging in this case is also same as in Case $S_{1_1}$. \\

{\it CASE $S_2$}: There does not exist any $w \in T$ such that $w \in T_u^{2} \bigcup T_v^{2}$.\\

Clearly, $ T_u^{2} \bigcup T_v^{2} \subset T_u^{1} \bigcup T_v^{1}$ and $T \subset T_E \bigcup (E \setminus \{u, v\})$.
Note that whenever, we are talking of set relations like union ,containments etc. we are considering 
the trees, edges etc as sets of appropriate vertices from $S$ which make them.
First we establish the following claim.\\

{\it Claim}: $ \exists t \in (E_1 \setminus \{u, w_1\}) \bigcup (E_2 \setminus \{w_1, w_2\})$ such that
$ t \in T_E$. \\

We have $ k > 0$ therefore, both $E_1$ and $E_2$ exist and since $H_1$ is $r$-uniform 
$|E_1| = |E_2| = r$. 
Also $ (E_1 \setminus \{u, w_1\}) \bigcap (E_2 \setminus \{w_1, w_2\}) $ is empty otherwise 
there will be a cycle in $H_1$ which is not possible as $H_1$ is a hypertree \cite{hbc1, berge}.      
Therefore, 

\bigskip

$|(E_1 \setminus \{u, w_1\} \bigcup (E_2 \setminus \{w_1, w_2\})|
           = |(E_1 \setminus \{u, w_1\}| + |E_2 \setminus \{w_1, w_2\}|
           = (r-2) + (r-2) = 2r-4$. \\
\bigskip

Also $|E| =r $ implying that $|E \setminus \{u,v\}| = (r-2)$.

It is clear that $ u, v \notin (E_1 \setminus \{u,w_1\}) \bigcup (E_2 \setminus \{w_1, w_2\})$.
\bigskip

$|(E_1 \setminus \{u,w_1\}) \bigcup (E_2 \setminus \{w_1, w_2\})| - |E \setminus \{u,v\}|
=(2r-4)-(r-2) = r-2 \geq 1 $ since $ r \geq 3$.
\bigskip

Also $(E_1 \setminus \{u,w_1\}) \bigcup (E_2 \setminus \{w_1 , w_2\}) 
\subset T \subset T_E \bigcup (E \setminus \{u,v\})$,
\bigskip

therefore, by Pigeonhole principle \cite{pigeon},
\bigskip

 $ \exists t \in (E_1 \setminus \{u,w_1\}) \bigcup (E_2 \setminus \{w_1,w_2\})$

and

$t \in T_E (\notin (E \setminus \{u,v\}))$.

Hence our claim is true. \\  

Now we have $t \in (E_1 \setminus \{u,w_1\}) \bigcup (E_2 \setminus \{w_1,w_2\})$ such that $t \in T_E$.
Since $t \in T_E$, by the definition of $T_E$ it is clear that there must exist $w \in E \setminus \{u,v\}$
such that $ t \in T_w$, the sub-hypertree in $H_2$ rooted at $w$ except for the branch containing $E$.
Depending on whether $t \in E_1 \setminus \{u,w_1\}$ or $t \in  E_2 \setminus \{w_1,w_2\}$,
we break this case in to several subcases and futher in sub-subclasses depending on the part in $H_1$
where $w$ lie. \\

\begin{figure}
\resizebox*{0.9\textwidth}{0.9\textheight}{\includegraphics{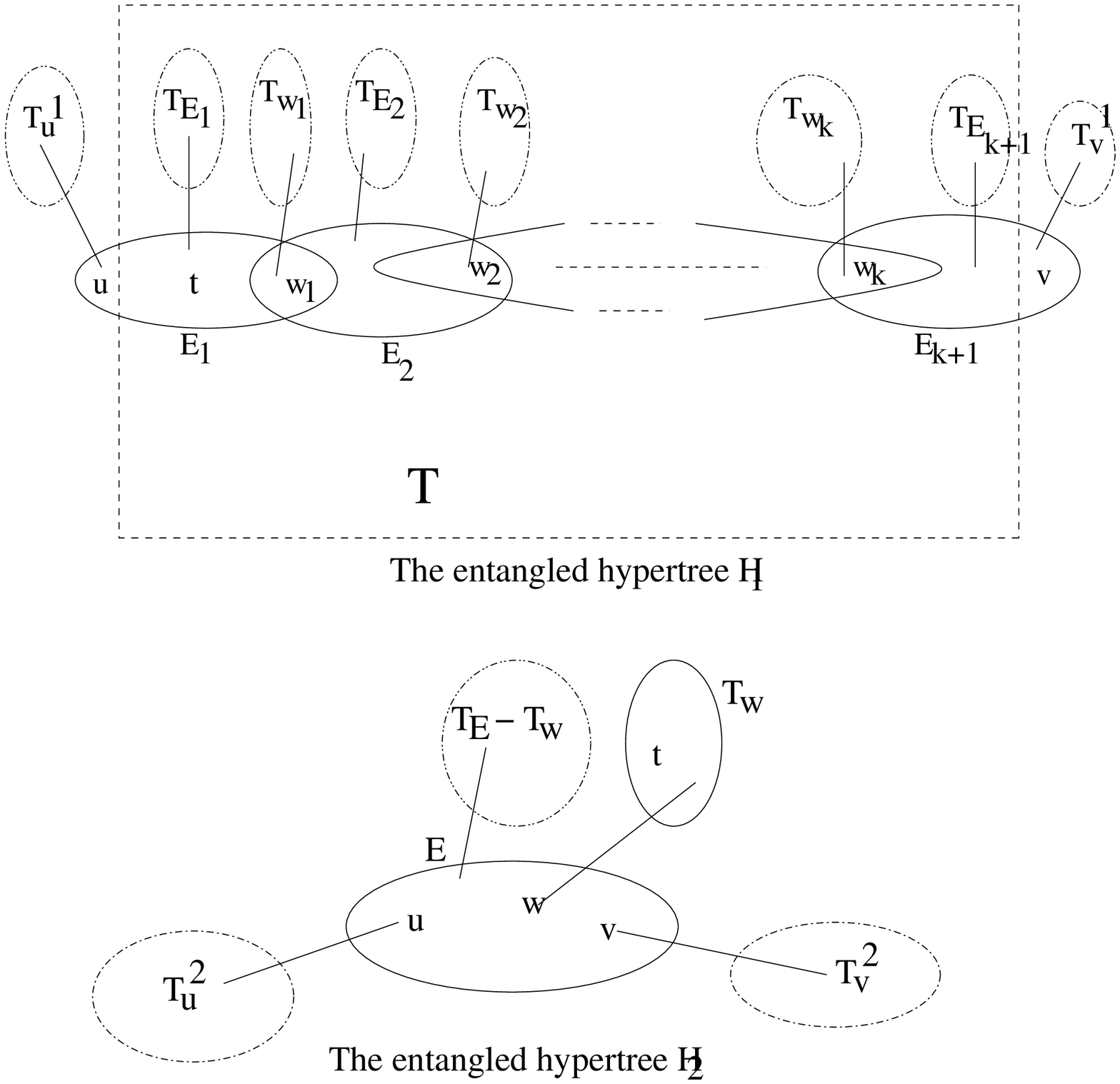}}
\caption{CASE $S_2$: CASE $1$}
\label{figure313}
\end{figure}

{\it CASE $S_{2_1}$}: $ t \in E_1 \setminus \{u,w_1\}$ (Figure \ref{figure313}). \\

{\it CASE $S_{2_{1_1}}$}: $ w \in T_u$.

Do the bicolored merging where $u$ and the vertices in $T_u$ are assigned the color $A$ and 
the rest of the vertices from $S$ are given the color $B$. \\    
   
{\it CASE $S_{2_{1_2}}$}: $ w \in T_v $.

Bicolored merging is done where $v$ as well as all the vertices in $T_v$ are assigned the color 
$B$ and rest of the vertices from $S$ are given the color $A$. \\

{\it CASE $S_{2_{1_3}}$ }: $w \in T $.

Here in this case, depending on whether $w$ is in $T_t$ or not there can be two cases.

{\it case $S_{2_{1_3}}^{1}$}: $ w \in T_t$.

Bicolored merging is done where all the vertices in $T_t$ are given the color $A$ and
rest of the vertices are assigned the color $B$.    

{\it case $S_{2_{1_3}}^{2}$}: $ w \notin T_t$.

$w \notin T_t$ implies that either $w \in E_i$ for some $i$ or $ w \in T_q$ where $ q \in E_i$ 
for some $i$ and $ q \neq t$. In any of these possibilities the bicolored 
merging is same and is done as follows.

Assign the color $A$ to $u$ as well as all vertices in
 
$T_u \bigcup E_1 \bigcup T_{E_1} \bigcup  T_{w_1} \bigcup \cdots \bigcup E_{i-1} \bigcup T_{E_{i-1}} 
\bigcup T_{w_{i-1}} \bigcup (E_i \setminus \{q,w,w_i\}) \bigcup (T_{E_i} \setminus T_q)$ 

and rest of the vertices are assigned the color $B$. \\

\begin{figure}
\resizebox*{0.9\textwidth}{0.9\textheight}{\includegraphics{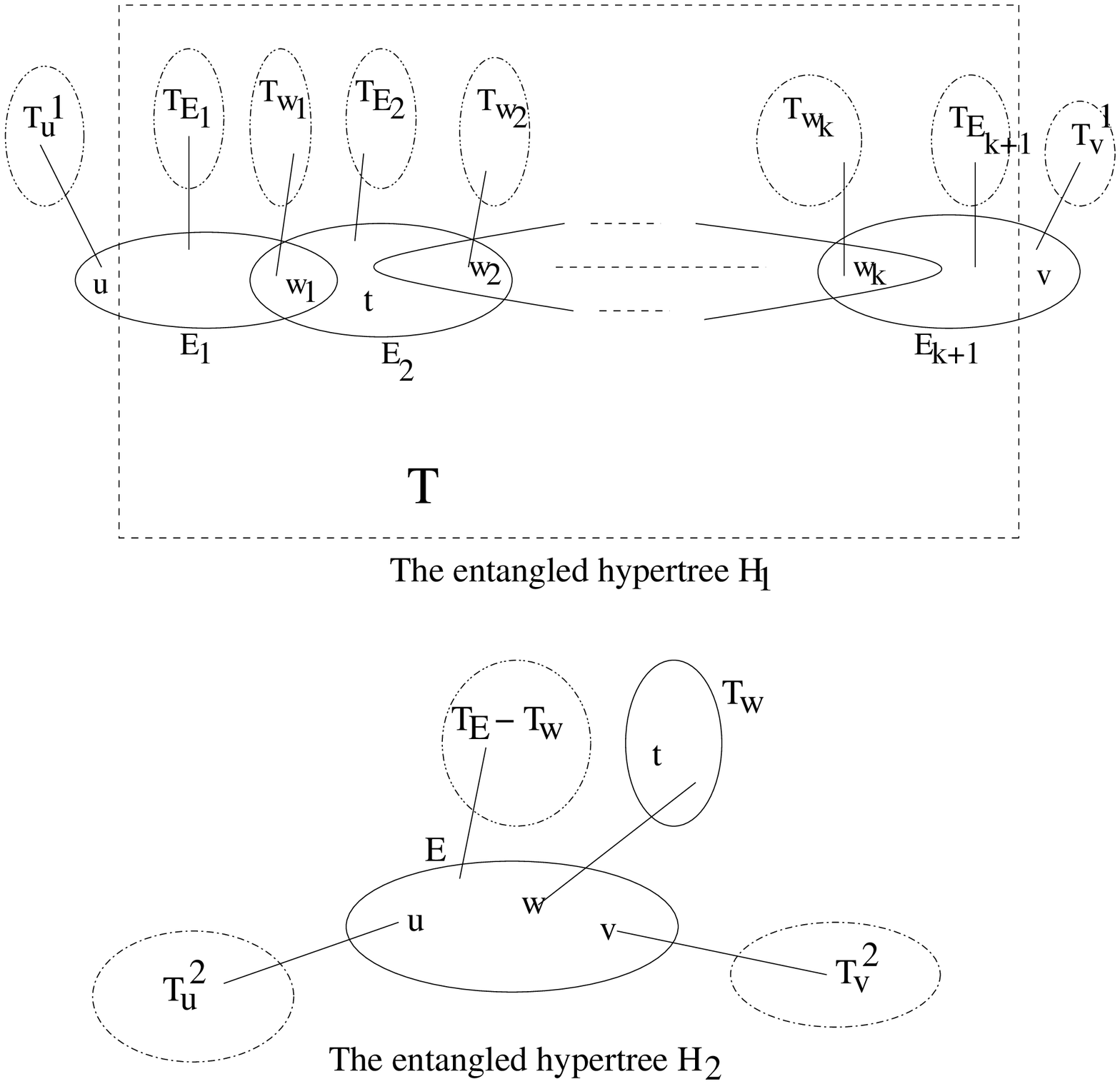}}
\caption{CASE $S_2$: CASE $2$}
\label{figure314}
\end{figure}

{\it CASE $S_{2_2}$}: $ i \in E_2 \setminus \{w_1,w_2\}$ (Figure \ref{figure314}).\\

{\it CASE $S_{2_{2_1}}$}: $ w \in T_u \bigcup E_1 \bigcup T_{E_1} \bigcup T_{w_1}$.

Do the bicolored merging where all the vertices in $T_u \bigcup E_1 \bigcup T_{E_1} \bigcup T_{w_1}$
including $u$  
are given the color $A$ and rest of the vertices are assigned the color $B$. \\

{\it CASE $S_{2_{2_2}}$}: $ w \in T_v \bigcup T_{E_{k+1}} \bigcup E_{k+1} \bigcup T_{w_k} \bigcup \cdots \bigcup T_{E_3} \bigcup E_3 \bigcup T_{w_2}$.

In bicolored merging give the color $B$ to all the vertices (including $v$) in 

$ T_v \bigcup T_{E_{k+1}} \bigcup E_{k+1} \bigcup T_{w_k} \bigcup \cdots \bigcup T_{E_3} \bigcup E_3 \bigcup T_{w_2}$
 
and color $A$ to the rest of the vertices. \\ 

{\it CASE $S_{2_{2_3}}$}: $ w \in E_2 \bigcup T_{E_2}$. 

In this case depending on whether $w \in T_t$ or $w \notin T_t$ the bicolored merging will be different.

{ \it case $S_{2_{2_3}}^{1}$}: $w \in T_t$.

Bicolored merging is done where all the vertices in $T_t$ are given the color $A$ and 
rest of the vertices are assigned the color $B$.

{\it case $S_{2_{2_3}}^{2}$}: $ w \notin T_t$.

$ w \notin T_t $ implies that either $w \in E_2$ or $ w \in T_q $ for some $q ( \neq )  \in E_2$.
In any case do the bicolored merging where the color $A$ is assigned to all the vertices in

$ T_u \bigcup E_1 \bigcup T_{E_1} \bigcup T_{w_1} \bigcup T_{E_1} \bigcup (E_2 \setminus \{w,q,w_2\})
\bigcup (T_{E_2} \setminus T_q)$ 

and rest of the vertices are assigned the color $B$.\\

Now that we have exhausted all the cases and shown by clever method of bicolored merging that
the $r$-uniform entangled hypertree $ H_1$ can not be LOCC converted to the $r$-uniform entangled 
hypertree $H_2$, the same arguments also work for showing that $H_2$ can not be LOCC converted to
$H_1$ by interchanging the roles of $H_1$ and $H_2$. 
Hence the theorem follows. \hfill \qed \\





Before ending our section on LOCC incomparability of multi-partite states 
represented by EPR graphs and entangled hypergraphs we note that 
Bennett et. al  partial entropic criteria \cite{bennet2000} which gives a sufficient condition for 
LOCC incomparability of multi-partite states do not capture
the LOCC-incomparability of spanning EPR tree 
or spanning entangled hypertrees in general. Consider two spanning EPR trees $T_1$ and $T_2$
on three vertices (say $1, 2, 3$). $T_1$ is such that the vertex pairs $ 1, 2$ and $1,3$ are forming 
the two edges where as in $T_2$ the vertex pairs $1,3$ and $2,3$ are forming the two edges.  
It is easy to see that $T_1$ and $T_2$ are not marginally isentropic.  \\

\section{Quantum Distance between Multi-partite Entangled States} 

In the proof of theorem \ref{twoeprtrees} we have utilized the fact that
there exists at least two vertices which are connected by an edge in $T_2$ 
but not in $T_1$ which follows because $T_1$ and $T_2$ are different as well 
as they have equal number of edges (namely $n-1$ if there are $n$ vertices).
In fact, in general there may exist several such pair of vertices depending on
the structures of $T_1$ and $T_2$. Fortunately there is some nicety in 
the number of such pair of vertices and it gives rise to a metric on the set 
of spanning (EPR) trees with fixed vertex set and hence a concept of distance 
\cite{ndeo}.
The distance between any two spanning (EPR) trees $T_1$ and $T_2$ denoted
by $QD_{T_1,T_2}$ on the same vertex set is defined as 
the number of edges in $T_1$ which are not in $T_2$.         
Let us call this distance to be the {\it quantum distance} between $T_1$ and $T_2$.
Now that we have proved in theorem \ref{twoeprtrees} obtaining $T_2$ from $T_1$ 
is not possible just through LOCC, we need to do quantum communication.
The minimum number of qubit communication required for 
this purpose should be an interesting parameter related to state transformations
amongst multi-partite states represented by spanning EPR trees ; let us
denote this number by $q_{T_1,T_2}$.
We can note that $ q_{T_1,T_2} \leq QD_{T_1,T_2}$. This is because 
each edge not present in $T_2$ can be created by only one qubit communication.
The exact value of $q_{T_1,T_2}$ will depend on the structures of $T_1$ and $T_2$ 
and , as we can note, on the number of edge disjoint paths in $T_1$ between the 
vertex pairs which form an edge in $T_2$ but not $T_1$. \\

\begin{figure}
\resizebox*{0.9\textwidth}{0.9\textheight}{\includegraphics{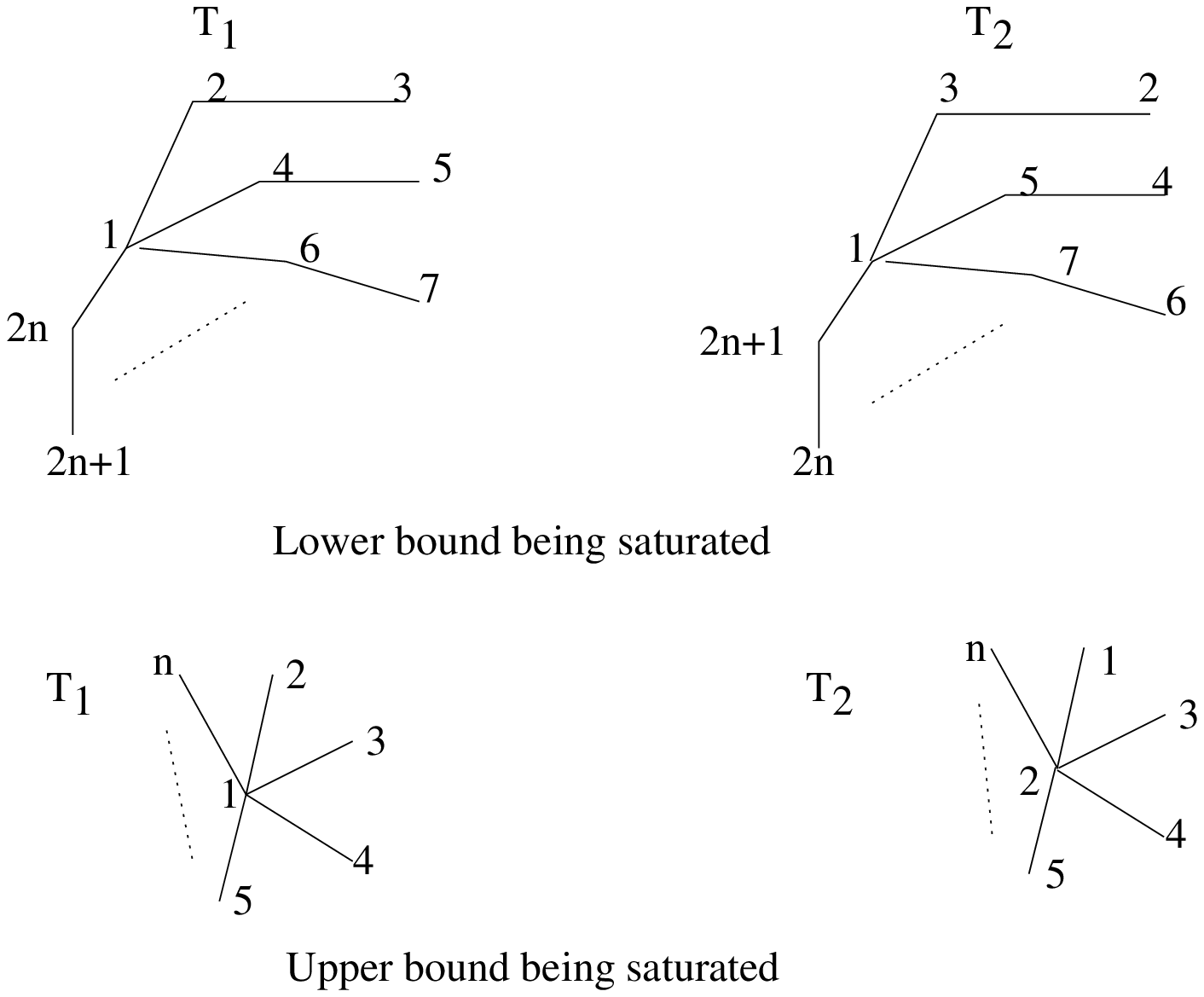}}
\caption{}
\label{figure315}
\end{figure}

But this is not all about the quantum distance.
Let us recall the theorem \ref{treecat} where we prove that 
$n-1$ is a lower bound on the number of copies of $n-CAT$ 
to prepare a spanning EPR tree by LOCC. Can we obtain some lower bound like this
in the case of two spanning EPR trees and relate it to the quantum distance?
Answer is indeed yes. Let $C_{T_1,T_2}$ denote the minimum number of copies of
the spanning EPR tree $T_1$ to obtain $T_2$ just by LOCC. 
We claim that $ 2 \leq C_{T_1,T_2} , C_{T_2,T_1} \leq QD_{T_1,T_2} + 1 $.
The lower bound follows from theorem \ref{twoeprtrees}.
The upperbound is also true becasue of the following reason.
$QD_{T_1,T_2}$ is the number of (EPR pairs) edges present in $T_2$ but not in 
$T_1$. For each such edge in $T_2$ if $u, v$ are the vertices forming the edge,
while converting many copies of $T_1$ to $T_2$ by LOCC,
an edge between $u$ and $v$ must be created. 
Since $T_1$ is a spanning tree and therefore connected , there must be a path
between  $u$ and $v$ in $T_1$ and this path can be well converted (using entanglement 
swapping) to an edge between 
them ( i.e. EPR pair between them) only using LOCC. Hence one copy each will suffice to
create each such edges in $T_2$. Thus $QD_{T_1,T_2}$ copies of $T_1$ will be sufficient 
to create all such $QD_{T_1,T_2}$ edges in $T_2$. One more copy will supply all the edges 
common in $T_1$ and $T_2$.  
Even more interesting point is that both these bounds are saturated. This means to say that
there do exist spanning EPR trees satifying these bounds (Figure \ref{figure315}).
  
\section{Appendix}



{Proof of Theorem \ref{lemmaruht}:} We first establish the following claim.\\

{\it claim}: $ \exists E_1 \in F_1 , E_2 \in F_2$ such that 
$ E_1 \neq E_2 , E_1 \bigcap E_2 \neq \phi$ and
$ E_2 \notin F_1 \bigcap F_2$. \\

{P roof of the claim:} We first show that on same vertex set, the number of 
hyperedges in any $r$-uniform hypertree is always same. 
Let $n$ and $m$ be the number of vertices and hyperedges in 
a $r$-uniform hypertree then we show by induction on $m$ that $n = m*(r-1) + 1$.

For $ m = 1$, $ n = 1 *(r-1) + 1 = r$ which is true because all possible
vertices (since no one can be isolated) fall in the single edge and it 
has exactly $r$ vertices. 

Let us assume that this relation between $n$ and $m$ for a fixed $r$ holds for all 
values of the induction variable up to $m-1$ then we are to show that it holds good 
for $m$. 

Now take a $r$-uniform hypertree with $m$ hyperedges. 
Remove any of the hyperedges to get another hypergraph (which may not be connected)  
which has only $m-1$ edges. This removal may introduce $k$ connected components 
(sub-hypertrees) where $ 1 \leq k \leq r$.
Let these components has respectively $m_1, m_2, \cdots , m_k$ number of
hyperedges. Therefore, $ \sum_{i=1}^{k} m_i = m-1$.
Total number of vertices in the new hypergraph (with the $k$ sub-hypertrees as components),
$n_1 = \sum n_i $ where $n_i$ is the number of vertices in the component $i$.

Therefore, $n_1 = \sum n_i = \sum_{i=1}^{k} \{m_i (r-1) +1\}  = (m-1)(r-1) + k$.

Now the number of vertices in the original hypertree,
$n = n_1 + (r - k)$ because $k$ vertices were already covered, one each in the $k$ components.
Therefore, $ n = (m-1)(r-1) + k + (r-k) = (m-1)(r-1) + r = (m-1)(r-1) + (r-1) + 1 = m(r-1) + 1$.
The result is thus true for $m$ and hence for any number of hyperedges by induction.
This result implies that any $r$-uniform hypertree on the same vertex set will always 
have the same number of hyperedges. \\

Let $ F = F_1 \bigcap F_2$ and $ m = |F_1| = |F_2| $.
Obviously $ m > |F| $ otherwise $H_1 = H_2$ implying that
$\exists E \in F_2 $ such that $E \notin F$. 

Let $ U = \bigcap_{A \in F} A $ is the vertex set on which the common hyperedges
relies. Depending on the characteristics of $U$ we break the proof in two cases.\\

{\it CASE $1$}: $ \exists w \in E $ such that $ w \notin U$.  

Since $ w \notin U$ we get $E \notin F$.
Now since $H_1$ is a hypertree $w$ can not be an isolated vertex and therefore there exists 
an edge say $E^1 \in F_1 $ but $ E^1 \notin F $ (otherwise $w \in U$). 
Take $E_2 = E$ and $ E_1 = E^1$. 

{\it CASE $2$}: $A \subset U \forall A \in F_2$     

We have $|E| = r $ and $ E \subset U $, say $ E = \{e_1, e_2, \cdots , e_r\}$.

No pair of $(e_i, e_j)$ can be connected by the hyperedges only in $F$ ( i.e. through 
common hyperedges) otherwise $ e_i$ (path in $F$) $e_j E e_i$ will be a cycle in $H_2$
(mean to say that there will be two paths between $e_i$ and $e_j$ in $H_2$ one in $F$ and another 
one $e_i E E_j$ both being in same hyperedge $E$) which is absurd given that $H_2$ is a hypertree.

Now $H_1$ is a hypertree, therefore must be connected and there must be a path between $e_i$ and 
$e_j$ in $H_1$. Say this path be $ e_i G_1 G_2 \cdots G_l e_j$ where $ G_k \in F_1$ for $ 1 \leq k \leq l$.
Take $E_1 = G_1$ and $E_2 = E$ 

Thus in all cases, we have proved that $ \exists E_1 \in F_1 and E_2 \in F_2$ such that
$ E_1 \neq E_2$, $ E_1 \bigcap E_2 \neq \phi$ and $ E_2 \notin F_1 \bigcap F_2$ and hence follows our claim.\\

Now switch over to prove the theorem. Choose $E_1$ and $E_2$ so as to satify the above claim.
Let $E_1 = \{ u_1, u_2, \cdots , u_l, w_{l+1}, w_{l+2}, \cdots , w_r \}$ and 
$E_2 = \{u_1, u_2, \cdots , u_l, v_{l+1}, v_{l+2}, \cdots , v_r\}$. 
Since $ E_1 \bigcap E_2 \neq \phi $ , $ l \geq 1$ and $E_1 \neq E_2$ implies that $ l \leq r-1$.
Hence $ 1 \leq l \leq r$.  
Now based on the value of $l$ we consider different cases below. \\

{\it CASE $1$ }: $ l > 1$ \\

{\it case $1_1$}: $\exists v_i$ such that $u_1$ and $v_i$ are not in same hyperedge in $H_1$.

Take $u=u_1$ and $v = v_i$ in the statement of the theorem. 

{\it case $1_2$}: All $v_i$ are respectively in some hyperedges in $H_1$ in which $u_1$ also lies.

None of these $v_i$ can belong to the same hyperedge as of $u_2$ in $H_1$.
This is because if say $v_j$ happen to be in same hyperedge as of $u_2$ in $H_1$ then 
$u_1 u_2 v_j u_1$ will be a cycle in $H_1$ which is absurd as $H_1$ is a hypertree.
Note that at least one such $v_i$ must exist as $l < r$. 

Take $u=u_2$ and $v =$ any $v_i$.\\

{\it CASE $2$}: $ l = 1$ \\

{\it case $2_1$}: $ \exists v_i$ such that $u_1$ and $v_i$ are not in same hyperedge in $H_1$.

Take $ u = u_1$ and $ v = v_i$.

{\it case $2_2$}: All $v_i$ are respectively in some hyperedges in $H_1$ in which $u_1$ also lies.

Since there are $r-1$ $v_i$ in number and $E_2 \notin F_1 \bigcap F_2$, these $v_i$ will be distributed in
at least two different hyperedges in $H_1$ in which $u_1$ also lies. 
Therefore, $ \exists v_i, v_j$ such that they  are in same hyperedge in $H_2$ (namely in $E_2$)but 
in necessarily different edges in $H_1$ otherwise (i.e. if they lie in the same hyperedge in $H_1$)
$ u_1 v_i v_j u_1 $ will be a cycle in $H_1$ which is absurd as $H_1$ is a hypertree.
Also note that both $v_i$ and $v_j$ will exist as $r \geq 3$. 

Take $u = v_i$ and $v=v_j$. \\   

Thus we have proved the above thorem in all possible cases. \hfill \qed 

We would like to point out that this very result might be some direct implication of standard results in combinatorics
however for sake of completeness we have proved it in our own way, moreover keeping in appendix section! \\


\chapter{Unconditionally Secure Multipartite Quantum Key Distribution}

\section{Introduction}
With the growing use of the internet and other forms of electronic communication,
the question of secure communication becomes one of considerable importance.
Modern cryptographic techniques, based on the availability of ever increasing computational
power, and the invention of public key cryptography, provide practical solutions
for information security in various situations. But invariably these techniques 
are only computationally-- and not unconditionally-- secure, that is, they depend
on the (unproven) hardness of certain mathematical problems. As a result, it cannot
be guaranteed that future advances in computational power will not nullify their
cryptographic protection. Nevertheless, there does
exist a form of encryption with unconditional security: the use of one-time
key pads. These are strings of random numbers added by the information sender to
encode the message, to be subtracted by the receiver to decode.
Provided that the key material is truely random
and used only once, this system is unbreakable 
in the sense described by Shannon in the 1940s \cite{sha45}. It is critically
important that the pad is only used once, i.e., an encryption key can never be used 
twice. This restriction translates into the practical one of
key distribution (KD). This need to securely distribute the key between the users
makes it impractical in
many applications. Where it is used in real life (eg., in confidential communications
between governments), the one-time key pads are actually delivered in person by
some {\em trusted third party}, an arrangement prohibitively expensive 
for common usage and moreover not truely secure. Fortunately, recent
advances in quantum information theory have shown that unconditionally secure
key distribution is possible in principle.

That quantum information can, on account of quantum uncertainty and the no-cloning
principle \cite{wooters82}, be used to distribute cryptographic
keys was realized two decades ago \cite{bb84,ekert91}. More rigorous and
comprehensive proofs of this task, generally called quantum key distribution
(QKD), taking into consideration source, device and channel noise
as well as an arbitrarily powerful eavesdropper,
have been studied by various authors
\cite{may,bih,lochau,ina,sp00}. Recently, the issues of efficiency \cite{hwa03},
security in the presence of an uncharacterized source \cite{koa03} and
high bit-error rate tolerance \cite{wang} of QKD have been considered.
In particular, Lo and Chau \cite{lochau} showed that, given fault-tolerant
quantum computers, quantum key distribution over an arbitrarily long distance
of a realistically noisy channel can be made unconditionally secure.
This is a heartening development, since QKD is, among quantum
information applications, relatively easy to implement, and some large scale
implementations have already been achieved \cite{gis02,zei03}. The above mentioned
works consider QKD between two parties (ie., 2-QKD). It is of interest to
consider its extension to more than two parties (ie., $n$-QKD). 

The problem of $n$-QKD is to determine how $n$ parties, who are able to
communicate quantally, may share an identical and unconditionally secure, 
secret key among themselves in the presence of eavesdroppers. 
(A different generalization of 2-QKD gives multipartite quantum secret-sharing 
\cite{scarani01}, which we do not consider here). 
In this work, we propose a protocol for this purpose and prove its unconditional security.
We note that a simple extrapolation of 2-QKD to $n$-QKD would suggest that the agents
should begin by sharing an $n$-partite entangled state. However, this proposition
suffers from two drawbacks: from a practical viewpoint, preparing $n$-partite
entanglement is no easy task; from a theoretical viewpoint, proving the
security of secure extraction of $n$ separated copies of a bit-string 
may not be simple, even given the existing proof of security of the bipartite case.
Our main result is that it is sufficient if some pairs of agents share
{\em bipartite} entanglement along any spanning tree connecting the $n$ agents,
who are taken to be vertices on a graph. 
In this way, $n$-QKD is reduced to a 2-QKD problem. Existing 2-QKD protocols 
\cite{may,bih,lochau,ina,sp00,hwa03,koa03,wang}
can be invoked to prove the unconditional security
of sharing nearly perfect Einstein-Podolsky-Rosen (EPR) pairs between two parties. 
They prescribe procedures for reliably sharing EPR pairs, and thence sharing
randomness, by virtue of fault-tolerant quantum computers, quantum error
correction and suitable random sampling. In the interests of brevity, we will
not elaborate these protocols here, and only point them out as subroutines for
the general $n$-QKD task.
Our protocol is fairly simple in the sense that the proof of its security
is built on top of the already proven security of the bipartite case. 
However, it is important to know the necessary and sufficient conditions 
on the network topology for our proposed protocol to work.

\section{Classical Reduction of $n$-KD to 2-KD}\label{reduc}
As in 2-KD, the goal of $n$-KD is to show that 
$n$ trustful parties can securely share random, secret classical bits, even
in the presence of noise and eavesdropping. 
It is assumed that the $n$ agents can share authenticated
classical communication. It is convenient to treat the problem graph theoretically 
as in Chapter 2.
The $n$ agents $A_i$ $(1 \le i \le n)$ are considered as the vertices (or nodes) 
of an undirected graph. An instance of a secure bi-partite channel being shared
between two parties is considered as an undirected edge between the two corresponding
vertices. A graph so formed is called a {\it security graph}.
It is obvious that if the security graph has a star topology (a hub vertex with
all edges radiating from it to the other vertices), a simple $n$-KD protocol
can be established. The agent at the hub vertex (say, called, Lucy) generates
a random bit string and transmits it to every other agent along the edges
to each of them. This will create a secure, identical random bit
string with each agent.

In real life situations, because of practical and geographical constraints, 
the $n$ agents may not form a security graph with
star topology. We describe a simple protocol that allows for more general
secure bi-partite connectivity between the agents. 
In particular, from among the secure bipartite channels suppose a spanning
tree (a graph connecting all vertices without forming a loop) 
can be constructed. This construction can be formalized in order to determine an 
optimal spanning tree. Some useful definitions are given below.

\begin{defni}
Weighted security graph:
Given $n$ parties treated as nodes on a graph, 
we extend the definition of a security graph 
to the {\em weighted security graph}. 
A weight is associated with every edge and is defined 
to be some suitable measure of the cost of communicating by means of the
channel corresponding to the edge. 
\end{defni}


\begin{defni}
Minimum spanning security tree: 
Consider the weighted security graph $G=(V,E)$. A spanning
tree selected from $G$, given by $G_1=(V,E_1), E_1 \subseteq E$ is called the
minimum spanning security tree if it minimizes the total weight of the graph.
\end{defni}

Minimum spanning security tree need not be
unique and can be obtained 
using Kruskal's or Prim's algorithm \cite{clr1990}. 
The minimum spanning security tree
minimizes the resources needed in the protocol as well as the
size of the sector eavesdroppers can potentially control.

\begin{defni}
Terminal agent: An agent that corresponds to a vertex of degree one 
(ie., with exactly one edge linked to it). On the other hand,
an agent that corresponds to a vertex of
degree greater than one is called a non-terminal agent.
\end{defni}

We now present a classical subroutine that allows $n-1$ pair-wise shared random bits
to be turned into a single random bit shared between the $n$ parties.
\begin{enumerate}
\item 2-KD: Along the $n-1$ edges of a minimum spanning security tree, $n-1$ 
random bits are securely shared by means of some secure 2-KD protocol.
\item Each non-terminal agent $A_i$ announces his 
{\em unformly randomized record}:
this is the list of edges emanating from the vertex along with the corresponding
random bit values, to all of which 
a fixed random bit $x(i)$ is added.
\item This information is sufficient to allow every player, in conjunction with
her/his own random bit record, to reconstruct the random bits of all parties.
The protocol leader (say, Lucy) decides randomly on the terminal agent whose
random bit will serve as the secret bit shared among the $n$ agents.
\end{enumerate}
This subroutine consumes $n-1$ pair-wise shared random numbers to
give a one-bit secret key shared amongst the $n$-parties. To generate an
$m$-bit string shared among the $n$ agents, the subroutine is repeated $m$ times.

Given that the initial bipartite sharing of randomness is secure, we will
show that the above protocol subroutine allows some randomness to be shared between
the $n$ agents without revealing anything to an eavesdropper. It
involves each non-terminal agent announcing his uniformly randomized
record. Suppose one such, $A_i$, has the random record 0,1,1 on the three edges 
linked to his vertex. He may announce 0,1,1 (for $x(i)=0$) or 
1,0,0 (for $x(i)=1$). All the three agents linked
to him can determine which the correct string is by referring to their shared
secret bit. It is a straightforward exercise to see that each of other agents
linked to these three can determine the right bit string. Therefore, 
each agent can determine the random bits of all others.
Eavesdroppers, on the other hand, lacking knowledge of any
of the $n-1$ shared random bits, 
can only work out the {\em relative} outcomes of all parties. 
The result is exactly two possible configurations for each secret bit, 
which are complements of each other. The eavesdropper ``Eve" is thus maximally
uncertain about which the correct configuration is. 
Hence, Lucy's choice of a party to fix the secret bit reveals little to Eve.
Insofar as the $n$-parties are able to communicate
authenticated classical messages, the subroutine protocol is 
{\em as secure as} the underlying procedure for 2-KD.

It is obvious that the above protocol works for any spanning
security tree. Clearly, a sufficient condition for turning shared 
bipartite randomness into randomness shared between $n$ parties is that
the weighted security graph should contain at least one spanning
tree. On the other hand, if the security graph is disconnected, one easily
checks that it is impossible to arrive at a definite random bit securely
shared between both the disconnected pieces.
Therefore, the existence of at least one spanning tree in the weighted 
security graph is both a necessary and sufficient condition for the required task.

The amount of securely shared randomness may be quantified by the length
of shared
random bit string multiplied by the number of sharing agents. In the above protocol,
the $n-1$ instances of pair-wise shared randomness is consumed to produce
exactly one instance of a random bit shared between the $n$ parties. 
We can then define the `random efficiency' of the above protocol by 
$\eta = (n\times 1)/((n-1)\times2)$, 
which tends to (1/2) as $n \rightarrow \infty$.
Unconditional security of the above subroutine can in principle
only be guaranteed in a protocol which includes in step 1 a
quantum sub-routine that implements 2-QKD. In the following Section, we will
present one such, based on the Shor-Preskill protocols \cite{sp00}, as
an example.

\section{Quantum Protocol}

As in 2-QKD, the goal of the proposed $n$-QKD protocol is to show that 
$n$ trustful parties can securely distil random, shared, secret classical bits, whose
security is to be proven
inspite of source, device and channel noise and of Eve, an eavesdropper
assumed to be as powerful as possible, and in particular, having control over all
communication channels. From the result of the preceding Section, it follows that 
a quantum protocol is needed only in step 1 above. It will involve
establishing 2-QKD along a minimum spanning tree in order to 
securely share pair-wise randomness along spanning 
tree's edges and thence proceed to $n$-QKD. We assume as given
the security of establishing pair-wise randomness along a spanning tree by means of 
a quantum communication network, based on a secure 2-QKD protocol
\cite{may,bih,lochau,sp00,ina,hwa03,koa03,wang}. In principle, these protocols guarantee
security under various circumstances. 

In an $n$-QKD scheme, the insecurity of even one of the players
 can undermine all. Hence
additional classical processing like key reconciliation and privacy
amplification of the final key may be needed at the $n$-partite level. 
In the full $n$-QKD protocol that we present below,
following Ref. \cite{sp00} we exploit the connection of
error correction codes \cite{mac} with key reconciliation and privacy amplification.
These procedures have been 
extensively studied by classical cryptographers \cite{gis02}, and other
possibilities exist.

In particular, we adopt a quantum protocol wherein pair-wise randomness is
created by means of sharing EPR pairs (this follows the pattern set by the
Ekert \cite{ekert91}, Lo-Chau \cite{lochau} and Modified Lo-Chau \cite{sp00} protocols, but 
entanglement is not necessary, as seen in the original BB84 protocol).
The basic graph theoretic definitions introduced above apply also for the quantum
case, except that now the security channels correspond to shared EPR pairs.
In place of a secure bipartite channel, an instance of EPR pair shared
between two parties is considered as an undirected edge between the two corresponding
vertices. A graph so formed is called an EPR graph (Chapter 2).
The analog of the weighted security graph is the
{\em weighted EPR graph}, and that of the minimum spanning security tree is the
{\em minimum spanning EPR tree}. 
Let us enumerate the $n$ parties as $A_1,A_2,\cdots,A_n$.
Suppose that only $A_{i_1},A_{i_2},\cdots,A_{i_s}$ $(i_1,i_2,\cdots,i_s\in
\{1,2,\cdots,n \})$ are capable of producing EPR pairs and
$S \equiv \{A_{i_1},\cdots,A_{i_s}\}$ is the set of all such vertices,
with $S\ne \emptyset$.
We construct a weighted undirected graph $G=(V,E)$ as one whose every edge
must contain a vertex drawn from the set $S$, as follows:
$V\equiv \{A_i;~i=1,2,\cdots,n\}$ and
$E \equiv \{(A_i,A_j) ~\forall~ A_i \in S ~{\rm and}~ \forall~ A_j \in V;~
i\neq j\}$. And the weight of edge $(A_i,A_j)$ is defined to be 
$w_{i,j} \propto$ number of quantum repeaters \cite{repeater}
(more generally: entanglement distilling resources \cite{cirnat}) 
required to be put between $A_i$ and $A_j$. 
Usually, the larger the distance between two
agents, the larger is the weight.
 The minimum spanning EPR tree minimizes the number 
of quantum repeators needed, and, in general, the resources needed in the protocol
(EPR pairs, etc.) subject to the constraint of available EPR sources. Apart from 
improving efficiency in terms of costs incurred,
this optimization is also important from the security perspective
in that it minimizes the size of the sector that Eve can potentially control.

Let ${\cal C}$ be a classical $t$-error correcting $[m,k]$-code \cite{mac}. 
We now present a protocol that consumes $n-1$ pair-wise securely shared 
sets of EPR pairs to create random bits shared between the $n$ parties with
asymptotic efficiency $\eta = (1/2)k/m$, where $k/m$ is the rate of the code.
The classical subroutine described in the previous Section is adapted to
include key reconciliation and privacy amplification at the $n$-partite level,
that uses the group theoretic properties of ${\cal C}$.
\begin{enumerate}
\item EPR protocol:
Along the $n-1$ edges of the minimum spanning EPR tree, EPR pairs
are shared (using eg., the Lo-Chau \cite{lochau}
or Modified Lo-Chau protocols \cite{sp00}). 
Let the final, minimum number of EPR pairs distilled along any edge of the 
minimum spanning EPR tree be $2m$. A projective measurement in the computational 
basis is performed  by all the parties on their respective qubits to obtain
secure pair-wise shared randomness along the tree edges
(making due adjustments according to whether the entangled spins are correlated or
anti-correlated).
\item Classical subroutine of Section \ref{reduc}:
All non-terminal vertices announce their 
{\em unformly randomized outcome record}.
This information in principle allows every party, in conjunction with
her/his outcome, to reconstruct the outcomes of all other parties, save for some errors
of mismatch. 
\item For each set of $n-1$ shared EPR pairs, 
protocol leader Lucy decides randomly on the terminal party whose outcome
will serve as the secret bit.
\item Lucy decides randomly a set of $m$ bits to be used as
check bits, and announces their positions.
\item All parties announce the value of their check bits. If too few of
these values agree, they abort the protocol.
\item Lucy broadcasts $c_i \oplus v$, where  
$v$ is the string consisting of  the remaining code (non-check) bits, and
$c_i$ is a random codeword in ${\cal C}$.
\item Each member $j$ from amongst the remaining $n-1$ parties 
subtracts $c_i \oplus v$ from his
respective code bits, $v\oplus\epsilon_j$, and corrects the result,
$c_i \oplus \epsilon_j$, to a codeword in ${\cal C}$. Here $\epsilon_j$
is a possibly non-vanishing error-vector.
\item The parties use $i$ as the key.
\end{enumerate}

A rigorous proof of the security of the $n$-QKD scheme
requires: (a) the explicit construction of a procedure such that 
whenever Eve's strategy has a non-negligible probability of
passing the verification test by the $n$ parties, 
her information on the final key will be exponentially small. 
(b) the shared, secret randomness is robust against source, device
and channel noise. By construction, our scheme combines a
2-QKD scheme to generate pair-wise shared randomness and a classical
scheme to turn this into multipartite-shared randomness. The security
of the latter (in its essential form)
was proven in Section \ref{reduc}. Therefore the security of
the protocol with respect to (a) and (b) reduces to that of the 
2-QKD in step 1.
For various situations, 2-QKD can be secured, as proven in Refs
\cite{may,bih,lochau,ina,sp00}. 
For example, Lo and Chau \cite{lochau} and Shor and Preskill \cite{sp00}
have proved that EPR pairs can be prepared to be nearly perfect, even in the
presence of Eve and channel noise. 
Their proofs essentially relies on the idea
that sampling the coherence of the qubits allows one to place an upper bound on the
effects due to noise and information leakage to Eve.
Yet, subject to the availability of high quality
quantum repeaters and fault-tolerant quantum computation, in principle 
2-QKD can be made unconditionally secure \cite{lochau}.

In regard to the key reconciliation part:
in step (3), each non-terminal vertex party announces his uniformly randomized
outcome record. Here this consumes $m$ instances of $n-1$ pair-wise shared random
bits into $k$ random bits shared between the
$n$ agents while revealing little to Eve. The random efficiency is given by
\mbox{$\eta = (k \times n \times 1)/(m \times (n-1) \times 2)$}, which tends to
$(1/2)k/m$ as $n \rightarrow \infty$.
The check bits, whose positions and values are announced in steps (4) and (5), 
are eventually discarded.
Steps (7) and (8) involve purely local, classical operations.
If security of step (1) against Eve is guaranteed, the string $v$,
and thereby the string $c_i \oplus v$ announced by Lucy in step (6), 
are completely random, as far as Eve
can say. So, she (Eve) gains nothing therefrom.
Hence her mutual information with any of the $n-1$ (sets of)
random bits does not increase beyond
what she has at the end of the EPR protocol. 

Finally, step (5) permits with high probability to determine whether the key
can be reconciled amongst the $n$ players.
The check bits that the parties measure behave
like a classical random sample of bits \cite{sp00}. 
We can then use the measured error rates
in a classical probability estimate. For any two parties,
the probability of obtaining
more than $(\delta + \epsilon)n$ errors on the code bits and fewer than 
$\delta n$ errors on the check bits is asymptotically less than
$\exp[-0.25\epsilon^2n/(\delta-\delta^2)]$. Noting that the errors
on the $n$ check vectors 
are independent, it follows that probability that the check vectors
are all scattered within a ball of radius $\delta n$ but one or more code
vectors fall outside a scatter ball of radius $(\delta+\epsilon)n$ is 
exponentially
small, and can be made arbitrarily small by choosing sufficiently small $\delta$.
The decision criterion adopted in step (5) is calculated so that 
the Hamming weight of the error vectors $\epsilon_j$ estimated in the above fashion
will be less than $t$ with high probability. Hence all parties correct 
their results to the same codeword $c_i$ in step (8) with high probability.
This completes the proof of unconditionally security of $n$-QKD.
\chapter{Generalized Quantum Secret Sharing}

\section{Introduction}\label{intro}

Suppose the president of a bank, Alice, wants to give access to a vault to two 
vice-presidents, Bob and Charlie, whom she does not entirely trust.
Instead of giving the combination to any one of them, she may desire to
distribute the information in such a way that no vice-president alone has any 
knowledge of the combination, but both of them can jointly determine the combination.
Cryptography provides the answer to this question in the form of
{\it secret sharing} \cite{schneier96,gruska97}.
In this scheme, some sensitive data is distributed among a
number of parties such that certain authorized sets of parties can 
access the data, but no other combination of players.
A particularly symmetric variety of secret splitting (sharing) is
called a {\it threshold scheme}: 
in a $(k,n)$ classical threshold scheme (CTS),
the secret is split up into $n$ pieces (shares),
of which any $k$ shares form a set {\em authorized} to reconstruct the secret, while any set of 
$k-1$ or fewer shares has no information about the secret.
Blakely \cite{blakely79} and Shamir \cite{sha79} showed that CTS's exist
for all values of $k$ and $n$ with $n \geq k$. By concatenating threshold schemes,
one can construct arbitrary access structures, subject only to the condition of
monotony (ie., sets containing authorized sets should also be authorized)
\cite{ben}. Hillery {\em et al.}
\cite{hil00} and Karlsson {\em et al.} \cite{kar00} proposed methods for
implementing CTSs that use {\em quantum} information
to transmit shares securely in the presence of eavesdroppers.

Subsequently, extending the above idea to the quantum case,
Cleve, Gottesman and Lo \cite{cle00} proposed
 a $(k,n)$ {\it quantum} threshold scheme (QTS) as a method to split up 
an unknown secret quantum state $|S\rangle$ into $n$ 
pieces (shares) with the restriction that 
$k > n/2$ (for if this inequality were violated, two disjoint sets of players
can reconstruct the secret, in violation of the quantum no-cloning theorem \cite{wooters82}). 
The notion of QTS is based on quantum erasure correction \cite{cs,gra97}.
QSS has been extended beyond QTS to
general access structures \cite{got00,smi00}, but here the no-cloning theorem implies
that none of the authorized sets shall be mutually disjoint.
Potential applications of QSS include
creating joint checking accounts containing quantum money \cite{wiesner83}, 
or share hard-to-create ancilla
states \cite{got00}, or perform secure distributed quantum computation \cite{cre01}.

In conventional QSS schemes, it is often implicitly assumed that 
all share-holders carry quantum information. Sometimes it is possible to
construct an equivalent scheme in 
which some share holders carry only classical information and no
quantum information \cite{nas01}. 
Such a hybrid (classical-quantum) QSS that combines classical and
quantum secret sharing brings a significant improvement to the
implementation of QSS, inasmuch as quantum information is much more fragile than
classical information. Moreover, hybrid QSS can potentially avail of features
available to classical secret sharing such as share renewal \cite{herzberg},
secret sharing with prevention \cite{beltel} and disenrolment \cite{martin}.
The essential method to hybridize QSS is to somehow
incorporate classical information that
is needed to decrypt or prepare the quantum secret as classical shares. 
A simple instance of such classical information is the ordering information 
of the shares. In QTS, it is implicity assumed that the share-holders know the
the coordinates of the shares in the secret, i.e., they know who is holding the
first qubit, who the second and so on. This ordering
information is necessary to reconstruct the secret, without which successful
reconstruction of the secret is not guaranteed. If we wish to make use of this
ordering information in the above sense, then only quantum error correction 
based secret sharing where lack of ordering information leads to maximal
ignorance can be used. In particular, the scheme should be 
sensitive to the interchange of two or more qubits.
For example, let us consider a $(2,3)$-QTS. The secret here is an arbitrary qutrit
and the encoding maps the secret qutrit to three qutrits as:
\begin{equation}
\alpha|0\rangle + \beta|1\rangle + \gamma|2\rangle \longmapsto
\alpha(|000\rangle + |111\rangle + |222\rangle) + 
\beta(|012\rangle + |120\rangle + |201\rangle) + 
\gamma(|021\rangle + |210\rangle + |102\rangle),
\end{equation}
and each qutrit is taken as a share. While from a single share no information can
be obtained, two shares, with ordering information, suffice to reconstruct the
encoded state \cite{cle00}. However, the lack of ordering information does not
always lead to maximal ignorace about the secret. Note that the structure of the above
code is such that any interachange of two qubits  leaves an encoded $|0\rangle$
intact but interchanges $|1\rangle$ and $|2\rangle$. Thus, a secret like $|0\rangle$
or $(1/\sqrt{2})(|1\rangle + |2\rangle)$ can be entirely
reconstructed without the ordering
information. Therefore, only the subset of quantum error correction codes admissible
in QSS that do not possess such symmetry properties can be used if the scheme is
to be sensitive to ordering information.

Theoretically simpler but practically somewhat more difficult is 
an interesting idea proposed by Nascimento et al. \cite{nas01},
based on qubit encryption \cite{qcrypt}. 
In Sections \ref{sec:infla} and \ref{sec:compress_aqss},
we adopt this method to generate the relevant encrypting classical information.
However, in principle any classical data
whose suppression leads to maximal ignorance of the secret
is also good. Elsewhere, in Section \ref{sec:qts}, we consider another way.
Quantum encryption works as follows: suppose we have a 
$n$-qubit quantum state $|\psi\rangle$ and random sequence $K$ of $2n$
classical bits. Each sequential pair of classical bit is associated with a qubit
and determines which transformation $\hat{\sigma} \in \{\hat{I}, \hat{\sigma}_x,
\hat{\sigma}_y, \hat{\sigma}_z\}$ is applied to the respective qubit. If the
pair is 00, $\hat{I}$ is applied, if it is $01$, $\hat{\sigma}_x$ is applied,
and so on. The resulting $|\tilde{\psi}\rangle$ is a complete mixture and
no information can be extracted out of it because the encryption leaves
any pure state in a maximally mixed state, that is: 
$(1/4)(\hat{I}|S\rangle\langle S|\hat{I} +
\hat{\sigma}_x|S\rangle\langle S|\hat{\sigma}_x +
\hat{\sigma}_y|S\rangle\langle S|\hat{\sigma}_y +
\hat{\sigma}_z|S\rangle\langle S|\hat{\sigma}_z) = (1/2)\hat{I}$.
However, with knowledge of $K$
the sequence of operations can be reversed 
and $|\psi\rangle$ recovered. Therefore, classical data can be used
to encrypt  quantum data.

In analogy with classical secret sharing, it is customary to consider
that all quantum shares must be distributed among the players. A simple generalization
(which we call `assisted QSS' schemes),
that is helpful from both theoretical and practical considerations, is to allow
some shares to remain with the share dealer. 
This simple device will, rather surprisingly, enable us to implement
QSS schemes in which authorized sets may be mutually {\em disjoint}.
In Sections \ref{sec:aqss} and \ref{sec:compress_aqss}, we study such assisted schemes.

\section{Inflating Quantum Secret Sharing Schemes}\label{sec:infla}

In hybrid QSS, the quantum secret is split up into quantum and classical shares of
information. We call the former q-shares, and the latter c-shares. A player holding only
c-shares is called a c-player or c-member. Otherwise, she or he is a q-player
or q-member. 
\begin{defni}
A QSS scheme realizing an access structure $\Gamma = \{\alpha_1, \alpha_2,
\cdots,\alpha_r\}$ among a set of players ${\cal P} = \{P_1, P_2, \cdots, P_n\}$
is said to be compressible if fewer than $n$ q-shares are sufficient to implement it. 
\end{defni}
Here the $\alpha_i$'s are the minimal authorized sets of players.
Knowledge of compressibility
helps us decide how to minimize quantum  resources needed for 
implementing a given QSS scheme. 
As an example of compression by means of hybrid QSS, suppose we want to split a
quantum secret $|S\rangle$ among a set of players ${\cal P} = \{A, B, C, D, E, F\}$
realizing the access structure $\Gamma = \{ ABC, AD, AEF\}$. That is, the only sets 
authorized to reconstruct the secret are $\{A,B,C\}$, $\{A,D\}$ and $\{A,E,F\}$
and sets containing them, whilst any other set is unauthorized to do so. For distributing
the secret, we encrypt
$|S\rangle$ using the quantum encryption method (described above) with classical key
$K$ into a new state $|\tilde{S}\rangle$ and give $|\tilde{S}\rangle$ to $A$.
We then split up $K$ using a CSS scheme that realizes $\Gamma$. Player $A$ cannot
recover $|S\rangle$ from $|\tilde{S}\rangle$ because he cannot unscramble it without
$K$. Only the $\alpha_j$'s, and sets containing them, can recover the
classical key $K$, and thence decrypted secret state. In this way, by means
of a hybrid (classical-quantum) secret-sharing scheme, we can compress the original
QSS scheme into an equivalent one in which fewer players need to handle quantum
information. 

The question, how to augment or ``inflate"
a given QSS scheme keeping
the quantum component fixed, is considered herebelow. This is of practical relevance
if we wish to expand a given QSS scheme
by including new players who do not
have (reliable) quantum information processing capacity. To this end, we now define
an inflatable QSS.
\begin{defni}
A QSS($\Gamma$) scheme
realizing an access structure $\Gamma = \{\alpha_1, \alpha_2,
\cdots,\alpha_r\}$ among a set of players ${\cal P} = \{P_1, P_2, \cdots, P_n\}$
using a total of $m$ q-shares 
is inflatable if $n$ can be increased for fixed $m$ to form a new QSS($\Gamma^{\prime}$)
such that $\Gamma^{\prime}|_{\cal P} = \Gamma$,
where $\Gamma^{\prime}|_{\cal P}$ denotes the 
restriction of $\Gamma^{\prime}$ to ${\cal P}$.
\end{defni}
Clearly, inflation involves the addition of classical
information carrying c-players. The additional
shares required for them will be c-shares,
so that q-shares may remain fixed at $m$.
The following theorem answers the question when a QSS scheme
can be inflated.

\begin{thm}
\label{thm:alinf}
A QSS scheme realizing an access structure $\Gamma = \{\alpha_1, \alpha_2,
\cdots,\alpha_r\}$ among a set of players ${\cal P} = \{P_1, P_2, \cdots, P_n\}$
using a total of $m$ q-shares can always be inflated.
\end{thm}

\noindent
{\bf Proof.} Consider the addition of a single player, $P_{n+1}$.
The new set of players are ${\cal P}^{\prime} \equiv
\{P_1, P_2, \cdots,P_{n+1}\}$. A new access
structure $\Gamma^{\prime}$ can be obtained by arbitrarily adding $P_{n+1}$
to any of the $\alpha_j$'s. Clearly, $\Gamma^{\prime}$ will not violate the
no-cloning theorem \cite{wooters82}, since $\Gamma$ does not. As a result, the augmented
scheme can be realized as a conventional QSS scheme using (say) $m^{\prime} > m$ q-shares.
Therefore, by construction, there is one $P_i$
(namely, that for $i=n+1$ in the above case) 
such that $\Gamma^{\prime}|_{{\cal P}^{\prime}-P_i}$ does not violate
the no-cloning theorem,
where $\Gamma^{\prime}|_{{\cal P}^{\prime}-P_i}$ denotes the 
restriction of $\Gamma^{\prime}$ to ${\cal P}^{\prime} - P_i$.
Therefore, according to Theorem 1 of Ref. \cite{nas01}, the new QSS scheme obtained
by adding $P_{n+1}$ is compressible, meaning that $P_{n+1}$ can be a c-player.
Therefore, the new scheme QSS($\Gamma^{\prime}$) is an inflation of 
the given scheme QSS($\Gamma$). It is clear that the 
process of addition of new c-players can
be continued indefinitely without restriction. \hfill \qed
\bigskip

The above theorem only says that that any QSS scheme can be inflated in {\em some}
way. A specific problem is whether a given $(k,n)$-QTS  can be inflated.
This is considered in the following two theorems.

\begin{thm}
A $(k,n)$-QTS cannot be inflated at constant threshold.
\end{thm}

\noindent {\bf Proof.} Suppose $(k,n)$-QTS can be inflated at constant threshold. Then
there exists a $(k,n^{\prime})$-QTS, consistent with the no-cloning theorem
and with $n^{\prime}>n$, whose restriction leads to $(k,n)$-QTS.
Let $n^{\prime} - n \equiv \gamma$. The restriction of $(k,n^{\prime})$-QTS
by $\gamma$ players will lead to a $(k-\gamma,n)$-QTS \cite{nas01}, 
whose access structure
is different from $(k,n)$-QTS. This contradicts our original assumption. 
\hfill \qed
\bigskip

\begin{thm}
A $(k,n)$-QTS can be inflated conformally, ie., to newer 
threshold schemes having the form $(k+\gamma,n+\gamma)$,
 for any positive integer $\gamma$. \end{thm}

\noindent
{\bf Proof.} If the given $(k,n)$-QTS satisfies the no-cloning theorem, then
clearly so will the $(k+\gamma,n+\gamma)$-QTS. Further, according to Lemma 1
of Ref. \cite{nas01}, a restriction of the $(k+\gamma,n+\gamma)$-QTS by $\gamma$ players 
yields a $(k,n)$-QTS. Therefore, an expansion of a $(k,n)$-QTS to a
$(k+\gamma,n+\gamma)$-QTS is possible adding only ($\gamma$) c-players.
\hfill \qed
\bigskip

\section{Assisted Quantum Secret Sharing}\label{sec:aqss}
Because of the no-cloning theorem, 
secret sharing requires q-shares being converged to some site in order to 
reconstruct the secret, which could be the secret dealer or some designated
reconstructor for final processing. 
For example, in the first example, access is allowed by the
vault (which can be thought of as the dealer) if the secret reconstructed from
the vice-presidents' shares is the required password. In this case,
by definition, the secret dealer is a trusted party in the secret sharing. 
In such cases, there appears to be little 
loss of generality in leaving some shares obtained from
splitting the secret with the dealer,
the other shares being shared among the players, each of whom receives at least one
share. Where the dealer may be distinct from the reconstructor, as in multiparty secure
computation, the dealer transmits his shares to the latter, once the latter is identified.
When shares of an authorized set converge at 
the reconstructor, the latter simply adds his own
shares before processing the verification. 
In practice, the dealer can simply be a
computer program that stores passwords of bank accounts or a central 
quantum computer in a multi-party secure computation procedure. 
We refer to this generalization of quantum secret sharing as `assisted quantum
secret sharing' (AQSS).

In classical secret
sharing, such ``share assistance" (from the dealer) does not appear to offer any
new advantage. However, the situation is quite different in QSS.
First, as we point out below, 
the only restriction on the access structure $\Gamma$ in AQSS is monotony. The 
members of $\Gamma$ are {\em not} required to have mutual overlaps, in order to satisfy the
no-cloning theorem. Further, as we show later, it can considerably
reduce the amount of quantum communication and the number of quantum information
carrying players required in a QSS, in ways clarified below.
This is quite important from the viewpoint of implementation,
considering that quantum information processing is extremely difficult.
Shares which are dealt out to players are called `player shares';
that/those which remain(s) back with the dealer (to be transmitted to the
reconstructor directly, if necessary) is/are called `resident share(s)'.
A conventional QSS scheme is a special cases of the assisted scheme,
in which the set of resident shares is empty. We note that 
share assistance is needed only
when the members of an access structure do not have pairwise overlap. 

\begin{thm}
Given an access structure $\Gamma = \{\alpha_1, \alpha_2,
\cdots,\alpha_r\}$ among a set of players ${\cal P} = \{P_1, P_2, \cdots, P_n\}$,
an assisted quantum secret sharing scheme exists iff $\Gamma$ is monotone.
\end{thm}

\noindent
{\bf Proof:} 
It is known that if the members of $\Gamma$ all overlap, then there exists a 
conventional QSS to realize it \cite{got00}.
Suppose the members of $\Gamma$ do not overlap.
(It is instructive to look at the classical situation. Suppose $\Gamma = \{ABC, DE\}$,
which can be written in the normal form $\{(A \og B \og C) \ou (D\og E)\}$. The
\og gate corresponds to a $(|\alpha_j|,|\alpha_j|)$ 
threshold scheme, while \ou to a (1,2) threshold
scheme. By concatenating these two layers, we get a construction for $\Gamma$.)
In the conventional QSS, the above fails for two reasons, because by the
no-cloning theorem: the members of $\Gamma$ should not be disjoint and further there
is no $((1,2))$ scheme (here, following Ref. \cite{got00},
double (single) brackets denote the quantum (classical) scheme.)
However, we replace $((1,2))$ by a $((2,3))$ scheme, which corresponds to a majority
function of \ouh. In general, we replace a $((1,r))$ scheme by a $((r,2r-1))$
scheme. $r$ of the shares correspond to individual authorized sets in $\Gamma$,
and the other $r-1$ shares will remain as resident shares with the dealer. Any
authorized set, by combining its second layer (AND) shares can reconstruct its first
layer (OR) share, which, combined with the resident share, can reconstruct
the secret. Since the necessity of the resident share by itself fulfils
the no-cloning theorem, authorized sets are {\em not} required to be
mutually overlapping. \hfill \qed

Another way to view this is that given any arbitrary
$\Gamma$, including disjoint members, we systematically add the same player
to all authorized sets to obtain a new $\Gamma^{\prime}$ which is compatible
with the no-cloning theorem in the usual sense. Thus, we can turn $\Gamma 
= \{ABC, DE\}$ into $\Gamma^{\prime} = \{ABCX, DEX\}$, by adding member $X$,
whose share is the resident share deposited with the dealer. Thereby, the structure $\Gamma =
\Gamma^{\prime}|_{\overline{X}}$, which denotes a restriction of 
$\Gamma^{\prime}$ to members other than $X$,  is effectively
realized among the players (excluding the dealer).

For example, for the access structure $\Gamma = \{ABC, DE\}$, in 
the first layer, a $((2,3))$ scheme is employed to split $|S\rangle$ into
three shares, with one share designated to $ABC$ and the other to $DE$. 
The last remains with the dealer (cf. Eq. (\ref{eq:qz0}).
In the second layer, the first two block rows are $((|\alpha_j|,|\alpha_j|))$ schemes.
\begin{equation}
\label{eq:qz0}
((2,3)) \left\{ \begin{array}{ll}
((3,3)) : & A, B, C \\
((2,2)) : & D, E \\
((1,1)) : &  {\rm dealer}
	   \end{array} \right. 
\end{equation}
In contrast, without share assistance, the share corresponding to the last block would,
recursively, be split according to a maximal scheme containing $\Gamma$ (which should of 
course not contain any disjoint members), for which a pure state scheme
exists \cite{got00}. Note that either $ABC$ or $DE$ can reconstruct only one 
of the ((2,3)) shares. Thus the disjointness of these sets does not violate the
no-cloning theorem. When the share from an individual authorized set
is submitted, the secret can be
reconstructed at the reconstructor station, when combined with the resident share.

\begin{thm}
\label{thm:thresk}
For a $((k,n))$ scheme, with $1 \le k \le n$, there exists
an equivalent assisted quantum threshold scheme. If $k \le n/2$, the equivalent
assisted scheme is given by $((k+\gamma,n+\gamma))$, where $\gamma=n-2k+1$.
\end{thm}

\noindent
{\bf Proof:} If $k > n/2$, then the theorem stands proved by the known result
\cite{cle00} that a quantum erasure code exists equivalent to a $((k,n))$ scheme.
Suppose that $k \le n/2$. Consider a $((k+\gamma,n+\gamma))$ scheme, where
$\gamma$ is increased until the no-cloning condition $2(k+\gamma)
> n+\gamma$ is met, which is at $\gamma=n-2k+1$. 
The dealer employs this scheme, retains $\gamma$ shares as resident shares, giving one
of the remaining $n$ shares to each player. Any $k$ player shares, combined with
the resident share, suffice to reconstruct the secret. Hence this method
effectively realizes the required $((k,n))$ scheme over the players.
\hfill \qed \bigskip

Once again we note that there is no contradiction with the non-cloning
theorem because the procedure basically realizes a $((k+\gamma,n+\gamma))$
scheme in which the $\gamma$ shares with the dealer forms a 
common element in all  authorized sets.
As an example, suppose a $((2,10))$ assisted scheme is required. From 
Theorem \ref{thm:thresk} we find $\gamma = 7$,
so that the required assisted version is a $((9,17))$ scheme, which can be realized
as a quantum erasure code. Of these, $\gamma=7$ shares
are resident shares, the remaining ten each being given to a player. Any two players
can reconstruct the secret by submitting their two shares jointly to the reconstructor,
which adds its seven shares to reconstruct the secret. Consider an access structure
$\Gamma = \{ABC, AD, EFG\}$. The associated assisted structure is 
$\Gamma^{\prime} = \{ABCX, ADX, EFGX\}$, where share $X$ is designated to be a resident
share. A q-share $q$ is `important' if there is an unauthorized set 
$T$ such that $T \cup \{q\}$ is authorized.
The size of an important share of a conventional QSS that realizes $\Gamma^{\prime}$
cannot be smaller than the dimension of  the secret \cite{got00,ima03}. 
Generalizing this argument gives the following easy theorem.
\begin{thm}
The dimension of each important share of an assisted quantum secret sharing scheme must
be at least as large as the dimension of the secret.
\end{thm}

\section{Compressing Assisted Quantum Secret Sharing Schemes}\label{sec:compress_aqss}

In a conventional $((k,n))$ scheme, a compression of shares is possible only
if $2k > n+1$ \cite{nas01}, in which case, the scheme can be compressed into a 
$((k-\gamma,n-\gamma))$ scheme combined with a $(k,n)$ scheme. A general access structure
$\Gamma = \{\alpha_1, \alpha_2,\cdots, \alpha_r\}$ can be realized by a first layer of 
$(1,r)$-threshold scheme. In the quantum case, since this violates the no-cloning
theorem, it is replaced by the majority function $(r,2r-1)$-QTS \cite{got00}.
This, again, is incompressible. However, in the second layer of the construction, 
the $((|\alpha_i|,|\alpha_i|))$ schemes can be replaced with a $((1,1))$ schemes combined
with $(|\alpha_i|,|\alpha_i|)$ schemes. 
As seen from the results below, further saving on q-players and q-shares becomes
possible under share assistance. Here compression refers to the player shares and {\em not}
the resident shares. Note that there is a trivial compression for
a share assisted scheme
realizing an access structure $\Gamma$, in which the entire secret simply remains with 
the dealer, and only the encryption information is split-shared according
to a classical scheme realizing $\Gamma$. This is ruled out by requiring that
every $\alpha_i$ must have at least one important q-share allocated to it. 

\begin{thm}\label{thm:minqp}
A conventional QSS scheme realizing an access structure $\Gamma = \{\alpha_1, \alpha_2,
\cdots,\alpha_r\}$ among a set of players ${\cal P} = \{P_1, P_2, \cdots, P_n\}$ can 
always be compressed to an assisted QSS scheme requiring no more than $M \equiv |{\cal M}(\Gamma)|$ 
quantum players, where ${\cal M}(\Gamma)$ 
is the smallest hitting set for the collection $\Gamma$. Further compression is impossible.
\end{thm}

\noindent
{\bf Proof.} A `hitting set' ${\cal S}(\Gamma)$ for the collection of 
sets $\Gamma$ is a set of players such that ${\cal S} \cap \alpha_i \ne \emptyset 
~\forall~ i ~(1 \le i \le r)$. Let ${\cal M}(\Gamma)$ be the smallest
hitting set for $\Gamma$. ${\cal M}(\Gamma)$ may or may not be unique. Having found 
${\cal M}(\Gamma)$, designate its members to be q-players and all others as c-players.
This guarantees that there is at least one q-player in each $\alpha_j$ (which is
necessary in a quantum scheme).
In the first layer of QSS, a $((M,2M-1))$ majority function scheme 
is used to divide $|S\rangle$ into $2M-1$ shares. 
$M$ of these shares are encrypted using keys $K_i$ ($1 \le i \le M$) and
given one per q-member. The respective keys are shared classically using a 
$(|\alpha_j|,|\alpha_j|)$ scheme in each authorized set.
The remaining $M-1$ are retained by the dealer as resident shares. Together the players
of any $\alpha_j$ can reconstruct $K_j$. This is used to decrypt the q-share(s)
of q-player(s) in $\alpha_j$. The decrypted share,
in conjunction with the resident shares, suffices to reconstruct 
the secret. This proves that $M$ members, provided they belong to ${\cal M}(\Gamma)$,
are sufficient to effectively enact the QSS. However, no further compression is possible.
For, if we have fewer than $M$ q-players, then there is at least one $\alpha_j$
with no q-player in it, which would render reconstruction of the quantum secret impossible
for that member of $\Gamma$.
\hfill \qed
\bigskip

The signifiance of the theorem lies in saving quantum communication and
minimizing q-players, which are required to be only $M \le n$ in number.
Let us consider Theorem \ref{thm:minqp} applied to the access structure
$\Gamma = \{ABC, DE\}$, the example considered in Eq. (\ref{eq:qz0}).
The two authorized sets are disjoint, so $M = 2$. We choose ${\cal M} =
\{A, D\}$, who are the two required q-players (instead of five q-players,
required in the uncompressed version).
The first layer will employ a
$(2,3)$-QTS to split $|S\rangle$. One share is encrypted using $K_1$ and given 
to $A$, another using $K_2$ and given to $D$. $K_1$ is shared using a classical
$(3,3)$ scheme among $ABC$, and $K_2$ using a $(2,2)$ scheme among $DE$.
The remaining share remains resident at the dealer. This is depicted
in Eq. (\ref{eq:qz1}). 
The authorized set on any row suffices 
to unscramble and reconstruct one first layer share, which in conjunction with the
resident share permits reconstruction of the whole secret.
\begin{equation}
\label{eq:qz1}
((2,3)) \left\{ \begin{array}{lll}
A & \rightarrow (3,3) : & A, B, C \\
D & \rightarrow (2,2) : & D, E \\
{\rm dealer} &  &  
	   \end{array} \right. 
\end{equation}
If some $\alpha_j$'s have a common element, as for example in $\Gamma = \{ABC, DE, AFG\}$,
then some q-players are chosen to belong to more than one authorized set; in this case,
eg., ${\cal M} = \{A,D\}$ or ${\cal M} = \{A,E\}$.
If all members of $\Gamma$ happen to have one or more common players, then only
one q-share is needed, given to one of the common players, and no share assistance is required,
as in the case of $\Gamma = \{ABC, AD, AEF\}$ that we encountered earlier.
For a general access structure involving a large number of players, computing
${\cal M}$ is a provably hard problem (Infact its decision version is shown to 
be NP-Complete \cite{garey79}). 
Nevertheless, a particularly simple case is the symmetric
one of a threshold scheme.

\begin{thm}
\label{thm:thresholdsk}
An assisted $((k,n))$ scheme, with $1 \le k \le n$, can be maximally compressed to one 
requiring no more than $n-k+1$ quantum players.
\end{thm}

\noindent
{\bf Proof:} 
If $k>n/2$, we retain the scheme as such, but if $k\le n/2$,
we replace the $((k,n))$ scheme by $((k+\gamma,n+\gamma))$ scheme, where by Theorem
\ref{thm:thresk}, $\gamma=n-2k+1$ and $\gamma$ shares are resident at the dealer.
In either case, the requirement that any authorized set should have a q-share
implies that any set of $k$ players must have at least one q-player. 
This is possible only if
$M \ge n-k+1$. Minimally, ${\cal M}$ can be chosen to be any $n-k+1$ players, who are 
designated as q-players. Thus, a further $n - (n-k+1) = k-1$ shares are designated
to remain with the dealer, bringing a total of $\gamma+k-1$ resident shares. The
remaining $n-k+1$ shares are encrypted and
given one each to $n-k+1$ q-players. The encryption key is shared among the players
using a $(k,n)$ scheme. Any $k$ players will have at least one
q-share among them, which they can decrypt and transmit to the reconstructor. This will
suffice to reconstruct the secret using the $((k+\gamma,n+\gamma))$ scheme. If a set
of $k$ players has $y$ ($> 1$) q-shares among them, then reconstruction can proceed
in any of $2^y$ ways, by transmitting any subset of the $y$ shares to the reconstructor.
\hfill \qed

For example, we return to the earlier example to realize a $((2,10))$ assisted 
scheme via a $((9,17))$ scheme. The uncompressed version requires 7 resident shares
and 10 q-players. In the compressed scheme, only $10-2+1=9$ q-players are required,
with $17-9=8$ resident shares. The q-shares with the players are encrypted using
a classical $(2,10)$ scheme. Now, a $((k,2k-1))$ is incompressible without assistance
\cite{nas01}. 
Theorem \ref{thm:thresholdsk} implies that with assistance it can be 
further compressed to one
involving only $k$ q-shares (rather than $2k-1$ q-shares) among the players.
For example suppose a $((2,3))$ scheme is to be realized among players $A, B, C$.
The authorized sets are $\{AB, BC, AC\}$. Going by Theorem \ref{thm:thresholdsk},
we require only 2 q-players, i.e., any 2 players fully construct ${\cal M}$. 
Let them be $A, B$. $C$ remains a c-player. 
The dealer $D$ encrypts the secret $|S\rangle$ 
using data $K$ and then splits the encrypted secret $|S^{\prime}\rangle$
according to $((2,3))$. 
He gives one of the resulting three shares to $A$, another to $B$, retaining the
third himself. Then he split-shares $K$ according to a $(2,3)$ scheme.
Any two members can reconstruct the secret by submitting the reconstructed
$K$ and taking share assistance of one q-share from the dealer, if necessary.
No fewer than two players can reconstruct the secret.
As members of ${\cal M}$, $A$ and $B$ do not require share assistance, but the other
two combinations do. 

\section{Twin-threshold Quantum Secret Sharing Schemes}\label{sec:qts}
In a conventional or compressed $(k,n)$-QTS, the threshold $k$ applies to
all members taken together. Now suppose that we have {\em separate} thresholds
for c-members and q-members, namely $k_c$ and $k_q$, with $k=k_c+k_q$. 
We now extend the definition of a conventional QTS 
to a $(k_c,k_q,n)$ {\it quantum twin-threshold scheme} (Q2TS) 
and a $(k_c,k_q,n,\mathbb{C})$
{\it quantum twin-threshold scheme with common set} (Q2TS+C), 
where a quantum secret $|S\rangle$
is split into $n$ pieces (shares) according to some pre-agreed procedure
and distributed among $n$ players. 
These $n$ share-holders consist of members
of set $\mathbb{Q}$ of q-players
and set $\bar{\mathbb{Q}}$ of c-players. 
We denote $q \equiv |\mathbb{Q}|$, so that $|\bar{\mathbb{Q}}| = n-q$.
Obviously, in a quantum scheme, $\mathbb{Q} \ne \emptyset$.

\begin{defni}
A QSS scheme is a $(k_c,k_q,n)$ quantum twin-threshold scheme (Q2TS) 
among $n$ players, of which $q$ are q-players and the remaining are c-players,
if at least $k_c$ c-players and at least $k_q$ q-players are necessary to
reconstruct the secret.
\end{defni}

\begin{defni}
A QSS scheme is a  $(k_c,k_q,n,\mathbb{C})$ quantum 
twin-threshold scheme with common set (Q2TS+C)
among $n$ players, of which $q$ are q-players and the remaining are c-players,
if: (a) at least $k_c$ c-players and at least $k_q$ q-players are necessary to
reconstruct the secret; 
(b) All members of the set $\mathbb{C}$ are necessary to reconstruct the secret. 
\end{defni}

The idea behind distinguishing between the classical threshold $k_c$
and the quantum threshold $k_q$ is to obtain a 
simple generalization that combines the properties of the
CTS and QTS. 
Practically speaking, it is best to minimize $k_q$, at fixed $k$. 
However, one can in principle consider situations of potential use for a
twin-threshold scheme, when a sufficiently large number of members are 
able to process quantum information safely.
Further, some of the share-holders, while not entirely trust-worthy, may yet
be more trust-worthy than others. The share-dealer (say Alice) may prefer
to include all such share-holders during any reconstruction of the secret.
This is the requirement that motivates the introduction of set $\mathbb{C}$.
In general, $\mathbb{C}$ can contain members drawn from $\mathbb{Q}$
and/or $\bar{\mathbb{Q}}$ or may be a null set. 
By definition, Q2TS+C with $\mathbb{C} = \emptyset$ is Q2TS.

In the following sections we present two methods to realize
in varying degrees the generalized quantum secret splitting scheme.
The first of these is the general version of Q2TS+C. The second, while
more restricted, is interesting
because it is not directly based on quantum erasure correction, but on
information dilution via homogenization, in contrast to current proposals of QSS. 

\subsection{Quantum Error Correction and Quantum Encryption}

We now give protocols that realizes the twin-threshold scheme based on 
quantum encryption.

{\em Scheme 1.} Protocol to realize $(k_c,k_q,n)$-Q2TS.
 
{\em Distribution phase.} (1) Choose a random classical encryption $K$. Encrypt the
quantum secret $|S\rangle$ using the encryption algorithm described in Section
\ref{intro}. The encrypted state is denoted $|\tilde{S}\rangle$; (2) Using a conventional
$(k_q, q)$-QTS, split-share $|\tilde{S}\rangle$ among the members of $\mathbb{Q}$;
to not violate no-cloning, $q$ should satisfy $k_q > (q/2)$;
(3) Using a $(k_c,n-q)$-CTS, split-share $K$ among the members of $\bar{\mathbb{Q}}$.

{\em Reconstruction phase.} (1) Collect any $k_q$ q-shares from members of
$\mathbb{Q}$ and reconstruct 
$|\tilde{S}\rangle$; (2) Collect any $k_c$ shares from members of $\bar{\mathbb{Q}}$
and reconstruct $K$;
(3) Reconstruct $|S\rangle$ using $|\tilde{S}\rangle$ and $K$.

Now consider the case $\mathbb{C}
\ne \emptyset$ and the Q2TS scheme becomes the more general Q2TS+C scheme.
We now give a protocol that realizes this more general twin-threshold scheme.
We denote $\lambda_q \equiv |\mathbb{Q} \cap \mathbb{C}|$ and 
$\lambda_c \equiv |\bar{\mathbb{Q}} \cap \mathbb{C}|$. Clearly, $\lambda_c +
\lambda_q = |\mathbb{C}|$. 
If there are no q-players in $\mathbb{C}$, set $\lambda_q
= 0$, and if there are no c-players in $\mathbb{C}$, set $\lambda_c = 0$.
Note that by definition, q-players may also carry classical information, but
c-players don't carry quantum information.

{\em Scheme 2.} Protocol to realize $(k_c,k_q,n,\mathbb{C})$-Q2TS+C.
 
{\em Distribution phase.} (1) Choose a random classical encryption $K$. Encrypt the
quantum secret $|S\rangle$ using the encryption algorithm described in Section
\ref{intro}. The encrypted state is denoted $|\tilde{S}\rangle$; 
(2) Using a $(2,2)$-QTS, divide $|\tilde{S}\rangle$ into two pieces, say
$|\tilde{S}_1\rangle$ and $|\tilde{S}_2\rangle$; 
(3) Using a $(\lambda_q,\lambda_q)$-QTS, split $|\tilde{S}_1\rangle$ 
among the q-members in $\mathbb{C}$;
(4) Using a conventional
$(k_q-\lambda_q, q-\lambda_q)$-QTS, split $|\tilde{S}_2\rangle$ among the 
q-members not in $\mathbb{C}$;
to not violate no-cloning, $q$ should satisfy $(k_q-\lambda_q) > (q-\lambda_q)/2$;
(5) Using a (2,2)-CTS, divide $K$ into two shares, say $K_1$ and $K_2$;
(6) Part $K_1$ is split among the members of $\mathbb{C}$ using a
$(|\mathbb{C}|,|\mathbb{C}|)$-CTS. Alternatively, it can be split using a
$(\lambda_c,\lambda_c)$-CTS among the c-players in $\mathbb{C}$;
(7) Using a $(k_c-\lambda_c,n-q-\lambda_c)$-CTS, 
split $K_2$ among the members of $\bar{\mathbb{Q}}-\mathbb{C}$.

{\em Reconstruction phase.} 
(1) Collect all $\lambda_q$ shares from all members of $\mathbb{Q} \cap \mathbb{C}$
and reconstruct $|\tilde{S}_1\rangle$; 
(2) Collect any $k_q-\lambda_q$ q-shares from
$\mathbb{Q} -\mathbb{C}$ to reconstruct $|\tilde{S}_2\rangle$;
(3) Combining $|\tilde{S}_1\rangle$ and $|\tilde{S}_2\rangle$, reconstruct
$|\tilde{S}\rangle$;
(4) Collect all $|\mathbb{C}|$ c-shares 
from members of $\mathbb{C}$ and reconstruct $K_1$.
Alternatively, collect all $\lambda_c$ c-shares 
from members of $\bar{\mathbb{Q}} \cap \mathbb{C}$ and reconstruct $K_1$; 
(5) Collect any $k_c-\lambda_c$ shares from $\bar{\mathbb{Q}}-\mathbb{C}$
and reconstruct $K_2$; 
(6) Combining $K_1$ and $K_2$, reconstruct $K$.
(7) Reconstruct $|S\rangle$ using $|\tilde{S}\rangle$ and $K$.

\subsection{Quantum Twin-threshold Scheme Based on Information Dilution via Homogenization}
The second, more restrictive scheme, is based on the procedure for information 
dilution in a system-reservoir 
interaction, proposed by Ziman {\it et al.} \cite{zim02}. The novelty of the
scheme lies in the fact that it is not directly based on an quantum error-correction
code. However, it is applicable only to QSS with $\mathbb{C} \ne \emptyset$.
Ref. \cite{zim02} present a {\it universal quantum homogenizer}, a machine
that takes as input a system qubit initially in the state 
$\rho$ and a set of $N$ reservoir qubits initially 
prepared in the identical state $\xi$.
In the homogenizer the system qubit sequentially interacts with the
reservoir qubits via the partial swap operation.
The homogenizer realizes, in the limit sense, 
the transformation such that at the
output each qubit is in an arbitrarily small
neighborhood of the state $\xi$ irrespective of
the initial states of the system and the reservoir qubits.
Thus the information contained in the unknown system state
is distributed in the correlations amongst the system and 
the reservoir qubits.
As the authors  point out, this process can be used 
as a {\it quantum safe with a classical combination}.
Now we show how this particular feature can be turned into
a special case of the $(k_c,k_q,n,\mathbb{C})$ threshold scheme,
subject to the restriction that $\mathbb{Q} \subseteq \mathbb{C}$, so that
$k_q = q$, i.e. all q-players must be present to reconstruct the secret.

The homogenization is reversible and the original state 
of the system and the reservoir qubits can be unwound.
Perfect unwinding can be performed only when 
the system particle is correctly identified from among the 
$N+1$ output qubits, and it and the reservoir qubits
interact via the inverse of the original partial swap operation.
Therefore, in order to unwind the homogenized system, the classical 
information (denoted $K$) about the sequence of the qubit interactions 
is essential. Now, of the $(N+1)!$ possible orderings,
only one will reverse the original process.
The probability to choose the system qubit correctly is 
$1/(N+1)$. Even when the particle is choosen successfully,
there are still $N!$ different possibilities in choosing 
the sequence of interaction with the reservoir qubits.
Thus, without the knowledge of the correct ordering,
the probability of successfully unwinding the homogenization
transformation is $1/((N+1)!)$, which is exponentially small in $N$ \cite{zim02}.
So for sufficiently large value of $N$, hardly any
information about the system qubit can be deduced without this classical
information.

If $K$ is split up
among the $q$ members holding the system and
reservoir qubits according to a $(q,q)$-CTS,
it is easy to observe that this realizes a $(q,q)$-QTS 
not based directly on a quantum error-correction code. 
In terms of the generalized definition, this
corresponds to a $(k_c,k_q,n,\mathbb{C})$-scheme in which 
$k_c=0$, $\mathbb{Q} = \mathbb{C}$
and $n=k_q=q$.
The classical layer of information sharing is
necessary in order to strictly enforce the threshold: if prior ordering
information were openly available, then for example
the last $q-1$ participants could collude to obtain a state close to $\rho$.
We now present the most general twin-threshold scheme possible based
on homogenization. It will still be more restricted than that obtained via
quantum encryption, requiring that $\mathbb{Q} \subseteq \mathbb{C}$, so
that $k_q=q$. If $n$ is not
too large, it is preferable for prevention of partial information leakage
to choose the number $N$ of
reservoir qubits such that $N \gg n$. The general protocol is executed 
recursively as follows.
Alice takes $N$ ($\gg 1$) reservoir qubits, 
where $N+1 = \sum_i m_i$ and integers $m_i \ge 1$ ($\forall~ i: 1 \le i\le n$), and 
performs the process of homogenization to obtain states $\xi_0,\xi_1,
\cdots \xi_N$ on the system qubit and the $N$ reservoir qubits.

{\em Scheme 3}: Protocol to realize a a restricted $(k_c,k_q,n,\mathbb{C})$-Q2TS+C,
with $k_q=q \le n$. 

{\em Distribution phase:} (1)
Any $m_i$ qubits from $N+1$ qubits are given to the $i$th member of 
$\mathbb{Q}$; (2) $K$ is divided into two parts, $K_1$ and $K_2$, according 
to a (2,2)-CTS;
(3) Let $\lambda_c \equiv |\bar{\mathbb{Q}} ~\cap~ \mathbb{C}| \ge 0$.
$K_1$ is further split among the members of $\mathbb{Q}$ 
and $\bar{\mathbb{Q}} ~\cap~ \mathbb{C}$ using a 
$(q+\lambda_c,q+\lambda_c)$-CTS; (4) $K_2$
is split among the members of $\bar{\mathbb{Q}}-\mathbb{C}$ using a 
$(k_c-\lambda_c,n-q-\lambda_c)$-CTS.

{\em Reconstruction phase:} (1) Collect all q-shares from members of $\mathbb{Q}$;
(2) Collect all $|\mathbb{C}|$ c-shares from members of $\mathbb{C}$ and 
reconstruct $K_1$; (3) Collect any $k_c-\lambda_c$ shares from members of
$\bar{\mathbb{Q}}-\mathbb{C}$ to reconstruct $K_2$; (4) Using $K_1$ and $K_2$,
reconstruct $K$; (5) Using the q-shares and $K$,
unwind the system state to restore the secret $|S\rangle$.
\chapter{A Step Towards Combining the Features of QKD and QSS}
\section{Introduction}
Regarding the cryptographic use of multipartite entanglement,
three broad issues may be discerned. 
In the first case, the parties may choose to simply obtain random 
bits shared between pairs of the $n$ parties. 
In the second case, the entanglement may be used to generate random shared 
bits between the $n$ trustful parites. Finally,
the $n$ parites may not be entirely trusting and we may wish to use entanglement 
to obtain a secret sharing protocol. First case is essentially QKD between 
two trustful parties and the later two cases are 
respectively $n$-QKD and QSS on which we have discussed a great 
deal in the last two chapters. 
$n$-QKD involves sharing a random key amongst $n$ trustworthy parties 
where as QSS splits quantum information amongst untrusted parties. 
It would be an interesting extension to consider situations where some kind of mutual
trust may be present between sets of parties while parties being individually mistrustful.
This way it could be possible to combine the essential features of 
QKD and QSS. In this chapter, we discuss two such extensions. \\

Firstly, 
we consider the problem of secure key distribution between two trustful groups
where the invidual group members may be mistrustful. The two groups 
retrieve the secure key string, only if all members should cooperate with one
another in each group. That is, how the two groups one of 
size $k$ and the other of size $n-k$ may share
an identical secret key among themselves while an evesdropper may co-operate with 
several(of course not all) dishonest members from any of the groups.
This task is trivially a classical secret sharing scheme if we involve a trusted third party,
say, Lucy. Lucy will simply generate a random classical bit string. Since it is just
a classical information she makes two copies of it and split one each 
amongst the two groups. Principles of quantum physics allows us, as in the case of 2-QKD,
to do away with the third party. 
Adopting an idea similar to that in Chapter 4, we present a quantum key distribution protocol
for this purpose based on entanglement purification, which can be proven
secure by reducing the problem to the biparitite case using combinatorics developed 
in Chapter 2.  \\
  
We can observe that the above problem essentially seems to be a combination of 
\begin{enumerate}
\item $2$-QKD between the two groups, each group being considered as a single party and 
\item Secret sharing in each group amongst their parties.
\end{enumerate} 
In the second possible extension, we consider several such groups. Members of the same group
trust each other whereas members from different groups do not and the problem is to establish
a common shared random key amongst the $n$ untrustful parties. 
This problem, as above, could also be tackled by similar reduction to bipartite case.
We discuss a neccessary condition
for such schemes to exist and also present one such scheme. 
Another step in combining QKD and QSS could be the generalization of 
the first case with a setting just opposite 
to that in the second case. Members of the same group
do not trust each other whereas members from different groups combined together do trust
the members of other group combined together. The problem is to establish
a common shared random key amongst the different groups.     

\section{Quantum Key Distribution between Two Groups}

In this section we develop a simple protocol for QKD between two groups. 
Our protocol works in two broad steps. In
the first step, the $n$-partite problem is reduced to a two-party problem
by means of SKP-2 (Chapter 2, Protocol-II).  
This creates a pure $n$-partite maximally entangled state among $n$-parties,
starting from $n-1$ EPR pairs shared along a 
spanning EPR tree using only
${\cal O}(n)$ bits of classical communication. The SKP-2 exploits
the combinatorial arrangement of EPR pairs to simplify the task of
distributing multipartite entanglement.
In the second step, as in case of $n$-QKD, the Lo-Chau protocol \cite{lochau} or Modified
Lo-Chau protocol \cite{sp00} is invoked to prove the unconditional security
of sharing nearly perfect EPR pairs between two parties. \\

To this end, we will be using a state of the form:
\begin{equation}
\label{maxeq}
|\Psi\rangle = \frac{1}{\sqrt{2}}(|00\cdots 0\rangle + |11\cdots1\rangle),
\end{equation}
a maximally entangled $n$-partite state, represented in the computational basis.\\

Our protocol is motivated by a simple mathematical property possessed by
multi-partite states, unlike EPR pairs, which forces them to behave differently when measured 
in computational or diagonal basis. Here below we develop this mathematics. \\

Let $H$, $\bigoplus$ and $\bigotimes$ denote the Hadamard gate, the XOR operation and 
the tensor product respectively then (assume the presence of 
a proper normalizing factor in each expression),\\

$H^{\bigotimes n}$ $ |1\rangle^{\bigotimes n}$ $ = \sum_{x_1,x_2,\cdots,x_n} 
(-1)^{x_1\bigoplus x_2\bigoplus \cdots \bigoplus x_n}$ $ |x_1 x_2 \cdots x_n\rangle $

and \\

$H^{\bigotimes n}$ $ |0\rangle^{\bigotimes n}$ $ = \sum_{x_1,x_2,\cdots,x_n}$ 
$ |x_1 x_2 \cdots x_n\rangle $

therefore, \\

$H^{\bigotimes n} (|1\rangle^{\bigotimes n} + |0\rangle^{\bigotimes n} )
= \sum_{x_1\bigoplus x_2\bigoplus \cdots \bigoplus x_n = 0} 
 |x_1 x_2 \cdots x_n\rangle$ 

\bigskip

$= (\sum_{x_1\bigoplus x_2\bigoplus \cdots \bigoplus x_s = 0} |x_1 x_2 \cdots x_s\rangle )
    (\sum_{x_{s+1} \bigoplus x_{s+2} \bigoplus \cdots \bigoplus x_n = 0} |x_{s+1} x_{s+2} \cdots x_n\rangle )$

\bigskip

$ + (\sum_{x_1\bigoplus x_2\bigoplus \cdots \bigoplus x_s = 1} |x_1 x_2 \cdots x_s\rangle )
    (\sum_{x_{s+1} \bigoplus x_{s+2} \bigoplus \cdots \bigoplus x_n = 1} |x_{s+1} x_{s+2} \cdots x_n\rangle )$.

\bigskip

We can observe by symmetry that the above factoring can be infact done for any two groups of sizes
$s$ and $n-s$ respectively. \\ 

We are now ready to develop our protocol which involves the following steps:
\begin{description}
\item{(1)} EPR protocol:
Along the $n-1$ edges of the minimum spanning EPR tree, EPR pairs
are created using Lo-Chau \cite{lochau} or Modified Lo-Chau protocol
\cite{sp00} (by leaving out the final measurement step). 
This involves pairwise quantum and classical
communication between any two parties connected by an edge.
Successful completion ensures 
that each of the two parties across a given edge share a nearly perfect singlet
state $\frac{1}{\sqrt{2}}(|01\rangle - |10\rangle)$. 
At the end of the run, 
let the minimum number of EPR pairs distilled along any edge of the 
minimum spanning EPR tree be $2m$. 
\item{(2)} The $2m$ instances of the singlet state are then converted to the 
triplet state $\frac{1}{\sqrt{2}}(|00\rangle + |11\rangle)$,
by the Pauli operator $XZ$ being applied by the second party (called ${\cal Y}$) on
his qubit.
\item{(3)} For each edge, the party ${\cal Y}$ intimates the protocol leader 
(say ``Lucy") of the completion of step (2). Lucy
is the one who starts and directs the SKP protocol (Chapter 2, Protocol-II) used below. Note that Lucy can 
be from any of the two groups.
\item{(4)} SKP protocol: Using purely local operations and classical communication 
(LOCC), the $n$ parties execute the SKP protocol, which consumes the $n-1$ EPR pairs
to produce one instance of the state (\ref{maxeq}) shared amongst them.
\item{(5)} A projective measurement in the {\it diagonal} 
 basis is performed  by all the parties on their respective qubits.
\item{(6)} Lucy decides randomly a set of $m$ bits to be used as
check bits, and announces their positions.
\item{(7)} All parties from a group assist(coperate) to get one cbit corresponding to each check
 bit position by XORing their corresponding check bits. This gives an effective 
check bit corresponding to each ckeck bit position. The two group then announce the value of 
their effective check bits. If too few of
these values agree, they abort the protocol. We can note from the mathematics developed 
above that the effective check bits should agree after the diagonal basis measurement.
Effective non-check bits are also calculated as above by XORing the non-check bits of 
the goup members. 
\item{(8)} Error correction is done as in for quantum key distribution between 
two trustful parites. 
\end{description}

\paragraph{Proof of unconditional security:}
The proof of unconditional security of the above protocol is almost the same as for 
$n$-QKD (Chapter 4). However, we would like to stress the role of fault tolerant 
quantum computation and of quantum error correcting codes 
during the execution of SKP protocol for the following reason:
Suppose the probability of error on a bit is $p$. 
Then the probability of an error on the bit obtained by XORing all 
the $s$ group members' bits may be larger, given by: \\
	$P = \sum_{r=1,3,5...} C(s,r) p^r (1-p)^{s-r}$, where $C(s,r)$ is 
the number of all possible way selecting $r$ elements from a set of $s$ distinct 
elements.  \\

If $P$ is too close to $0.5$, then the effective channel capacity $Ch$ 
for the protocol (given by $Ch = 1 - H(P)$, where $H(.)$ is Shannon entropy) 
will almost vanish. Therefore, the quantum part of the protocol 
implementation should be very good to ensure that $P$ is not too close to $0.5$.\\

Of the XORed $2m$ raw bits, $m$ bits are first used for getting an estimate of $P$,
 by obtaining the Hamming distance $\delta$ between each group's $m$-bit check string. 
If they are mutually too distant, the protocol run is aborted. If $\delta$ is 
not too great (that is $2\delta + 1 <= d$), it can be corrected with a classical 
code $C(m,k,d)$, where $m$ is block length, $d$ is (minimum) code distance 
and $k/m$ is code rate. Each group decodes its XOR-ed $m$-bit string to the 
nearest codeword in $C(m,k,d)$. This $k$-bit string is guaranteed with high 
probability to be identical between the two groups. An binary enumeration 
of the $2^k$ codewords of $C(m,k,d)$ can be used as the actual key shared 
between the two groups.

\section{$n$-QKD amongst  Untrustful Parties}
We now consider the second case where there are trustful groups but different
set of these groups may not trust each other.  
For example, suppose there are $10$ parties $\{1,2,3,4,5,6,7,8,9,10\}$. 
$\{1,2,3\}$ trust each other but any of $\{1,2\}$ do not trust any of the parties
$\{4,5,6,7,8,9,10\}$. 
$\{3,4,6,7\}$ trust each other but $6$ does not trust any other party. 
$\{4,5,7,8,9,10\}$ trust each other but any of $\{5,8,9,10\}$ do not trust
any of $\{1,2,3,6\}$. We mean to say that there can be several trustful groups 
and they may have some common parties. 
Within a group every one trust each other but a party does not
trust another party who is not in any group he belongs.
In the above example the trustful groups are $\{1,2,3\}$,$\{3,4,6,7\}$ 
and $\{4,5,7,8,9,10\}$. Our aim in this case will be to
obtain an unconditionally secure QKD scheme amongst the $n$-parties.\\

Using our $n$-QKD scheme for trustful parties presented in Chapter 4 and hypergraph
combinatorics developed in Chapter 2 the above problem, 
along with necessary and sufficient conditions, can be easily 
tackled  as follows:

Let $S$ denotes the set of parties.
Then the trustful groups can be represented as subsets of $S$. Now let 
these groups are $E_1, E_2, \cdots,$ and $E_m$ and $F$
is collection of these subsets. 
This structure can be represented by a hypergraph $H=(S, F)$ and we call it a 
{\it security hypergraph} of the $n$ parites. 
This representation is inline with that of entangled hypergraph 
in Chapter 2. The necessary and sufficient condition is 
same as dictated by Theorem \ref{entanhyperiff}, that is, the QKD between the $n$
parties can be successfully done if 
and only if the security hypergraph is connected.\\ 

The protocol is to first reduce the security hypergraph to a simple { \it security 
graph} (Chapter 4) and then to apply a slight modification of the scheme  
for trustful $n$-QKD. In the reduced simple security graph only those vertices 
will survive which belongs to at least two hyperedges (that is, two trustful groups) 
and edges will be between those vertices that trust each other.

\chapter{Open Research Directions}

We conclude brifly with some open questions based on our research. \\

\begin{enumerate}

\item {\it Lower bound on classical communication complexity for preparing multi-partite 
entanglement from bi-partite entanglement under LOCC.} \\

In Chapter 2,  we observed that, 
all the schemes for creating a pure $n$-partite maximally entangled state from
the distributed network of EPR pairs amongst $n$ agents, require 
$O(n)$ cbits of communication. 
An obvious open problem is to determine whether there is an $\Omega (n)$ lower bound 
on the cbit communication complexity for preparing 
a pure $n$-partite maximally entangled state given a 
spanning EPR tree of $n$ agents. 
We hope that critaria using quantum information theory may help settle this issue.\\
  
\item {\it Partial secret sharing} \\

Our Protocol-I in Chapter 2 was motivated by the dynamic and symmetric involvement 
of all the players in the preparation of GHZ state from two EPR pairs.
We pointed out there that it should be interesting to investigate 
the ramifications of the protocol. One possible approach in this direction
could be a secret sharing scheme with relaxed conditions as discussed below.\\ 

In usual secret sharing schemes we have mainly two constraints 
as specified in their definition:

1. Any authorized set will have the complete information about the secret.
 
2. Any unauthorized set will have {\it no} information about the secret, that is,  
any unauthorized set is maximally uncertain about the secret. 

Now let us relax the second constraint. Let the unauthorized sets have partial
information about the secret but not the complete information, that is,
they may not be maximally uncertain about the secret though they 
may have partial knowledge of the secret. How much information an 
unauthorized set should have could be specified as a part of the specification 
of the scheme and could be better given in terms of Shannon's or Von-Neumann entropy 
respectively for classical and quantum case. 
For example, one way to relax the threshold scheme (k, n) may be as
follows: The ignorance about the secret wich any $m$ members will have 
is given by Shannon entropy $H(m)= 0$ for $m \geq k$ and $(k-m)/k$ for $m < k$.  
We could call such schemes to be {\it partial} secret sharing schemes (PSS). 
There may be some practical applications of such schemes especially 
in the quantum case for say multi-party secure computation.
For example, let there are two players $B$ and $C$. The dealer $A$ wants 
that $B$ and $C$ should be able to do full proof computation only when both 
of $B$ and $C$ co-operate. 
She also wants to allow a computation by $B$ up to a phase 
factor and that by $C$ up to a bit-flip factor. Such a scheme may be obtained 
by relaxing a $((2,2))$ QTS. This is where we might use the above mentioned 
symmetric teleportation circuit of Chapter 2 (Protocol-I). 
Such schemes may also be important in the 
situation wherein some members die (in computer network language these are 
called faulty nodes), we have partial information and can do  
reliable (probabilistic) computation. 
It should also be interesting to investigate the compression and 
inflation of partial secret sharing schemes and to investigate
how the relaxation constraint relates with such compression and 
inflation.\\

Quantitatively, for PSS, one would have to show that the weakening does not 
compromise the SS itself. This means that we have to obtain a lower 
bound on the number of dishonest colluders (who can sabotage the protocol 
for the remaining players) as a function of the weakening 
(quantified using some parameter). 
One could then say that the protocol is $Y$-level secure at $X$-degree weakness, 
or suchlike. \\

Another approach to implement PSS could be like our QSS scheme (Chapter 5)
based on method of information 
dilution via homogenization, where we had to scramble the ordering information 
in order to prevent partial information gain. 
Not scrambling this information 
would appear to allow partial information reconstruction in a way that seems 
compatible with PSS. 
Therefore, perhaps this method affords a possible approach to PSS.

\item {\it Distributed quantum error correction and  multi-party problems based on secure
execution of SKP protocol}\\

In Chapter 6, the protocol for 2-group QKD utilized SKP protocol
as a subroutine.
If we have fault tolerant quantum computers, then from the 
security proof of the protocol (similar as for $n$-QKD in Chapter 4), 
it can be noted that a pure $n$-partite 
maximally entangled state could be securely prepared. 
Given the slow pace in practical realization of quantum computers than that of 
quantum communication systems, it would be very 
pragmatic to do away with the requirement of the fault tolerant quantum computers
in the secure execution of SKP protocol.
An interesting way of doing this may be to do {\it quantum} error correction in 
a distributed manner while determining syndromes using local measurements. 
In the bipartite case, this has been acheived by Modified Low-Chou protocol.
In the case of multi-partite states, however it seems to be a subtle 
task to acheive and remains an unresolved problem for future research. \\
 
Nevertheless,  we could also get a simpler protocol for $n$-QKD which did not use SKP protocol 
at all, though it enjoyed the spanning tree combinatorics.
Therefore, it should also be interesting to investigate the class of
secure multi-party computational and cryptographic problems which can be or 
can not be executed without secure execution of SKP. \\

\item {\it Combining the features of QKD and QSS} \\

Chapter 6 takes the first step towards expoliting the essential features
of QKD and QSS together, however, we have only dealt with two cases, namely, 
$2$-group QKD and QKD amongst untrustful parties. It remains 
to obtain more interesting situations, such as a simple generalization of
$2$-group QKD to $n$-group QKD, which could essentially combine the
features of QKD and QSS. \\       

\item {\it Generalizing the concept of bicolored merging} \\ 

In Chapter 3, we have developed the idea of bicolored merging and
utilized it to show various results on the possible or impossible
state transformations of the multi-partite states 
represented by EPR graphs and entangled hypergraphs by utilizing only
the monotonicity postulate for appropriate entanglement measures.
However, we would also like to stress that any kind of reduction,
which leads to the violation of {\it any} of
the properties of a potential entanglement measure, 
is pertinent to show the impossibility
of many multi-partite state transformations under LOCC. 
Since the bipartite case has been extensively studied,
such reductions can potentially provide many ideas about multi-partite case by 
just exploiting the results from bipartite case. 
In particular, the definitions of EPR graphs and entangled hypergraphs
could also be suitably extended to capture more types of multi-partite pure states 
and even mixed states and a generalization of the idea of bicolored merging as a
suitable reduction for this case could also be worked out. 
It would be interesting to investigate such issues.
For example, one possible variation of bicolored merging could be 
to relax the third constraint, that is the constraint on the number 
of EPR pairs in the BCM EPR graphs. One could relax this condition such
that the BCM EPR graphs need only to be distinct and use entropic
criterion (restrictions directly put only by monotonicity postulate
may not help here) to show possibility or impossibility of various state transfomations.   

\end{enumerate}

\end{document}